\documentclass[11pt,a4paper]{article}
\pdfoutput=1
\usepackage{jheppub}
\usepackage[utf8]{inputenc}
\usepackage{color,hyperref,dsfont,slashed,multirow}

\usepackage{amssymb,amsmath,amsfonts,comment,float,latexsym,graphicx,epstopdf, float}
\graphicspath{{plots/}}
\usepackage{color}
\usepackage{fancyvrb}
\usepackage{chngcntr}
\usepackage{etoolbox}
\usepackage{booktabs}
\usepackage{bbm,epsfig}
\usepackage{soul}
\usepackage{pdfpages}
\usepackage{array}
\usepackage{lipsum}
\usepackage{amssymb}
\usepackage{graphicx}
\usepackage{hepunits}
\usepackage{slashed}
\usepackage{ccaption}
\usepackage{pdfpages}
\usepackage{scrextend}
\usepackage{hyperref}
 \usepackage{makecell}
\usepackage{cellspace} 
\usepackage[titletoc,toc,title]{appendix}

\hypersetup{bookmarks=true,unicode=true,pdftoolbar=true,pdfmenubar=true,
  pdffitwindow=false,pdfstartview={FitH},pdftitle={},
  pdfauthor={}, pdfsubject={Subject},
  pdfcreator={Creator},pdfproducer={Producer}, pdfkeywords={}{Beyond the Standard Model Physics},
  pdfnewwindow=true,colorlinks=true,
  linkcolor=blue,citecolor=magenta,filecolor=magenta,urlcolor=cyan}
 
\newcommand{\be}{\begin{equation}}
\newcommand{\ee}{\end{equation}}
\def\bsp#1\esp{\begin{split}#1\end{split}}

\newcommand\bpm{\begin{pmatrix}}
\newcommand\epm{\end{pmatrix}}

\def\sectionautorefname~#1\null{Sec.~#1\null}
\def\subsectionautorefname~#1\null{sub--Sec.~#1\null}
\def\figureautorefname~#1\null{Fig.~#1\null}
\def\tableautorefname~#1\null{Table~#1\null}
\def\equationautorefname~#1\null{Eq.~(#1)\null}

\date{\today}

\title{Radion-Higgs Mixing in 2HDMs}

\author[a]{Marco Merchand}
\author[a]{\!, Marc Sher}
\author[a]{\! and Keith Thrasher}

\emailAdd{mamerchandmedi@email.wm.edu}
\emailAdd{mtsher@wm.edu}
\emailAdd{rkthrasher@email.wm.edu}

\affiliation[a]{High Energy Theory Group, College of William and Mary,
Williamsburg, VA 23187, USA}

\abstract{
  We study the custodial Randall-Sundrum model with two Higgs doublets localized in the $\TeV $ brane. The scalar potential is CP- conserving and has a softly broken $Z_2$ symmetry. In the presence of a curvature-scalar mixing term $ \xi_{ab} \mathcal{R} \Phi_a^\dagger \Phi_b$ the radion that stabilizes the extra dimension now mixes with the two CP-even neutral scalars $h$ and $H$. A goodness of fit of the LHC data on the properties of the light Higgs is performed on the parameter space of the type-I and type-II models. LHC direct searches for heavy scalars in different decay channels can help distinguish between the radion and a heavy Higgs. The most important signatures involve the ratio of heavy scalar decays into $b$ quark pairs to those into $Z$ pairs, as well as the decay of the scalar (pseudoscalar) into a $Z$ plus a pseudoscalar  (scalar).}

\begin{document}
\maketitle


\section{Introduction}
\label{sec:intro}

The electroweak scale set by the vacuum expectation value (VEV) $v \approx 246 \ \GeV$ of the Higgs field is very sensitive to physics at high scales. This sensitivity appears in loop  corrections to the Higgs mass and is known as the hierarchy problem. Randall and Sundrum \cite{Randall:1999ee} proposed a solution to this puzzle by considering an extra dimensional model with the extra dimension being spatial in nature and compactified into a $S_1 / Z_2$ orbifold. In this model there are two 4D manifolds, called ``3-branes", separated by a distance $y_c = \pi r_c$ in the extra dimension where $r_c$ is the "radius of compactification". The brane at $y=y_c$ is called the $\TeV$-brane or  IR-brane and the brane at $y=0$ is usually called the UV- or Planck brane. A fine tuning is required between the $5D$ cosmological constant and the brane tensions in order to achieve a static flat solution which corresponds to a vanishing effective $4D$ cosmological constant. The solution to Einstein equations gives the $5D$ metric 
 \begin{equation}
 ds^2 = e^{-2 A}\eta_{\mu \nu}dx^\mu dx^\nu - dy^2,  \label{metric1}
 \end{equation} 
where $A = k  |y|$ is the warp factor and $k$ is the $AdS$ curvature scale. This solution corresponds to a slice of $AdS_5$ space between the two branes. The result of their seminal work can explain the hierarchy of scales by warping down the Planck scale  \footnote{ We use the value $M_{pl} = 10^{19}\GeV$} to the $\TeV$ scale, i.e. $M_{\TeV} = M_{Pl} e^{- k y_c}$, therefore requiring that $k y_c \approx 37$. 

In the original Randall and Sundrum (RS) model, it was assumed that the SM fields live in the visible brane,  and only gravity propagates in the bulk of the extra dimension. In Ref. \cite{Davoudiasl:1999jd} the phenomenology of the KK gravitons was studied.  Shortly after the RS model appeared, several extensions with SM fields propagating in the bulk were found. Bulk gauge bosons were first considered in \cite{Pomarol:1999ad, Davoudiasl:1999tf} where the KK mass spectrum as well as their localization were derived.  In  \cite{Huber:2000fh} a complete analysis of the Higgs mechanism for bulk gauge bosons was done for both a bulk and a brane Higgs boson. Fermions in the bulk were introduced in \cite{Grossman:1999ra}. The whole SM was placed in the bulk in \cite{Chang:1999nh}. In \cite{ Gherghetta:2000qt} bulk fields and supersymmetry were studied. Perhaps the most attractive reason to consider placing fermions in the bulk is that one can explain the mass hierarchy and flavor mixing with parameters of $\mathcal{O}(1)$ \cite{Gherghetta:2000qt, Huber:2000fh}. Several works with bulk fermions have appeared \cite{Csaki:2003sh, Gherghetta:2003hf,Iyer:2013hca, Gherghetta:2010cj, Kim:2002kkb, Huber:2000ie, Csaki:2007ns, Chacko:2014pqa}.   

 One inconvenience in RS models with gauge and matter fields propagating in the bulk are large contributions to electroweak precision observables (EWPO) \cite{Csaki:2002gy} that push the KK scale far beyond the reach of accelerators. A possible cure can be implemented by imposing a gauge $SU(2)_L\times SU(2)_R \times U(1)_{X}$ symmetry in the bulk that is spontaneously broken to provide custodial protection \cite{Agashe:2003zs} for the $S$ and $T $ parameters and this reduces the bound on the KK scale to $m_{KK} \gtrsim 3 \  \TeV$. This custodial protection also protects the $Z b \bar{b}$ vertex from large corrections \cite{Agashe:2006at}.

Scalar fluctuations in the RS  metric give rise to a massless scalar field called the radion and in order to fix the size of the extra dimension, the radion needs to have a mass. Goldberger and Wise  \cite{Goldberger:1999uk} were the first to consider a model with a scalar field propagating in the bulk of $AdS_5$ and solved for its profile functions and KK masses. Later they showed in \cite{Goldberger:1999un} that by choosing appropriate bulk and boundary potentials for the scalar one can generate an effective $4D$ potential for the radion and therefore were able to stabilize it without requiring fine tuning of the parameters. This became known as the Goldberger Wise (GW) mechanism. However in the GW mechanism they used an ansatz for the metric perturbations that do not satisfy Einstein equations and did not include the radion wavefunction and the backreaction of the metric due to the stabilizing field. In the paper of Csaki et al \cite{Csaki:2000zn} these effects were included by using the most general ansatz \cite{Charmousis:1999rg} and the superpotential method \cite{DeWolfe:1999cp} to solve for the backreaction. Then they considered the small backreaction approximation to solve for the coupled scalar-metric perturbation system and found the radion mass to be $m_{r} \sim l \ \TeV$ where $l$ parametrizes the backreaction and its value is model dependent on the specifics of the scalar VEV profile. Therefore the radion could have a mass of few hundred of $\GeV$ and is the lightest particle in the RS model.

Since the radion field emerges as the lightest new state the possibility of being experimentally accessible and its  effects on physical phenomena must be investigated.  In general, when a scalar is propagating on the brane one can include, by arguments of general covariance, in the four dimensional effective action terms involving the Ricci scalar $\mathcal{L} \supseteq M \mathcal{R}(g) \phi - \xi \mathcal{R}(g) \phi^2$. In this way a scalar can couple non-minimally to gravity. If the brane scalar is a Higgs boson, gauge invariance implies $M=0$ and from dimensional analysis one expects $\xi$ to be an $\mathcal{O}(1)$ number with unknown sign. Particular attention has been placed on the curvature-Higgs term $\mathcal{R}\  \Phi^\dagger \Phi$ since after expanding out the radion field around its VEV this term induces kinetic mixing between the radion field and the Higgs, therefore requiring a non-unitary transformation to obtain the canonically normalized degrees of freedom. After diagonalization the physical fields become mixtures of the original non-mixed radion and Higgs boson. The phenomenological consequences of a non-zero mixing $\xi \neq 0$ have been studied extensively in the literature  \cite{Giudice:2000av,Csaki:2000zn, Hewett:2002nk, Dominici:2002jv, Chakraborty:2017lxp,  Chaichian:2001rq, Datta:2003kf, deSandes:2011zs, Kubota:2012in, Desai:2013pga, Boos:2014xha, Boos:2015xma, Frank:2016oqi}

The radion interacts with matter via the trace of the energy-momentum tensor and the form of these interactions is very similar to those of the SM Higgs boson but are multiplied by $v/\Lambda$ where $\Lambda \sim \mathcal{O}(\TeV)$ is a normalization factor. In the case $\xi =0$, there is no Higgs-radion mixing and the branching ratios of the radion become very similar to those of the SM in the heavy mass region, being dominated by vector bosons while for the low mass region the $gg$ mode is dominant. Due to its large, anomaly induced, coupling to two gluons a radion can be produced through gluon fusion.

 The parameter space coming from the curvature-Higgs mixing scenario consists of four parameters, viz., the bare mass terms $m_h$ and $m_r$, the mixing parameter $\xi$ and the normalization scale $\Lambda$. However in some of the above references, the Higgs boson had been discovered \cite{Aad:2012tfa, Chatrchyan:2012xdj}  and their parameter space is reduced to $(m_r, \xi, \Lambda)$. The $\xi-m_r$ parameter space is very constrained by direct searches for additional scalars at the LHC \cite{Frank:2016oqi} leaving only small experimentally and theoretically allowed windows for $\Lambda = 3 \ \TeV$ and these windows open up as one increases $\Lambda$. The bounds on the parameter $\Lambda$ are dependent the mass the first KK excitation $m_{KK}$ and the curvature scale $k$ as was shown in \cite{Ahmed:2015uqt}.

Despite the model differences in the  analyses that have appeared on Higgs-radion mixing, the overall conclusion is that there is possibility that the measured Higgs boson could  be in fact a mixture of the radion with the Higgs doublet that is consistent with experimental data. However the constraints mentioned in the previous paragraph will be pushed further if a radion signal is not seen in the coming future  and it would be interesting  to look at possible ways to relax these constraints.

In addition to the RS model, several Beyond the Standard Model (BSM) scenarios have appeared in the last several decades as promising candidates for new physics. One of the most studied and simplest extensions is the Two-Higgs-Doublet Model (2HDM) where a second Higgs doublet is added to the electroweak sector. The 2HDM was primarily motivated by minimal supersymmetry \cite{Martin:1997ns} and it has also been studied in the context of axion models \cite{Kim:1986ax}, the baryon asymmetry of the universe  \cite{Branco:2011iw, Dorsch:2014qja}, the muon $g-2$ anomaly \cite{Broggio:2014mna} and dark matter \cite{Berlin:2015wwa}.
 
In this work we will study how some of the constraints on the minimal Higgs-radion mixing may be relaxed or modified by having curvature scalar couplings of the form $\mathcal{L} \supseteq  \xi_{ab} \mathcal{R}(g_{ind}) \ \Phi_a^\dagger \Phi_b$ where $a,b = 1,2$ and a 2HDM is located on the $\TeV$ brane. The SM gauge bosons and fermions correspond to the zero modes of 5D bulk fields. In section \ref{1ww} we introduce some notation and we briefly describe the custodial RS model in section \ref{RS}.   A review of the radion field emergence in the RS model together with its interactions with SM particles is done in section \ref{s1}. The 2HDM is presented in subsection \ref{1wwA}.  The two-Higgs-radion mixing Lagrangian is discussed in section \ref{Mixing}. In section \ref{constraints} the predictions of the model are presented including constraints from electroweak precision data, from LHC data, collider signals and constraints and expectations from heavy Higgs searches. A summary of the interactions of the Higgs eigenstates and the radion with SM particles before mixing is given in appendix  \ref{interactions}, LHC data used in the fits is in appendix  \ref{appendixB} and some formulae involving electroweak precision observables are in appendix  \ref{STapp}.

\section{ Model Description }   \label{1ww}

\subsection{ The Custodial RS Model}  \label{RS}

We first review the RS model with a custodial \cite{Agashe:2003zs} gauge symmetry $  SU(2)_L \times SU(2)_R \times U(1)_X \times P_{LR}$ in the bulk where $P_{LR}$ is a parity symmetry that makes left and right gauge groups equal to each other. 
 In our notation Latin letters denote  $5D$ indices $M= (\mu , 5)$ and Greek letters denote $4D$ indices $\mu = 0,1,2,3$.  
The background metric is that of equation \eqref{metric1} and we use the convention for the flat space Minkowski tensor $\eta_{\mu \nu} = diag(+1,-1,-1,-1)$. We will introduce fluctuations around the background later. 
The $5D$ action of the model is given by 
\begin{align}
S = & \int d^5x \sqrt{g} \left[- 2 M^3 \mathcal{R}(g)  +  \mathcal{L}_{\phi} + \mathcal{L}_{gauge} +\mathcal{L}_{fermion} \right] \nonumber \\ 
   &   + \int d^4x \sqrt{g_{ind }(y=y_c)} \left[ \mathcal{L}_{H} +\mathcal{L}_{Y}  - V_{IR}(\phi) \right]  -  \int d^4x \sqrt{g_{ind }(y=0)}   V_{UV}(\phi)  
 \end{align}
where the first term corresponds to the Einstein-Hilbert action where $M$ is the $5D$ Planck scale and $\mathcal{R}$ the Ricci scalar and $\mathcal{L}_{\text{Y}}$ and $\mathcal{L}_{\text{H}}$ are the SM Yukawa and Higgs Lagrangians respectively. The stabilization mechanism is contained in $\mathcal{L}_{\phi}$ together with its brane potentials $V_{IR}$ and $V_{UV}$. We do not discuss this sector and simply assume that stabilization is performed as in  \cite{Csaki:2000zn}. The gauge sector is given by 
\begin{equation}
\mathcal{L}_{gauge} = - g^{MO}g^{NP} \left[ \frac{1}{2}\mathrm{Tr}\{L_{MN}L_{OP} \}+ \frac{1}{2}\mathrm{Tr}\{R_{MN}R_{OP}\} + \frac{1}{4}X_{MN} X_{OP}  \right]
\end{equation}
where  $L_{MN}$, $R_{MN}$ and $X_{MN}$ are the gauge bosons associated with $SU(2)_L$, $SU(2)_R$ and $U(1)_X$ respectively. 
In the Planck-brane the symmetry is broken $ SU(2)_R \times U(1)_X \rightarrow U(1)_Y$ by appropriate BC's of the gauge fields to generate the SM gauge group. This BC's are given by \cite{Delgado:2007ne}
\begin{align}
\partial_5 L_\mu^a(x,0) =  & 0 ,\quad a=1,2,3, \nonumber \\
R_\mu^i(x,0) = 0&  \quad  i=1,2 \nonumber \\
g_X \partial_5 R_\mu^3 (x,0) +  g_R \partial_5 X_\mu(x,0)= 0  \nonumber \\
-g_R R_\mu^3 (x,0) + g_X X_\mu (x,0)=0 
\end{align}
where $g_L$, $g_R$ and $g_X$ are the $5D$ gauge couplings associated with the gauge fields $L_\mu^a$, $R_\mu^a$ and $X_\mu$ respectively. The SM gauge bosons $W^\pm$, $Z$ and the photon are embedded into the $5D$ gauge bosons. Calculation of the spectrum and profiles was performed in Ref. \cite{Delgado:2007ne, Casagrande:2010si} with different KK basis.

Boundary mass terms are generated by the Higgs VEV's
\begin{equation}
\mathcal{L}_{mass} = \frac{v_1^2 + v_2^2}{8}(g_L L_\mu^a - g_R R_\mu^a)^2  \delta(y-y_c),
\end{equation} 
where $v_1$ and $v_2$ are the vevs of the Higgs doublets. Therefore in the $\TeV$ brane the gauge symmetry is spontaneously broken down by the Higgs VEV's to the diagonal group, i.e.  $ SU(2)_L \times SU(2)_R \rightarrow SU(2)_V$ so that $SU(2)_V$ generates custodial protection for the $T $ parameter. The extra parity symmetry$P_{LR}: SU(2)_L \leftrightarrow SU(2)_R$  was introduced to protect the $Zb_L\bar{b}_L$ vertex from non universal corrections \cite{Agashe:2006at}. 

In the fermion sector all three generations are embedded in the same representation of the gauge group with the following transformation properties \cite{Casagrande:2010si, Albrecht:2009xr}
\begin{equation}
Q_L \sim (\textbf{2},\textbf{2})_{\textbf{2/3}},
\end{equation}
\begin{equation}
u_R \sim (\textbf{1},\textbf{1})_{\textbf{2/3}},
\end{equation}
\begin{equation}
d_R \sim (\textbf{1},\textbf{3})_{\textbf{2/3}} \oplus  (\textbf{3},\textbf{1})_{\textbf{2/3}},
\end{equation}
and this choice guarantees custodial protection for the $Zbb$ coupling and for flavor violating couplings $Zd_L^i d_L^j $ as well. Using appropriate BC one can ensure that only the SM quarks appear in the low energy theory. 

The motivation for the custodial symmetry came from requiring corrections to EWPO, parametrized by the Peskin-Takeuchi parameters $S$ and $T$, be sufficiently small.  The corrections have contributions from the KK excitations of the fermions and gauge bosons, from the 2HDM sector and from the radion.  As discussed in the introduction, an extended gauge custodial symmetry in the bulk keeps the corrections from the KK excitations under control \cite{Agashe:2003zs}.
In the absence of mixing, a custodially symmetric 2HDM potential has vanishing contributions to the T parameter \cite{Haber:2010bw} and the contributions of the radion are also small (see Csaki et al. \cite{Csaki:2000zn}).      However when one includes mixing, the radion and Higgs scalar couplings are modified and could result in large corrections depending on the values of the mixing parameters and masses.    The contributions in this model are discussed in Section 4.

  \subsection{The Two-Higgs Doublet Model} \label{1wwA}
    
 In this work we consider two Higgs doublets living in the visible brane. The most general parametrization  for the scalar potential \cite{Branco:2011iw,  Bhattacharyya:2015nca} is given by
  \begin{align}
V(\Phi_1 , \Phi_2) = & \bar{m}_{11}^2 \Phi_1^\dagger \Phi_1 + \bar{m}_{22}^2 \Phi_2^\dagger \Phi_2- \left( \bar{m}_{12}^2 \Phi_1^\dagger \Phi_2 + H.c. \right)  \nonumber \\
 &+ \frac{\lambda_1}{2}  (\Phi_1^\dagger \Phi_1)^2 + \frac{ \lambda_2}{2}  (\Phi_2^\dagger \Phi_2)^2  +\lambda_3 (\Phi_1^\dagger \Phi_1)(\Phi_2^\dagger \Phi_2) + \lambda_4 (\Phi_1^\dagger \Phi_2)(\Phi_2^\dagger \Phi_1)    \nonumber \\
& + \left[   \frac{\lambda_5}{2} (\Phi_1^\dagger \Phi_2)^2 + \lambda_6 (\Phi_1^\dagger \Phi_1)(\Phi_1^\dagger \Phi_2) + \lambda_7 (\Phi_2^\dagger \Phi_2)(\Phi_1^\dagger \Phi_2) + H.c.   \right],   \label{pot1}
 \end{align}
 where $m_{11}^2$, $m_{22}^2$, and $\lambda_{1,2,3,4}$ are real by hermiticity and $m_{12}^2$ and $\lambda_{5,6,7}$ are in general complex. In this expression there are fourteen parameters,  however the freedom in the choice of basis can be used to reduce this number down to eleven degrees of freedom that are physical.  

 To provide custodial protection for the $T$ parameter we promote the Higgs fields to bi-doublets $M_i = (\tilde{\Phi}_i, \Phi_i)$ (with $\tilde{\Phi}_i = i \sigma^2 \Phi_i^*$) of the gauge  group  $SU(2)_L \times SU(2)_R$ that transform in the representation $(\textbf{2},\bar{\textbf{2}})_0$ \cite{Pomarol:1993mu}
 \begin{equation}
M_i  \rightarrow U_L M_i U_R^\dagger,   \quad  i = 1,2.
\end{equation} 
where 
\begin{equation}
U_L \in SU(2)_{L}, \quad  U_R \in SU(2)_{R}.
\end{equation}
Using the three independent invariant quadratic forms $\mathrm{Tr}[M_1^\dagger M_1]$, $\mathrm{Tr}[M_2^\dagger M_2]$ and  $\mathrm{Tr}[M_1^\dagger M_2]$\footnote{For a basis independent treatment see Ref.  \cite{Haber:2010bw}} the most general expression that has all possible combinations of traces invariants is given by 
 \begin{align}
 V(M_1 M_2) = & \frac{\bar{m}_{11}^2}{2} \mathrm{Tr}[M_1^\dagger M_1] +\frac{\bar{m}_{22}^2}{2} \mathrm{Tr}[M_2^\dagger M_2] - \bar{m}_{12}^2  \mathrm{Tr}[M_1^\dagger M_2] + \frac{\lambda_{1}}{8} \mathrm{Tr}[M_1^\dagger M_1]^2    \nonumber \\
  &+ \frac{\lambda_{2}}{8} \mathrm{Tr}[M_2^\dagger M_2]^2 + \frac{\lambda_{3}}{4}\mathrm{Tr}[M_1^\dagger M_1]\mathrm{Tr}[M_2^\dagger M_2] +  \frac{ \lambda'_4}{2} \mathrm{Tr}[M_1^\dagger M_2]^2  \nonumber \\
  &+ \frac{\lambda'_{5}}{2} \mathrm{Tr}[M_1^\dagger M_1] \mathrm{Tr}[M_1^\dagger M_2]  + \frac{\lambda'_6}{2} \mathrm{Tr}[M_2^\dagger M_2] \mathrm{Tr}[M_1^\dagger M_2]    \label{pot2} 
 \end{align}
where all the parameters are real and the correspondence with the potential of equation \eqref{pot1} is 
 \begin{equation}
 \lambda'_4 \equiv \lambda_4 = \lambda_5, \quad \lambda'_5 \equiv \lambda_6, \quad  \lambda'_6 \equiv \lambda_7.
 \end{equation}
Thus by imposing the gauge $SU(2)_L \times SU(2)_R$ symmetry one immediately reduces the number of free parameters in the scalar potential down to nine. Also a custodially protected 2HDM potential is automatically CP conserving.

  The kinetic terms for the Higgs bi-doublets are given by  
   \begin{equation}
 \mathcal{L}_{H} \supseteq \sum_{i=1,2} g_{ind}^{\mu \nu}\frac{1}{2} \mathrm{Tr}[(D_\mu M_i)^\dagger D_\nu M_i ]  \label{kinetic term}
 \end{equation}
 where $g_{ind}^{\mu \nu}$ is the induced metric on the $\TeV$ brane and the covariant derivative is
 \begin{equation}
 D_\mu M_i = \partial_\mu M_i - i g_L \textbf{L}_{\mu } M_i + i g_R M_i \textbf{R}_{\mu }  
 \end{equation}
 and  $\textbf{L}_{\mu} = L^a_{\mu} \tau^a_{L} $ is the gauge boson associated with $SU(2)_{L}$. Therefore under the custodial gauge symmetry the gauge bosons transform as 
 \begin{equation}
 \textbf{L}_{\mu } \rightarrow U_L \textbf{L}_{\mu }U_L^\dagger  - \frac{i}{g_L} \partial_{\mu} U_L U_L^\dagger,
 \end{equation}
 \begin{equation}
 \textbf{R}_{\mu } \rightarrow U_R \textbf{R}_{\mu }U_R^\dagger  + \frac{i}{g_R} U_R  \partial_{\mu} U_R^\dagger.
 \end{equation}
 Of course one needs to also include the term corresponding to the gauge group $U(1)_X$ which violates the custodial symmetry.
 
In conventional 2HDM's one can avoid the presence of potentially dangerous flavor changing neutral currents (FCNC) by imposing a discrete $Z_2$ symmetry $\Phi_1 \rightarrow   \Phi_1 , \  \Phi_2 \rightarrow -\Phi_2,  
$ on the Higgs doublets.  The fermion mass in \eqref{fc} is generated either by $\Phi_1$ or $\Phi_2$ since the discrete $Z_2$ symmetry is extended to the fermion sector. This results in four different types of Yukawa interactions \cite{Aoki:2009ha}. In the type-I model all fermions couple to a single Higgs doublet, usually chosen to be $\Phi_2$. In the type-II model up-type quarks couple to $\Phi_2$ and d-type quarks and leptons couple to $\Phi_1$. In the lepton-specific model all leptons couple to $\Phi_1$ and all quarks couple to $\Phi_2$. Finally in the flipped model up-type quarks and leptons couple to $\Phi_2$ and d-type quarks couple to $\Phi_1$. In general, radion mediated FCNC can be present and this was analyzed in \cite{Azatov:2008vm}. For simplicity we don't consider flavor mixing in the bulk mass parameters, i.e., $c_{L,R}^{i,j} =c_{L,R}^{i,i} $ since we want to achieve minimal flavor violation \cite{Glashow-Weinberg} in the Yukawa sector.

In terms of bi-doublets this symmetry reads
\begin{equation}
M_1 \rightarrow M_1, \quad M_2 \rightarrow -M_2,
\end{equation}
and implies $\lambda'_5 = \lambda'_6 =0$ with $\bar{m}^2_{12} \neq 0$ remaining as a soft-violating term. 
 The Higgs doublets can be expressed as 
\begin{equation}
\Phi_a =\begin{pmatrix}
 \phi_a^+ \\ \frac{\bar{v}_a + \rho_a + i \eta_a}{\sqrt{2}}
\end{pmatrix},  \quad a=1,2  \label{doublets}
\end{equation} 
where $\bar{v}_a$ are the VEV of the scalars. The VEV's satisfy the relation $\bar{v}^2 = \bar{v}_1^2+ \bar{v}_2^2$ with $\bar{v}$ the localized Higgs VEV and should not be confused with the SM value $v = \bar{v} e^{- k y_c} = 246 \ \GeV$ since we still need to canonically normalize the Higgs doublets\footnote{We put a bar on mass parameters that are not yet redshifted down to the EW scale.}.

The fields appearing in the expression of the Higgs doublets \eqref{doublets} are not the physical scalars. To obtain the physical eigenstates one has to diagonalize the mass matrices that are constructed using equation \eqref{pot2} with the appropriate imposed symmetries. For a custodial and $Z_2$ symmetric scalar potential the mass matrix for the CP-odd state and for the charged Higgs fields are equal
\begin{equation}
  \begin{pmatrix}
\bar{m}_{11}^2 + \frac{\bar{v}_1^2 \lambda_1 + \bar{v}_2^2 \lambda_3 }{2} & - \bar{m}_{12}^2 + \bar{v}_1 \bar{v}_2 \lambda'_4   \\
- \bar{m}_{12}^2 + \bar{v}_1 \bar{v}_2 \lambda'_4   & \bar{m}_{22}^2 + \frac{\bar{v}_2^2 \lambda_2 + \bar{v}_1^2 \lambda_3 }{2} 
\end{pmatrix} =  \begin{pmatrix}
\bar{m}_{12}^2\frac{\bar{v}_2}{\bar{v}_1} - \lambda'_4 \bar{v}_2^2   & - \bar{m}_{12}^2 + \bar{v}_1 \bar{v}_2 \lambda'_4   \\
- \bar{m}_{12}^2 + \bar{v}_1 \bar{v}_2 \lambda'_4   & \bar{m}_{12}^2\frac{\bar{v}_1}{\bar{v}_2}  - \lambda'_4 \bar{v}_1^2 
\end{pmatrix} 
\end{equation}
where in the last equality $\bar{m}_{11}^2$ and $\bar{m}_{22}^2$ were eliminated using the minimization conditions of the potential. The matrix above has a zero eigenvalue corresponding to the Goldstone bosons $G^0$ and $G^{\pm}$ and the nonzero mass eigenvalue is given by 
\begin{equation}
\bar{m}_A^2 = \bar{m}^2_{H^{\pm}} = \bar{m}_{12}^2 \frac{\bar{v}^2}{\bar{v}_1 \bar{v}_2} - \lambda'_4 \bar{v}^2.
\end{equation}
The fact that the CP-odd field mass is degenerate with the charged Higgs bosons is a direct consequence of imposing a custodial symmetry in the scalar potential however this symmetry is not respected by the hypercharge gauge and Yukawa interactions, so we can only expect the masses to be approximately degenerate.
The diagonalization of the CP odd fields (as well as the charged scalars) is carried out by the orthogonal transformation 
\begin{equation}
\begin{pmatrix}
\eta_1 \\
\eta_2
\end{pmatrix} = \begin{pmatrix}
c_\beta & -s_\beta  \\
s_\beta  &   c_\beta
\end{pmatrix} \begin{pmatrix}
G^0 \\
A
\end{pmatrix}
\end{equation}
where $c_\beta=\cos \beta$, $s_\beta = \sin \beta$ and $\tan \beta = v_2/v_1$. $G^0$ is the neutral Goldstone boson and $A$ is the physical pseudoscalar.

The physical CP even scalars are obtained by the rotation
\begin{equation}
\begin{pmatrix}
\rho_1 \\
\rho_2 
\end{pmatrix} = \begin{pmatrix}
c_\alpha & -s_\alpha  \\
s_\alpha & c_\alpha
\end{pmatrix}\begin{pmatrix}
H \\
h
\end{pmatrix}  \label{scalars}
\end{equation}
where $h (H)$ corresponds to the lighter (heavier) scalar.

 Notice that there were $7$ real parameters in the Higgs potential to start with, namely $\{ \bar{m}_{11}^2, \bar{m}_{22}^2, \bar{m}_{12}^2, \lambda_1',  \lambda_2',  \lambda_3',  \lambda_4' \}$.  Using the two minimization conditions we can trade $\bar{m}_{11}^2$ and $\bar{m}_{22}^2$ for $v_1$ and $v_2$ and then use the relations $v^2=v_1^2+v_2^2$ and $\tan\beta=v_2/v_1$ to trade $v_1$ and $v_2$ for $v$ and $\beta$. Finally we can trade the soft breaking parameter and three lambdas for the three scalar masses and $\alpha$ ending up with the set $\{ \beta, \alpha, m_h,  m_H, m_A,  \lambda_4 \}$ (notice that $\lambda_4 = \lambda_4'$) where we fixed $v=246 \ \GeV$ therefore we only have to specify $6$ parameters.
 
   \begin{center}
\begin{table}[h]
\renewcommand*{\arraystretch}{0.6}
\scalebox{1}{
\begin{tabular}{ | c | c | c | c | c | c | c |c |c |c | }
  \hline			
               & $\xi_h^u$ &$  \xi_h^d$ &  $\xi_h^l$ & $ \xi_H^u$ & $ \xi_H^d$&  $\xi_H^l$ & $ \xi_A^u$& $ \xi_A^d $&  $\xi_A^l$  \\
   \hline
  Type-I & $c_\alpha/s_\beta$ & $c_\alpha/s_\beta$ & $c_\alpha/s_\beta$ & $s_\alpha/s_\beta$& $s_\alpha/s_\beta$ & $s_\alpha/s_\beta$ &$\cot\beta$ &$-\cot\beta$ & -$\cot\beta$\\ \hline
  Type-II & $c_\alpha/s_\beta$ & $-s_\alpha/c_\beta$ & $-s_\alpha/c_\beta$ &  $s_\alpha/s_\beta$ &$c_\alpha/c_\beta$& $c_\alpha/c_\beta$  & $\cot\beta$ & $\tan\beta$ & $\tan\beta$ \\
  \hline  
\end{tabular}}  \caption{Scalar couplings to pairs of fermions.} \label{fermioncouplings}
\end{table} 
\end{center}

 The  couplings of the scalars with the fermion fields can be written as \cite{Aoki:2009ha}
\begin{align}
\mathcal{L}_\phi^{ff} = & \sum_{f=u,d,l} \frac{m_f}{v} \left( \xi_h^f \bar{f}f h + \xi^f_H \bar{f}f H - i \xi_A^f \bar{f}\gamma_5 f A \right),  \nonumber \\
 &- \left\lbrace \frac{\sqrt{2}V_{ud}}{v} \bar{u} (m_u \xi_A^u P_L + m_d \xi_A^d P_R)d H^+ + \frac{\sqrt{2}m_l \xi_A^l}{v} \bar{\nu}_L l_R H^+ + h.c. \right\rbrace,
\end{align}
where the mixing factors are summarized in Table \ref{fermioncouplings}. Here the gauge bosons and fermions are the zero modes of the 5D bulk fields. Non-zero KK modes are presumed to be sufficiently heavy that they will not have a phenomenological impact.

The couplings of the scalars to a pair of gauge bosons are given by
\begin{equation}
\mathcal{L}_\phi^{WW,ZZ} = \left( h \sin{(\beta - \alpha)} +  H \cos{(\beta - \alpha)} \right) \left( \frac{2 m_W^2}{v}W_\mu^+ W^{\mu -}+\frac{m_Z^2}{v}Z_\mu Z^\mu  \right),
\end{equation}
\begin{equation}
\mathcal{L}_{\phi}^{ gg, \gamma \gamma} =  \sum_{\phi=h,H,A}- \frac{\phi}{4v} \left\lbrace \frac{\alpha_s}{2 \pi}   b^\phi_{QCD} G_{\mu \nu}^a G^{a\mu \nu}  +\frac{\alpha_{EM}}{2\pi} b^\phi_{EM}F_{\mu \nu}F^{\mu \nu}  \right\rbrace,    \label{loop}
 \end{equation}
where 
\begin{equation}
b^\phi_{QCD} = \xi_\phi^t \times \begin{cases}
F_f, \quad \phi = h,H, \\
f(\tau_t) \tau_t, \quad \phi = A,
\end{cases}
\end{equation}
 
\begin{equation}
b_{EM}^h=\left( \frac{8}{3}\ \xi_h^t F_f - \sin(\beta-\alpha)F_W  +  g_{h}F_H \right),
\end{equation}

\begin{equation}
b^H_{EM}= \left( \frac{8}{3}\ \xi_H^t F_f - \cos(\beta-\alpha)F_W  +  g_{H}F_H \right),
\end{equation}
\begin{equation}
b^A_{EM} = \frac{8}{3} \xi_A^t f(\tau_t)\tau_t,
\end{equation}

The form factor for the charged Higgs in the loop is \cite{Posch:2010hx, Akeroyd:2007yh} $F_H = -\tau_H \left( 1- \tau_H f(\tau_H)\right)$ and has limiting behaviors $F_H \rightarrow 1/3$ for $\tau >1$ and $F_H \rightarrow 0$ for $\tau <1$. The couplings multiplying the form factor are given by 
$g_{\phi} = - \frac{m_W}{g m^2_{H^{\pm}}} g_{\phi H^+ H^-}$ with $g_{\phi H^+ H^-}$ the tree level coupling that arises from the 2HDM potential.


 \subsection{ The Radion Field}  \label{s1}
For the background metric solution in the RS model, given by equation \eqref{metric1}, any value of the radius dimension $y_c$ is equally acceptable. Therefore a mechanism is needed to fix the value $y_c \sim 37/k$ so that the EW hierarchy is explained and this must be accomplished without severe fine tuning of parameters.  Here we simply assume that a GW bulk scalar is responsible for the stabilization and that the bulk and brane potentials are chosen by applying the method of the superpotential of Ref.\cite{DeWolfe:1999cp}. This method has the advantage of reducing the coupled non-linear second order Einstein equations to simple ordinary differential equations for a simple choice of superpotential. The backreaction of the background metric due to the scalar can be solved directly using this method. 

After the extra dimension is stabilized the radion field arises from the scalar fluctuations of the metric given  by the general ansatz \cite{Csaki:2000zn,Charmousis:1999rg}
\begin{equation}
ds^2 = e^{-2 A - 2F(x,y)}\eta_{\mu \nu}dx^\mu dx^\nu - (1+G(x,y))^2  dy^2,
\end{equation}
and since the background VEV for the bulk scalar also depends on the extra dimension one also has to include the fluctuations in the GW scalar namely: $\phi(x,y )=  \phi_0(y) + \varphi(x,y)$ where $\phi_0 $ is the background VEV and $\varphi$ denotes the fluctuation. By evaluating the linearized Einstein equations one is able to derive $G=2F$. To solve the system one linearizes the Einstein and scalar field equations to obtain coupled relations for $\varphi$ and $F$. In particular, by integrating the $(\mu,5)$ component of the linearized Einstein equations $\delta R_{\mu 5} = \kappa^2 \delta T_{\mu 5}$ with $\kappa^2 =1/2M^3$, one obtains 
\begin{equation}
\phi'_0 \varphi = \frac{3}{\kappa^2} (F' - 2A' F)   \label{fluctuations}
\end{equation}
where the prime indicates $d/dy$ and this equation implies that the fluctuations $\varphi$ and $F$ will have the same KK eigenstates but with different profiles. Using the Einstein equations together with \eqref{fluctuations} a single differential equation in the bulk for $F$ can be obtained  \cite{Csaki:2000zn}:
\begin{equation}
F'' - 2A' F' -4 A'' F - 2 \frac{\phi''_0}{\phi'_0}F' + 4 A' \frac{\phi''_0}{\phi'_0} F = e^{2 A }\Box F
\end{equation}
supplemented by the boundary conditions 
 \begin{equation}
 (F' - 2A' F)|_{y=0,y_c} =0, 
 \end{equation}
 where the boundary conditions are simplified in the limit of stiff boundary potentials of the bulk stabilizer $\partial^2 V_i/\partial \phi^2 \gg 1$ implying  $\varphi|_{y=y_i} =0$. In the system there are two integration constants and one mass eigenvalue $\Box F_n(x,y) = - m_n^2 F_n(x,y)$. One integration constant corresponds to an overall normalization while the other constant and the mass eigenvalue are determined by the boundary conditions. In Ref \cite{Csaki:2000zn}  this differential equation was solved in a perturbative approach in the limit of small backreaction of the metric due to the stabilizing scalar, and it was found to zero-order in the backreaction that the KK zero-mode can be approximated by 
\begin{equation}
F_0(x,y) \approx e^{2k|y|}R(x) + \mathcal{O}(l^2),
\end{equation}
 where $R(x)$ is the radion field. Using the boundary conditions the radion mass is \cite{Csaki:2000zn}
 \begin{equation}
 m_r \approx  0.1 \ l \ k e^{-k y_c} 
 \end{equation}
 where $l^2 \equiv \phi_P^2/4M^3$ is the backreaction and $\phi_P$ is the VEV of the bulk stabilizer field on the Planck brane. It should be noted that generically, the radion mass is always proportional to the backreaction independently of the stabilization mechanism. 
From the expression above, the radion mass is expected to be of  $\mathcal{O}(\TeV)$ scale. The canonical normalization of the radion comes from integrating out the extra dimension in the Einstein-Hilbert action
 \begin{equation}
 M^3 \int dy \sqrt{g} \mathcal{R}(\bar{g})  \supseteq \frac{6 M^3}{k}e^{2 k y_c} (\partial_\mu R(x))^2
 \end{equation}
 therefore a canonically normalized radion is obtained by writing 
 \begin{equation}
 R(x)=r(x)\frac{e^{-k  y_c}}{\sqrt{6}M_{Pl}}.
 \end{equation}
 It is explicitly proved in \cite{Csaki:2000zn} that the normalization is dominated by the gravitational contribution coming from the Einstein-Hilbert action against that coming from the kinetic term of the bulk stabilizer.

We now proceed to present the radion interactions with the SM fields. The induced metric on the $\TeV$ brane is given by
\begin{equation}
\bar{g}^{ind}_{ \mu \nu}(x) = e^{-2A(y_c)} e^{-2e^{2k y_c}R(x)}\eta_{\mu \nu} \equiv e^{-2k y_c} \Omega(r)^2 \eta_{\mu \nu},
\end{equation}
where we use $\bar{g}_{MN} $ to denote the metric with scalar perturbations included. After rescaling of the doublets $\Phi_a \rightarrow e^{ k y_c} \Phi_a$, the radion couplings to the Higgs sector are obtained from (including the possibility of adding extra scalars in the sum)
  \begin{equation}
 S_{H}   = \int d^4x   \left[ \sum_{a=1,2} \eta^{\mu \nu}\frac{1}{2} \mathrm{Tr}[(D_\mu M_a)^\dagger D_\nu M_a ] \Omega(r)^2  - V(M_1, M_2)\Omega(r)^4   \right],
 \end{equation}
 and all mass terms are redshifted accordingly. Expanding to linear order in the radion field $\Omega(r)\approx 1-r \frac{\gamma}{v} ,$ with $\gamma \equiv v/\Lambda$ and $\Lambda \equiv \sqrt{6}M_{Pl}e^{-k y_c}$, a straightforward calculation yields the coupling of the radion with the trace of the energy-momentum tensor 
\begin{equation}
\frac{\gamma }{v} r \ T^\mu_\mu \supseteq -  \sum  \frac{\gamma }{v} r  \left[ (\partial_\mu \phi)^2 - 2 m_\phi^2 \phi^2 \right], \label{trace}
\end{equation}
 with the sum performed over all physical scalars.

The couplings to the EW gauge sector are obtained from the kinetic terms of the Higgs doublets expanding to linear order in the perturbations 
\begin{equation}
S_{H}   \supseteq - \int d^4 x\  \frac{\gamma}{v} r(x) \ \eta^{\mu \nu}  \left\lbrace  2 m_W^2   W_\mu^{(0)+}(x) W_\nu^{(0)-}(x) + m_Z^2   Z^{(0)}_\mu(x) Z^{(0)}_\nu(x)  + ...   \right\rbrace
\end{equation}
where the dots represent higher KK excitations. In addition to the boundary terms there are tree level couplings of the radion coming from the kinetic term of the bulk gauge bosons \cite{Csaki:2007ns} 
\begin{equation}
S_{gauge} \supseteq - \int d^4x \frac{\gamma}{v} r(x) \left\lbrace  \frac{1}{ky_c} \frac{1}{4 } \eta^{\mu \nu} \eta^{\alpha \beta} V_{\mu \alpha}^{(0)}(x) V_{\nu \beta}^{(0)}(x)    +    \frac{m_V^4}{2 k^2}e^{2  k y_c} k y_c \eta^{\mu \nu} V_\mu^{(0)}(x) V_\nu^{(0)}(x)  \right\rbrace.
\end{equation}
where $V_{MN} = \partial_M V_N - \partial_N V_M$  is the usual field strength and $V=\{ \sqrt{2} W^{\pm}, Z, A \} $ and $m_V = \{ m_W, m_Z, 0 \}$. The coupling to the field strengths above becomes significant for momentum transfer much larger than the EW scale and the second term constitutes a correction of about $20 \%$ to the dominant $\TeV$-boundary coupling. In the case of the photon only the first term is present. A similar expression for gluons should be included.

Overall we can write 
\begin{equation}
\mathcal{L}_{r}^{ WW, ZZ} =  \frac{\gamma}{v}r \left\lbrace  2m_W^2 \left( 1- \frac{3 m_W^2 k y_c}{\Lambda^2}\right)W_\mu^+ W^{\mu-} + m_Z^2 \left( 1- \frac{3 m_Z^2 k y_c}{\Lambda^2}\right)Z_\mu Z^{\mu}  \right\rbrace. \label{rVV}
\end{equation}

For massless gauge bosons we have to include the contributions coming from the localized trace anomaly and from loop triangle diagrams in which the $W$ gauge boson and fermions in the case of the photon and only fermions in case of the gluons that induce couplings to the radion.

All these contributions can be written as \cite{Giudice:2000av, Kubota:2012in, Frank:2016oqi, Csaki:2007ns} \footnote{The Lagrangian takes into account only the leading order mass effects for the radion coupling to exactly two gauge bosons.}
\begin{equation}
\mathcal{L}_{r}^{ gg, \gamma \gamma} =  - \frac{\gamma}{4v}r \left\lbrace   \left( \frac{1}{k y_c} + \frac{\alpha_s b_{QCD}^r }{2 \pi} \right)G_{\mu \nu} G^{\mu \nu} + \left( \frac{1}{k y_c} + \frac{\alpha_{EM} b_{EM}^r }{2 \pi} \right)F_{\mu \nu} F^{\mu \nu}  \right\rbrace, \label{rGG}
\end{equation}
with $\alpha_s$($\alpha_{EM}$) being the strong (electroweak) coupling constant and 
\begin{equation}
b_{QCD}^r = 7 + F_f ,
\end{equation}
\begin{equation}
b_{EM}^r = -\frac{11}{3} + \frac{8}{3}F_f - F_W ,
\end{equation}
\begin{equation}
F_f = \tau_f \left(  1+ (1-\tau_f) f(\tau_f) \right) ,
\end{equation}
\begin{equation}
F_W = 2+3 \tau_W + 3 \tau_W( 2-\tau_W)f(\tau_W),
\end{equation}
\begin{equation}
f(\tau) = Arcsin^2(\frac{1}{\sqrt{\tau}}) \quad  \tau \geq 1,
\end{equation}
\begin{equation}
f(\tau) = -\frac{1}{4}\left( \log\frac{1+\sqrt{1-\tau}}{1-\sqrt{1-\tau}}-i\pi\right)^2, \quad  \tau<1,
\end{equation}
and $\tau_i = (\frac{2 m_i}{m_r})^2$, $m_i$ is the mass of the particle going around the loop. An important property of the kinematic functions is their saturation $F_f \rightarrow 2/3$, $F_W \rightarrow 7$, $\tau f(\tau) \rightarrow 1$ for $\tau >1$ and $F_{f,W} \rightarrow 0$ for $\tau <1$.

In this paper we do not consider the corrections to the couplings coming from excited KK modes of the top and W boson in the loop and simply assume that the above contributions are dominant. However we leave this issue for future work.

Fermions propagating in the bulk are characterized by a bulk mass parameter $c=m/k$ which specifies their location in the bulk. In addition, the boundary conditions of their profiles at the location of the branes force either the left- or the right-handed zero modes to be zero \cite{Grossman:1999ra}. Therefore for each SM fermion we need to introduce two different bulk fermions, one with bulk mass parameter $c_L$ and for which the right-handed zero mode vanishes and the other with a bulk mass parameter $c_R$ and for which the left-handed zero mode vanishes.

The couplings of the radion to SM fermions can be simplfyfied as \cite{Frank:2016oqi}
\begin{equation}
S \supseteq \int d^4 x \sum_{f=u,d,e} \frac{\gamma}{v} r(x)m_f \bar{f}f \times \begin{cases} 
       	1 & Planck \\
         (c_L-c_R)&  \TeV.
       \end{cases} \label{fc}
       \end{equation}
      with the lower option if the zero-mode profile is peaked towards the TeV brane $c_L <1/2, \ c_R > -1/2$ otherwise the localization is in the Planck brane and the upper option applies. Besides this couplings it seems that the boundary Yukawa couplings will have a direct contribution to the radion couplings to fermions. However, as shown in \cite{Csaki:2007ns}, these contributions get cancelled by induced wave function discontinuities obtained by carefully treating the boundary conditions.
      

\section{Two Higgs-radion Mixing}  \label{Mixing}

The most general term that will give rise to kinetic mixing between the Higgs doublets and the radion field is given by 
\begin{equation}
\mathcal{L}_\xi = \sqrt{\bar{g}_{ind}} \xi_{ab} \mathcal{R}(\bar{g}_{ind}) \frac{1}{2} \mathrm{Tr}[M_a^\dagger M_b] \label{curvature-scalar}
\end{equation}
where the indices $a,b=1,2$ are summed so that we have, in principle, four different mixing parameters. However the assumption of CP invariance forces $\xi_{12} = \xi_{21}$ and thus the pseudoscalar does not mix with the radion.
Evaluation of the Ricci scalar is straightforward and yields the following expression \cite{Csaki:2000zn}
\begin{equation}
\mathcal{L}_\xi = -6 \xi_{ab} \Omega^2 \left[ \Box \ln \Omega + (\nabla \ln{\Omega})^2    \right]\frac{1}{2} \mathrm{Tr}[M_a^\dagger M_b] 
\end{equation}
The warp factor disappears after we make the rescaling of the Higgs doublets.
Using the expression for the Higgs mass eigenstates \eqref{scalars} and expanding to linear order in the fields we can write
\begin{equation}
\mathcal{L}_\xi \supseteq -6 \left[ -\frac{\gamma}{v}\Box r + \frac{\gamma^2}{v^2} r \Box r \right] \left[ \frac{v^2}{2} K_r  + \frac{v}{2} K_h h + \frac{v}{2} K_H H \right],
\end{equation} 
where $\gamma \equiv v/\Lambda$ and we define the mixing parameters by
\begin{equation}
K_r = \xi_{11} c_\beta^2 + \xi_{22} s_\beta^2 +2  \xi_{21} s_\beta c_\beta,     \label{kr}
\end{equation}
\begin{equation}
K_h = 2(\xi_{22}s_\beta c_\alpha - \xi_{11}c_\beta s_\alpha) + 2 \xi_{12} \cos(\alpha + \beta),   \label{kh}
\end{equation}
\begin{equation}
K_H = 2(\xi_{11}c_\beta c_\alpha + \xi_{22}s_\beta s_\alpha) + 2\xi_{12}  \sin(\alpha + \beta).   \label{kH}
\end{equation}
 Adding the kinetic and mass terms of each field, the mixing Lagrangian can be expressed as 
\begin{equation}
\mathcal{L} = -\frac{1}{2}(1+6 \gamma^2 K_r)r \Box r - \frac{1}{2}m_r^2 r^2 + \sum_{\phi = h,H} \left\lbrace 3\gamma K_\phi \phi \Box r - \frac{1}{2}\phi (\Box + m_\phi^2)\phi \right\rbrace   \label{HR}
\end{equation}
The kinetic terms can be diagonalized by performing the transformation
\begin{equation}
r \rightarrow \frac{r'}{Z},  \quad \phi \rightarrow \phi' + \frac{3 \gamma K_\phi}{Z}r'   \label{transformation}
\end{equation}
with $\phi = h,H$ and
\begin{equation}
Z^2 = 1+6 \gamma^2 K_r - 9 \gamma^2 (K_h^2 + K_H^2 ), \label{Z}
\end{equation}
is the determinant of the kinetic mixing matrix and therefore should always satisfy $Z^2>0$ to avoid the presence of ghosts fields. This condition allows us to impose our first theoretical constraint on the mixing parameters after choosing appropriate values for $\alpha$, $\beta$ and  $\gamma$.
This transformation induces mixing in the mass terms. The mass matrix obtained can be written as
\begin{equation}
M =
\begin{pmatrix}
\omega_{rr}^2 & \omega_{rh}^2 & \omega_{rH}^2   \\
\omega_{rh}^2 & m_h^2 & 0   \\
\omega_{rH}^2 & 0 & m_H^2   \\
\end{pmatrix}, \label{massmatrix}
\end{equation}
where
\begin{equation}
\omega_{rr}^2 = \frac{m_r^2}{Z^2} + \frac{9 \gamma^2}{Z^2} \left( K_h^2 m_h^2 + K_H^2 m_H^2  \right),  
\end{equation}
\begin{equation}
 \omega_{r \phi}^2 = \frac{3 \gamma}{Z} K_\phi m_\phi^2.
\end{equation}
 
 The physical eigenstates are obtained by performing a three dimensional rotation
 \begin{equation}
 \begin{pmatrix}
 r' \\
 h' \\
 H'
 \end{pmatrix} = U   \begin{pmatrix}
 r_D \\
 h_D \\
 H_D
 \end{pmatrix}. \label{diagonalization}  
  \end{equation}
 The relation between the gauge eigenstates and the mass eigenstates can be written as 
 \begin{equation}
 r = \frac{U_{11}}{Z}r_D +  \frac{U_{12}}{Z}h_D  +  \frac{U_{13}}{Z}H_D,  \label{Dstate1}
 \end{equation}
 \begin{equation}
 h = \left( U_{21} + 3 \gamma \frac{K_h}{Z}U_{11} \right)r_D  +  \left( U_{22} + 3 \gamma \frac{K_h}{Z}U_{12} \right)h_D  +\left( U_{23} + 3 \gamma \frac{K_h}{Z}U_{13} \right)H_D,   \label{Dstate2}
 \end{equation}
 \begin{equation}
 H = \left( U_{31} + 3 \gamma \frac{K_H}{Z}U_{11} \right)r_D  +  \left( U_{32} + 3 \gamma \frac{K_H}{Z}U_{12} \right)h_D  +\left( U_{33} + 3 \gamma \frac{K_H}{Z}U_{13} \right)H_D.   \label{Dstate3}
 \end{equation}
 
 For later convenience we name the coefficients of this transformation as 
 \begin{equation}
 U_{rr} = \frac{U_{11}}{Z}, \quad \quad U_{rh} = \frac{U_{12}}{Z}, \quad \quad  U_{rH} = \frac{U_{13}}{Z},
 \end{equation}
 \begin{equation}
 U_{hr} = U_{21} + 3 \gamma K_h \frac{U_{11}}{Z}, \quad \quad U_{hh} =U_{22} + 3 \gamma K_h  \frac{U_{12}}{Z}, \quad \quad  U_{hH} = U_{23} + 3 \gamma K_h \frac{U_{13}}{Z},
 \end{equation} 
 \begin{equation}
 U_{Hr} = U_{31} + 3 \gamma K_H \frac{U_{11}}{Z}, \quad \quad U_{Hh} =U_{32} + 3 \gamma K_H  \frac{U_{12}}{Z}, \quad \quad  U_{HH} = U_{33} + 3 \gamma K_H \frac{U_{13}}{Z},
 \end{equation}
 which will be used in the next section for the predictions of the electroweak precision observables.
 
 The Higgs scalars-radion system is determined by the three mixing parameters of equation \eqref{curvature-scalar}, the two mixing angles of the Higgs sector, the scale $\gamma$ and the three scalar masses, giving a total of nine parameters. However one of the physical masses will be set to the Higgs mass value and only the set $(\xi_{11}, \xi_{12}, \xi_{22}, \alpha, \beta, \gamma, \lambda_r, \lambda_H )$ needs to be specified.
 
 Another important parameter in the study of RS models with bulk gauge bosons is the KK scale defined to be the mass of the first excited state  of the gauge bosons. Recall that this parameter is independent of the gauge symmetry and gauge couplings and is universal for all gauge bosons that satisfy the same BCs.  In particular, for gauge bosons satisfying Neumann BCs at both branes it is given by  \cite{Dominici:2002jv}
 \begin{equation}
 m_{KK} = 2.45 \frac{k}{\sqrt{6}M_{Pl}} \Lambda,
 \end{equation}
 so any bound on the KK scale will directly affect the allowed values of the curvature scale $k$ and $\Lambda$.

In Higgs-radion mixing scenarios there is a particular point in the parameter space called the ``conformal point"  \cite{Giudice:2000av, Frank:2016oqi, Chakraborty:2017lxp}, usually around $\xi = 1/6$ where the conformal symmetry is minimally violated by the Higgs VEV. At this point the tree-level couplings of the radion to the massive fermions and gauge bosons are very suppressed and the $gg$ decay mode dominates even in the large radion mass limit. 
In this work we do not attempt to calculate a conformal point due to the large number of parameters.

In what follows we will sometimes reduce the parameter space by assuming that the diagonal elements of the curvature-scalar mixing matrix are equal to each other, $\xi_{11}=\xi_{22} \equiv \xi_{1}$ and for simplicity we will refer to the off diagonal as $\xi_{12}\equiv \xi_{2}$. Relaxing this constraint will not radically alter the numerical results in the following sections.   However, we will primarily focus on $K_r, K_h, K_H$, which is independent of this assumption.

From now on we will drop the subindex $D$ for the diagonal eigenstates and simply write them as $r$, $h$ and $H$. Whenever we need to distinguish between the non-diagonal and physical states a clarification will be made.

\section{Model Predictions}  \label{constraints}
\subsection{Electroweak Precision Observables}

The motivation for the custodial symmetry came from requiring corrections to EWPO, parametrized by the Peskin-Takeuchi \cite{Peskin:1991sw} parameters $S$ and $T$, be sufficiently small.  The corrections have contributions from the KK excitations of the fermions and gauge bosons, from the 2HDM sector and from the radion.  As discussed in the introduction, an extended gauge custodial symmetry in the bulk keeps the corrections from the KK excitations under control \cite{Agashe:2003zs}.
In the absence of mixing, a custodially symmetric 2HDM potential has vanishing contributions to the T parameter \cite{Haber:2010bw} and the contributions of the radion are also small (see Csaki et al. \cite{Csaki:2000zn}).

 However when one includes mixing, the radion and Higgs scalar couplings are modified and could result in large corrections depending on the values of the mixing parameters and masses. This was first discussed by Csaki, et al.(CGK) \cite{Csaki:2000zn} and a paper dedicated entirely to electroweak precision constraints was written by Gunion et al. (GTW)\cite{Gunion:2003px}, we follow the notation of the latter.  Both showed that there are three types of contributions to the $S$ and $T$ parameters: $(1)$ with each scalar eigenstate going through the loop of the vacuum polarization graph of the vector bosons, $(2)$ anomalous terms coming from the conformal couplings of the radion when the theory is regulated and $(3)$ higher dimensional operators which arise after integrating out the heavy degrees of freedom, e.g. spin-2 graviton states.
 
 The first contribution comes from vacuum polarization graph loops.    Let us first consider the single Higgs case in which there is one $\xi$ term.   As shown above, this leads to kinetic mixing between the radion and Higgs.    Diagonalizing the kinetic mixing terms and then further diagonalizing the mass matrix gives \cite{Csaki:2000zn,Gunion:2003px}(with $h$ and $\phi$ being the mass eigenstates and $h_0$ and $\phi_0$ the geometric eigenstates):
 \be
 h_0 = c\ \phi + d\ h\qquad    \phi_0 = a\ \phi + b\ h
 \ee
where 
\be
a = -\frac{\cos\theta}{Z}\quad b=\frac{\sin\theta}{Z}\qquad c=\left(\sin\theta +\frac{6\xi v}{Z\Lambda_\phi}\cos\theta\right)\quad d=\left(\cos\theta -\frac{6\xi v}{Z\Lambda_\phi}\sin\theta\right) 
\ee
The terms with an explicit $\xi$ are obtained when the kinetic terms are diagonalized and the others arise when rotating to the mass basis from the geometric basis.  Here
\be
Z^2 = 1 + 6\xi(1-6\xi)v^2/\Lambda_\phi^2 \qquad   \tan 2\theta = 12\gamma \xi Z\frac{m^2_{h_0}}{m^2_{\phi_0} - m^2_{h_0}(Z^2-36\xi^2\gamma^2)}
\ee
Here, $m^2_{h_0}$ and $m^2_{\phi_0}$ are the Higgs and radion masses when the mixing vanishes.    Note that $\gamma = v/\Lambda_{\phi}$ is very small.

As shown in \cite{Csaki:2000zn,Gunion:2003px}, the contributions of the radion and Higgs to the electroweak $S$ and $T$ parameters are
\be
S_i = -\frac{g^2_i}{\pi}\left( B_0(M^2_Z,m^2_i,M^2_Z) - \frac{B_{22}(M^2_Z,m^2_i,M^2_Z}{M^2_Z} - B_0(0,m^2_i,M^2_Z) - \frac{B_{22}(0,m^2_i,M^2_Z)}{M^2_Z} \right)
\ee
\be
T_i = -\frac{g^2_i}{4\pi\sin^2\theta_W}\left( B_0(0,m^2_i,M^2_W) - \frac{B_{22}(0,m^2_i,M^2_W}{M^2_W} - \frac{B_0(0,m^2_i,M^2_Z)}{\cos^2\theta_W} - \frac{B_{22}(0,m^2_i,M^2_Z)}{M^2_W} \right)
\ee
where the $B$-functions are the Passarino-Veltman functions \cite{Passarino:1978jh}. 
There are contributions from the Higgs and the radion and the couplings are given by $g_h = d+b\gamma$ and $g_\phi= c+a\gamma$.

In the 2HDM-radion model, the expressions in the above paragraph for $S_i$ and $T_i$ are still present, but now one includes contributions from the radion and both $h$ and $H$ (note, additional contributions from charged Higgs masses and neutral scalar mass splittings will be discussed later).    However, the three $g_i$ are now different.   The diagonalization of the kinetic terms was given in Eqs. \eqref{Dstate1}-\eqref{Dstate3}.  The mass matrix in Eq. \eqref{diagonalization} is a $3 \times 3$ matrix and no similar analytic form is possible, so we must do the calculation numerically

\begin{figure}[h]
 \centering
\includegraphics[scale = 0.4]{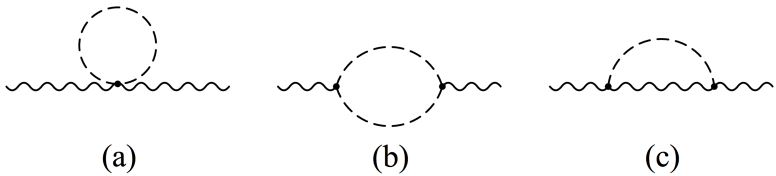} 
\caption{Feynman diagrams relevant for the contributions of the scalar sector to the oblique parameters. }
\label{diagrams}
\end{figure}

 The expressions for the $g_i$ can be read off from Eqs.\eqref{Dstate1}-\eqref{Dstate3}.
In addition, there are contributions from loops with two scalars, diagram (b) in Fig. \ref{diagrams}, including the charged scalar and the pseudoscalar with a neutral scalar.  The charged scalar can also contribute to $\Pi_{Z\gamma}$ and $\Pi_{\gamma\gamma}$. In this model, the contributions involving the physical fields $r$, $h$ and $H$ for diagrams of the type (b) and (c) in Figure \ref{diagrams} are listed in Appendix \ref{STapp}.

There are two other contributions in the Higgs-radion mixing case.     There are anomalous terms where the linear $\epsilon$-terms in dimensional regularization in the radion-matter interactions are elevated to finiteness by $1/\epsilon$ poles in the radion loops.     These are calculated in both CGK and GTW and turn out to be negligibly small, as noted explicitly in CGK.   In GTW, an additional term is shown to be present, but this term (as they state) makes a negligible contribution for a Higgs mass of $120$ GeV.     There are extra parameters in this model, but barring very unnatural values of these parameters, we expect the contribution to be completely negligible.      The other contribution comes from non-renormalizable operators which come from integrating out heavy states above the scale $\Lambda_\phi$.    These can affect the $T$ parameter since they can break isospin symmetry.   However, the value of these parameters at the scale $\Lambda_\phi$ is completely arbitrary.    Although CGK discussed these terms, they didn't include them in their calculation.  GTW assume they vanish at $\Lambda_\phi$, and consider the running of these terms down to $M_Z$.    In this section we are concerned with how this model differs from the single Higgs case, and since the terms are arbitrary (and likely to be different in the single and two Higgs cases), we will not include these terms here.

As a first case example, Case $1$, we consider the case of the exact alignment limit where $\cos(\beta - \alpha) = 0$ and $\xi_2 =0$. In this simplified scenario we only have one mixing parameter since the mixing coefficients of equations \eqref{kr}-\eqref{kH} reduce to $K_r = \xi_1$, $K_h = 2 \xi_1 $ and $K_H = 0$. Then the field $H$ doesn't mix with the radion and has vanishing tree-level couplings to pairs of gauge bosons therefore diagrams of type (c) with the $H$ running in the loop are absent however diagrams of type (a) and type (b) are still present.    This doesn't quite reduce to the single Higgs radion mixing case because of the presence of $H-A$ loop in diagram (b).

\begin{figure}[h]
 \centering
\includegraphics[scale = 0.23]{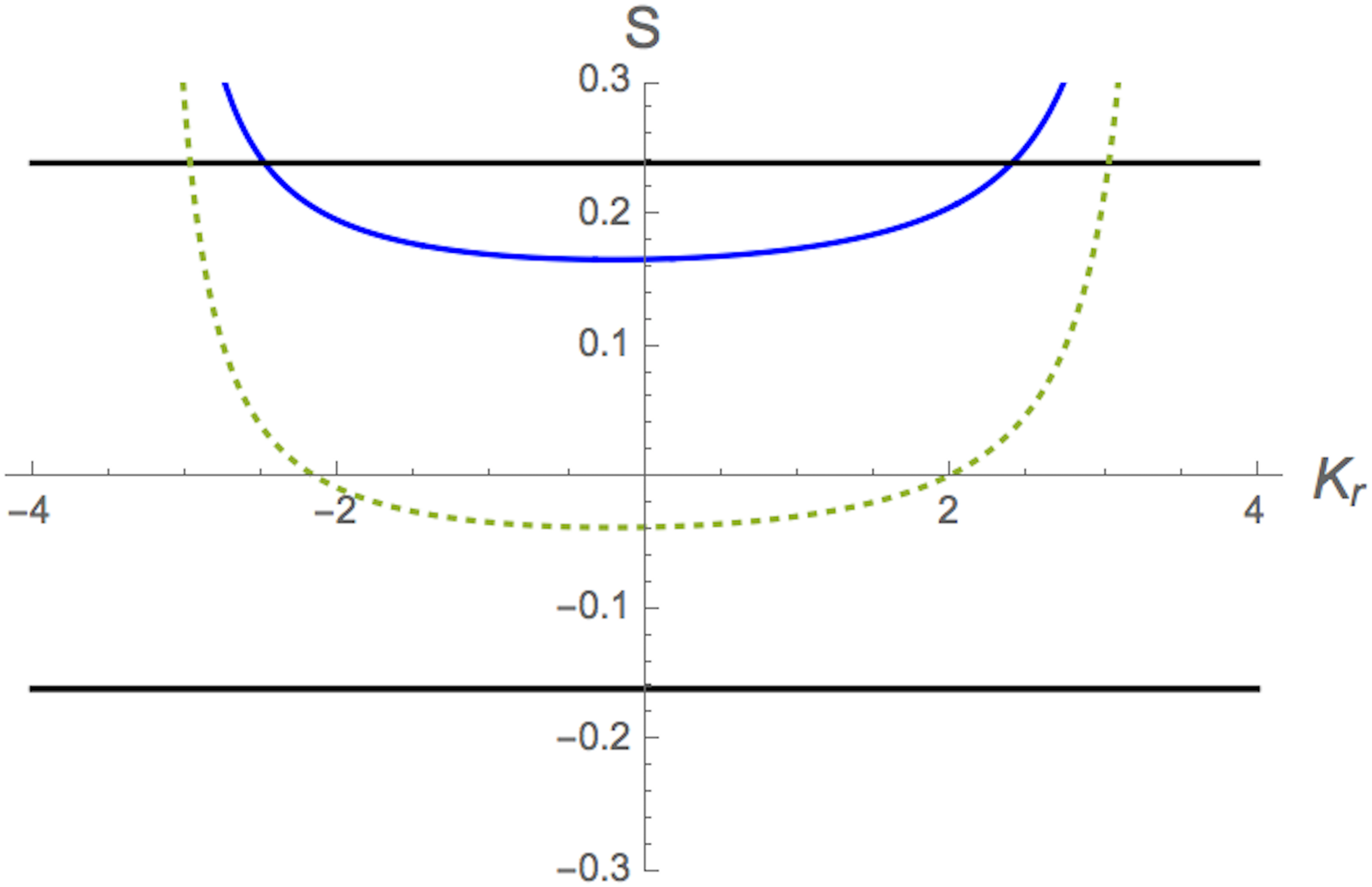} 
\includegraphics[scale = 0.299]{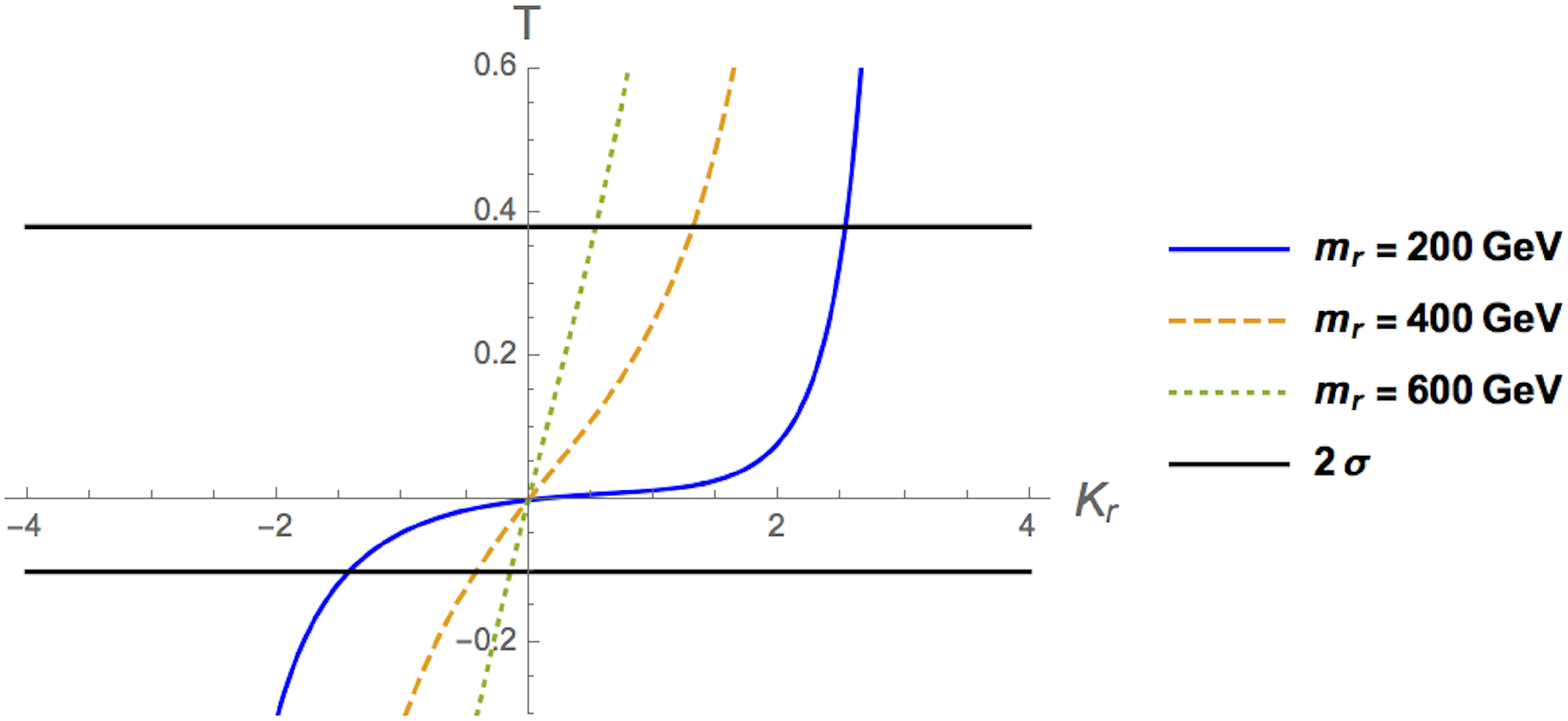} 
\caption{$S$ and $T$ parameter curves as a function of the mixing parameter $K_r$ for different values of the radion mass. The solid horizontal black line represents the $2\sigma$ upper bound from current value. The rest of the parameters chosen were $\cos(\beta - \alpha)=0$, $\xi_2=0$, $\tan\beta=1$, $m_h=125 \ \GeV$, $m_H=1000 \ \GeV$, $\Lambda = 5 \ \TeV$ and  $m_A = 500 \ \GeV$.   The dotted line in the S-plot corresponds to the single Higgs limit, without the $H-A$ loops included, for a radion mass of $200\ \GeV$ (it is very insensitive to the radion mass).}
\label{Scase1}
\end{figure}

The $S$ and $T$ parameters in the Case $1$ scenario are shown in fig.  \ref{Scase1}. As can be noticed the constraints from the $T$ on the mixing parameter become more stringent with increasing radion mass. The combined constraints are $-1.43<K_r<2.4$ for $m_r = 200 \ \GeV$, $-0.4<K_r<1.3$ for $m_r = 400 \ \GeV$ and $-0.14<K_r<0.5$ for $m_r = 600 \ \GeV$.     To compare with the single Higgs case, we include in the S-parameter plot the calculation without the additional $H-A$ loop (due to the custodial symmetry, the $T$ parameter is not changed).      We see immediately that the additional Higgs bosons increase the S parameter (this also occurs in the conventional 2HDM, of course), but the model is still acceptable.     Now, however, we relax the $\xi_2=0$ assumption (which gave $K_H=0$) and see how the EWPO contributions change.

We first will continue to set $\cos(\beta-\alpha)$ to zero, since it must be small, as shown by the fit of the model to the LHC Higgs data in the next subsection.  As $\xi_2$ is now nonzero, we will have $K_r, K_h$ and $K_H$ all nonzero.    The results will be plotted as curves in the $K_H, K_h$ plane, and we will see that the results are very insensitive to $K_r$.    The region of parameter-space in this plane allowed by the positivity of the determinant of the kinetic matrix, $Z^2 > 0$, is a circle in this plane (since the $K_r$ term is multiplied by $\gamma^2$ and thus is very small) as shown in Eq. \eqref{Z}.

\begin{figure}[H]
\centering
\includegraphics[scale = 0.27]{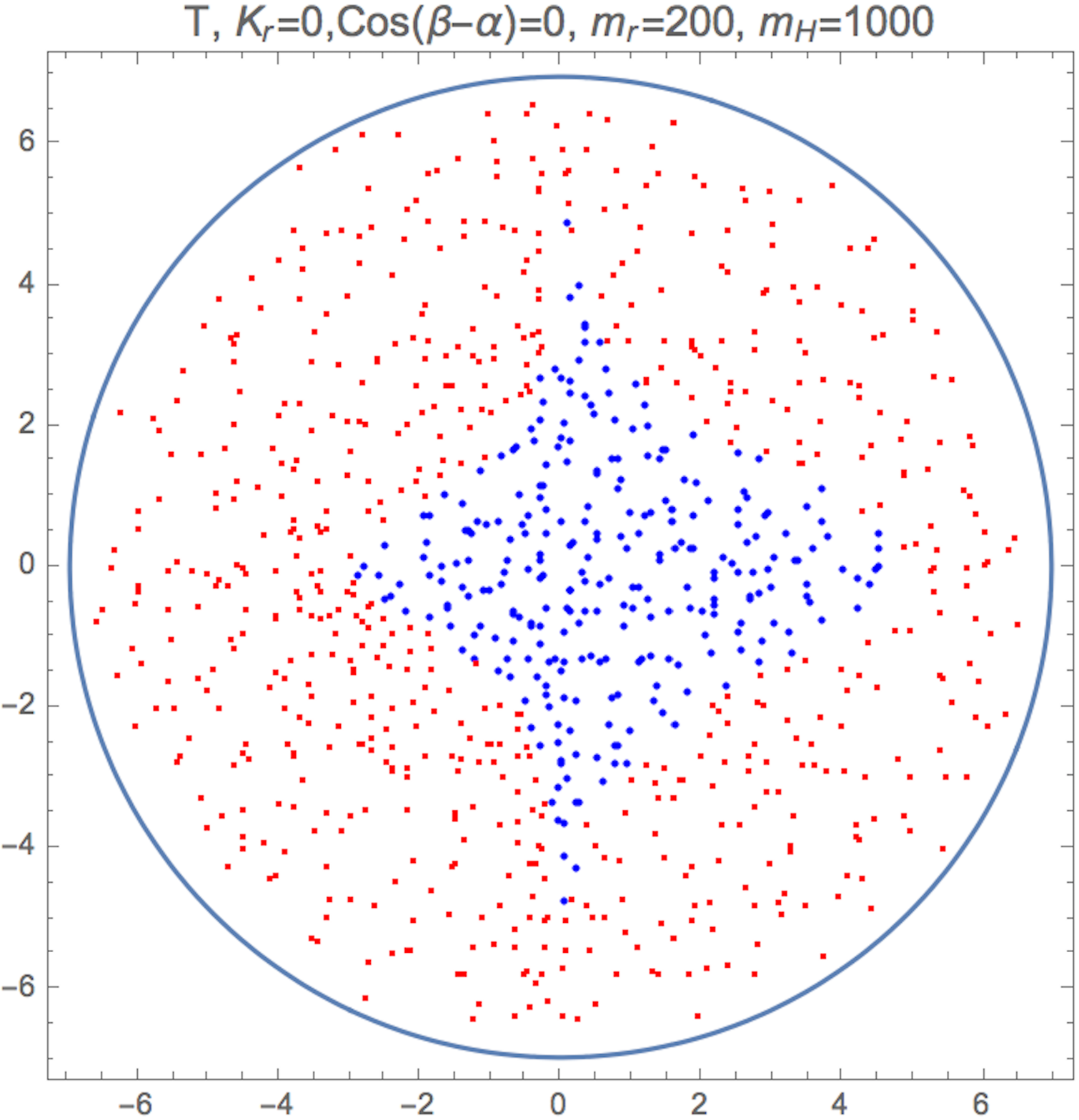} 
\hspace{4mm}
\includegraphics[scale = 0.27]{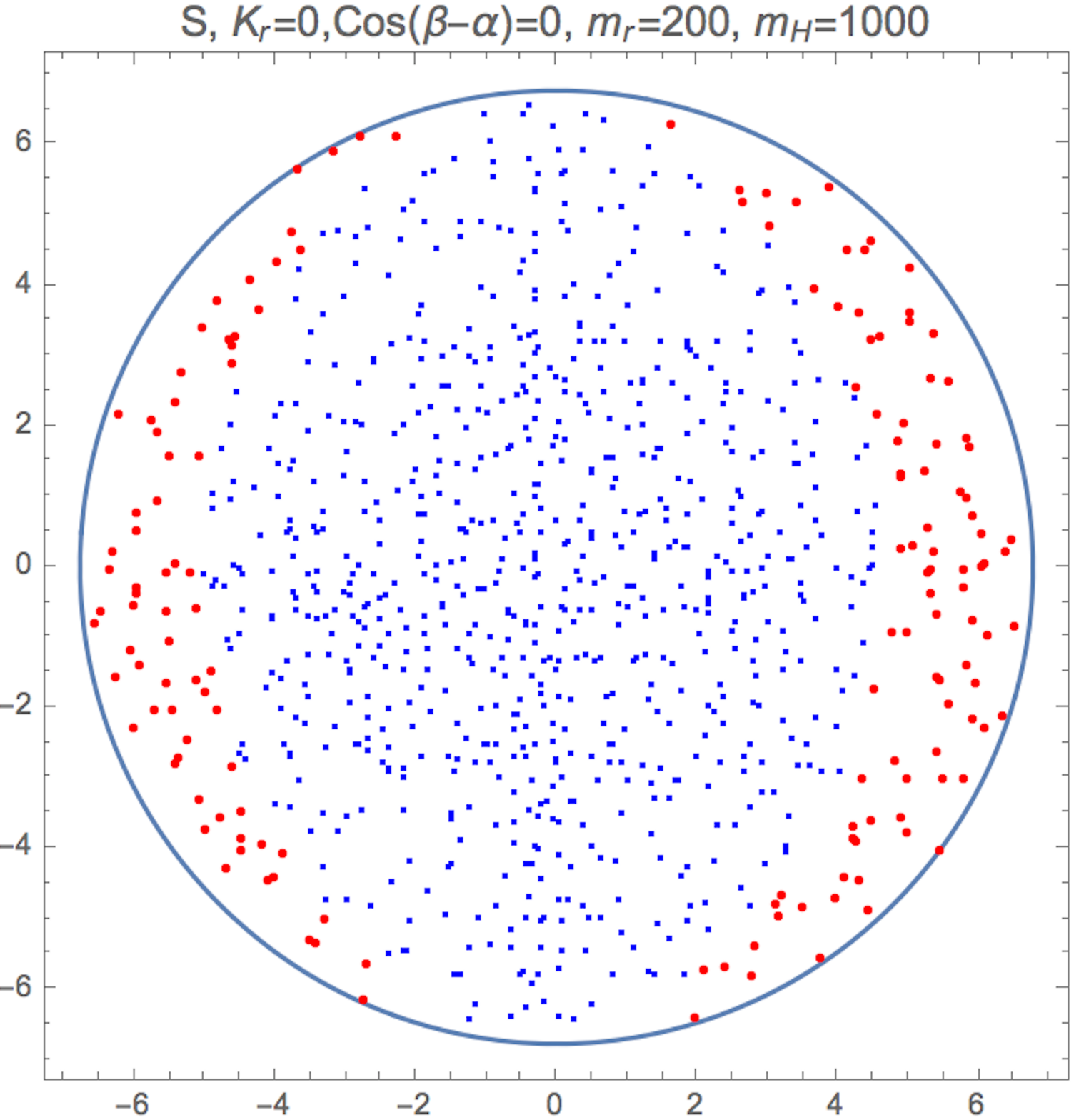} 
\hspace{4mm}
\centering
\includegraphics[scale = 0.27]{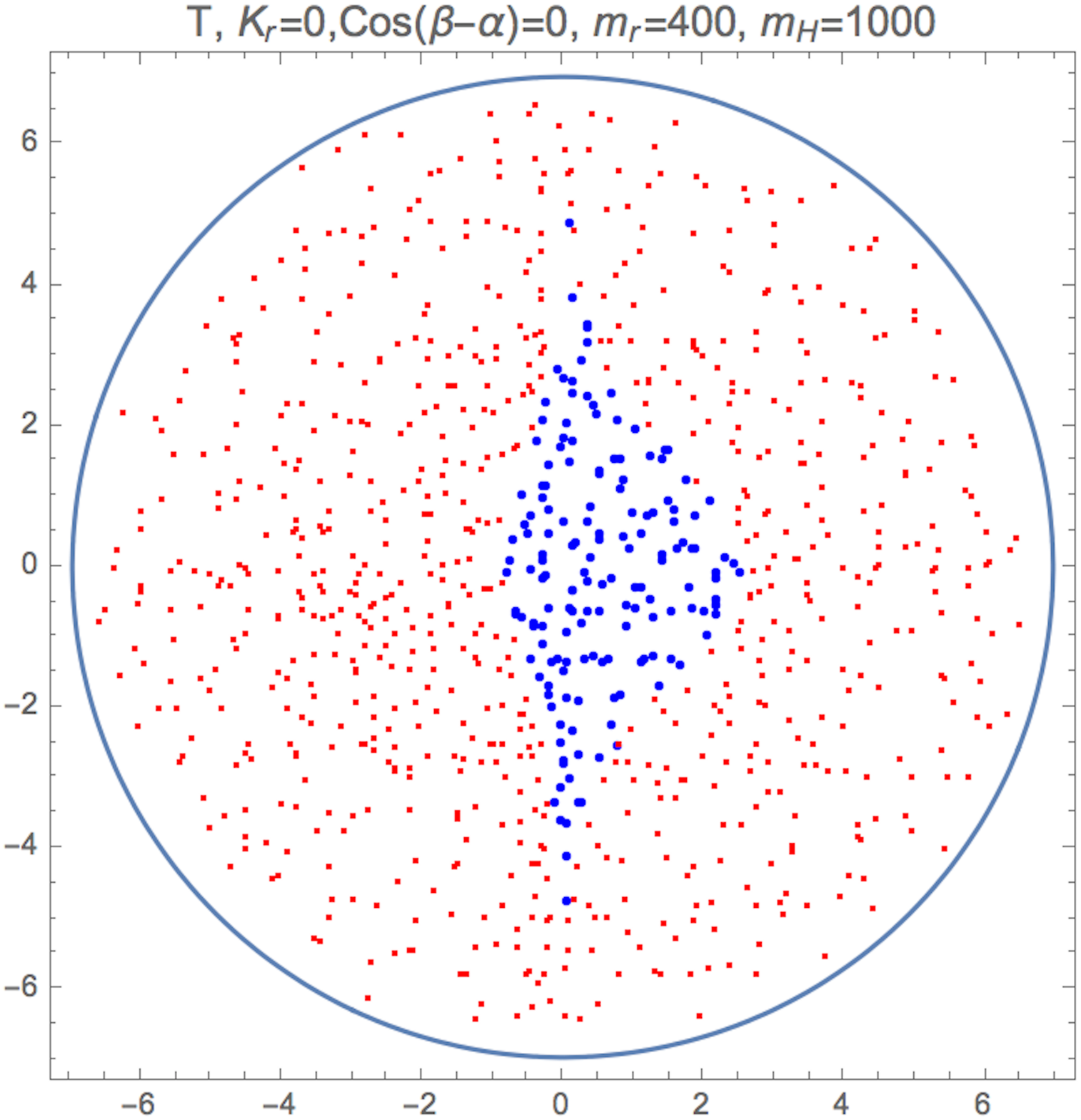} 
\hspace{4mm}
\includegraphics[scale = 0.27]{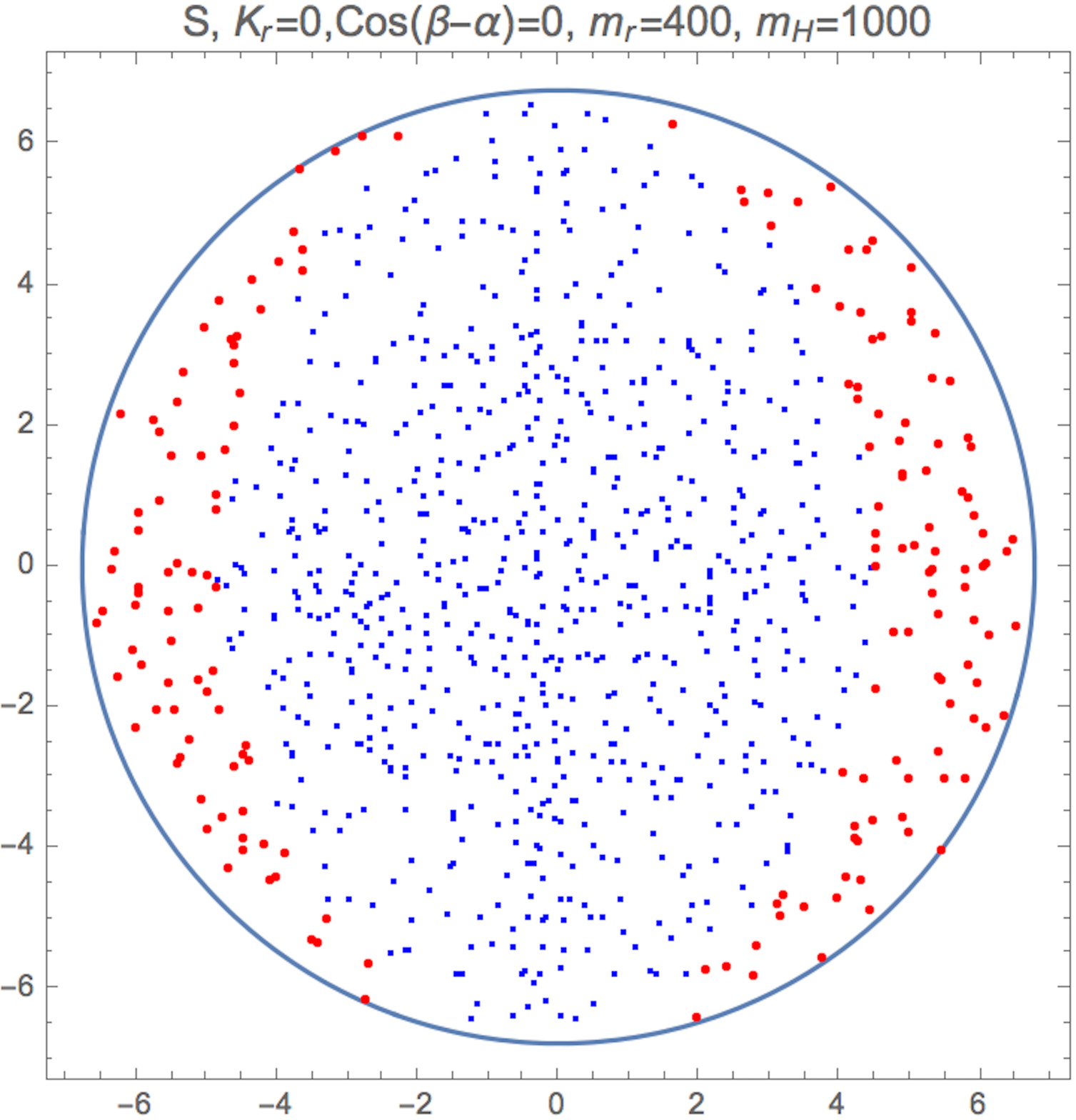}
\hspace{4mm}
\includegraphics[scale = 0.27]{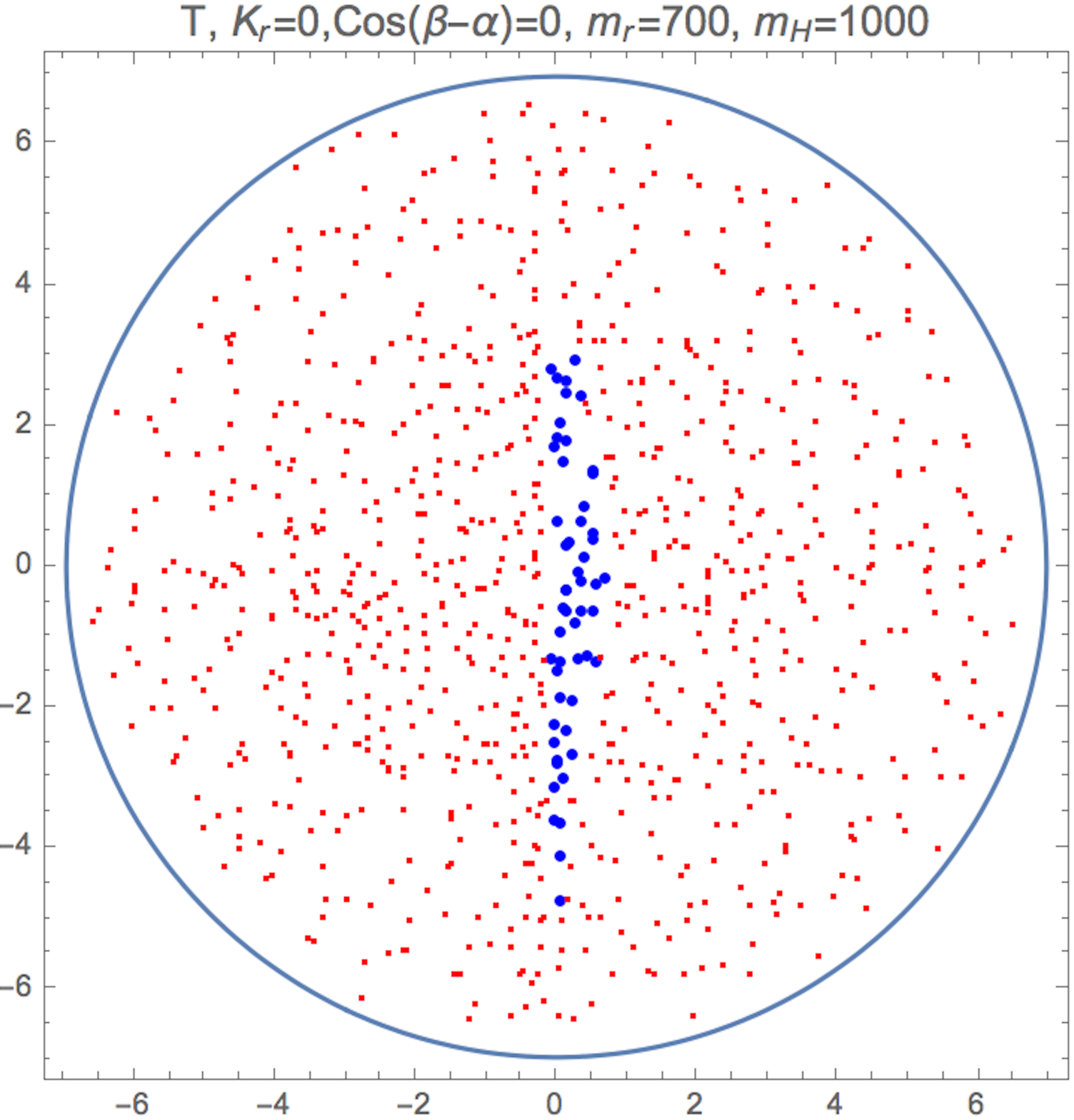} 
\hspace{4mm}
\includegraphics[scale = 0.27]{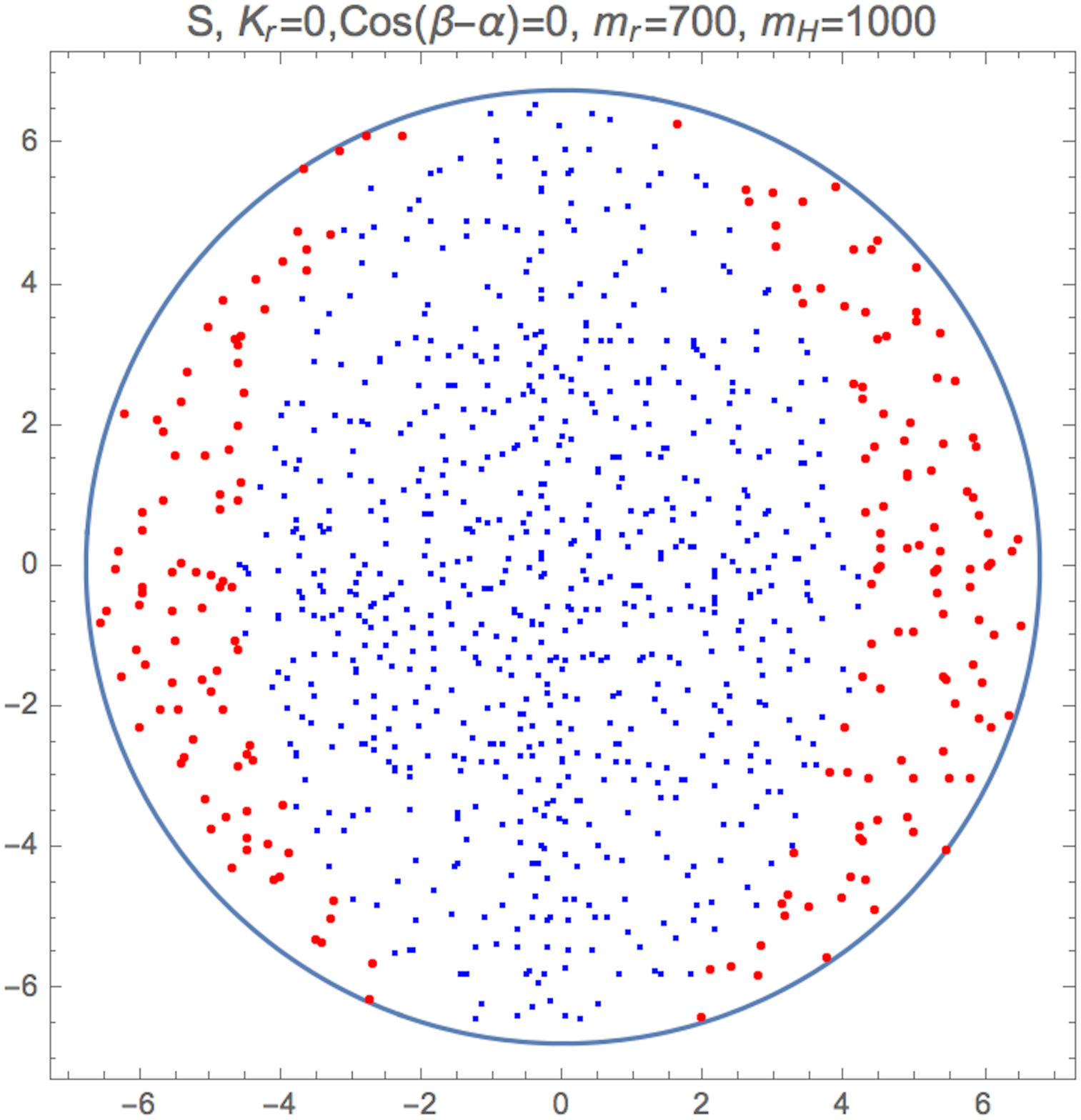} 
\caption{Constraints on  $K_H$ (y-axis) and $K_h$ (x-axis) from the $S$ and $T$ parameters.   The circle is the theoretically allowed region, the blue dots are allowed by the $S$ and $T$ parameter bounds, and the red dots are disallowed. The other parameters chosen are listed at the top of the figures with $m_A = 500 \ \GeV$ and $\Lambda = 5\TeV$.} \label{ST700}
\end{figure}

For  radion masses of $200$, $400$ and $700  \GeV$, the allowed region is shown in Fig. \ref{ST700}.   In these figures, the x-axis is $K_h$ and the y-axis is $K_H$.     We  see that the bounds for $S$ are fairly mild, but are much stronger for the $T$ parameter.   Note that the $K_H=0$ value gives results identical to the earlier result for $K_h$ (which in that limit is twice $K_r$).     We see that the largest allowed values occur when either $K_h$ or $K_H$ is small.   One can see that the parameter space does get squeezed for higher radion masses.   It turns out the results are almost unchanged if one chooses a nonzero value of $K_r$.

We thus see that, in the alignment limit of $\cos(\beta-\alpha)$, the parameter space in the 2HDM is restricted in a manner similar to the single Higgs case (with the exception of the increase in $S$ due to heavy Higgs loops) but is a two dimensional restriction, rather than a restriction on the single mixing parameter.  If either $K_h$ or $K_H$ is near zero, the results are similar, but differ if they are both nonzero.

\begin{figure}[h]
 \centering
\includegraphics[scale = 0.3]{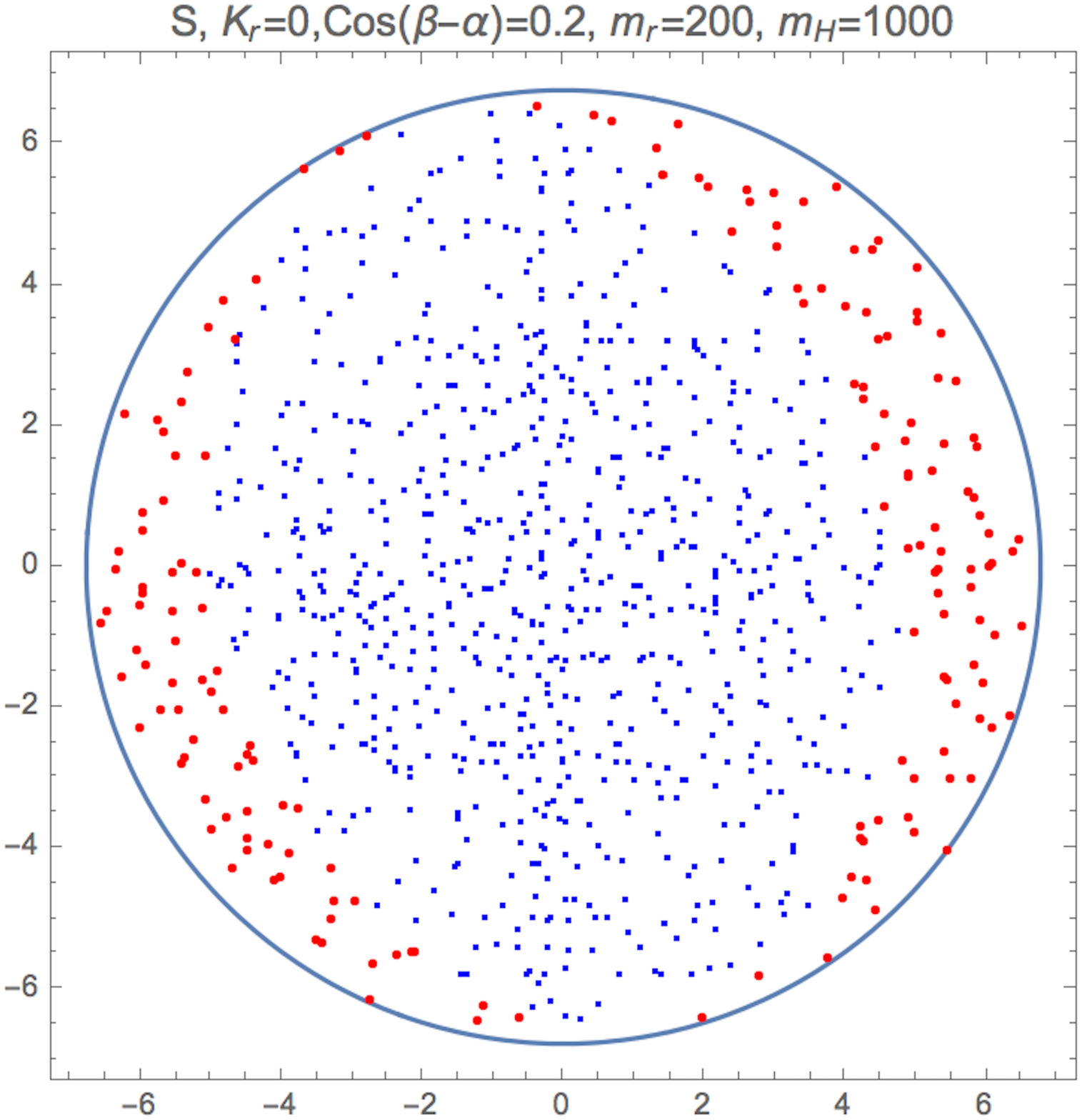} 
\hspace{2mm}
\includegraphics[scale = 0.3]{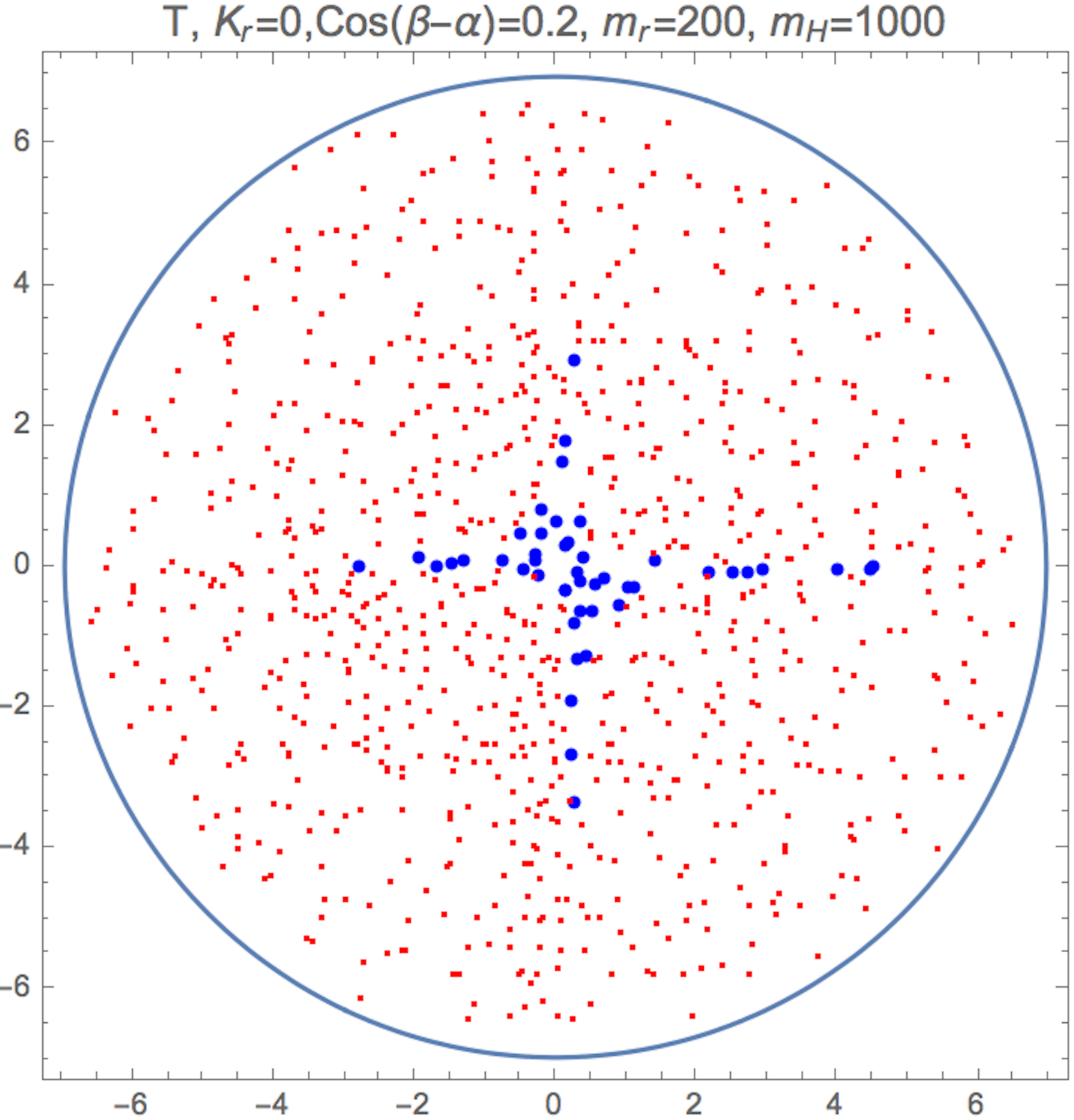} 
\caption{Constraints on  $K_H$ (y-axis) and $K_h$ (x-axis) from the $S$ and $T$ parameters, in the case in which one moves away from the alignment limit.   The circle is the theoretically allowed region, the blue dots are allowed by the $S$ and $T$ parameter bounds, and the red dots are disallowed.  The other parameters chosen are listed at the top of the figures with $m_A = 500 \ \GeV$, $\Lambda = 5\TeV$.}
\label{STcos}
\end{figure}

What if one moves away from the alignment limit?    Using $\cos{(\beta-\alpha)}$ = 0.2, we find the results in Fig. \ref{STcos}.     The $S$ parameter constraints are similar, but the $T$ parameter constraints become much more restrictive.   These features do not change much as the radion mass changes from $200$ GeV to $700$ GeV.   As we will see in the next section, though, the Type II model does not allow $\cos(\beta-\alpha)$ much larger than $0.1$, and thus this restriction will not be relevant, although it will be for the type I model.

One should consider these results with caution.   We have not included the non-renormalizable contributions since they are arbitrary at the cutoff scale, and those could affect the $T$ parameter, which gives the strongest constraints.  As shown in GTW, given certain assumptions, these can be substantial for large mixing and could broaden the parameter space. In addition, it is quite possible that the custodial symmetry will be broken on the Higgs brane, in which case the charged Higgs and pseudoscalar masses will not be degenerate.   Depending on which is heavier, the $T$ parameter can be substantially increased or decreased, which would drastically affect the bounds (this arbitrariness, of course, is not relevant in the single Higgs case).

\subsection{Constraints From Current LHC Higgs Data}

In the 2HDM the interactions of all the scalars to the SM fields are completely determined by the two mixing angles of the scalar sector $\beta$ and $\alpha$. In addition, the alignment limit is defined to be the limit in which one of the CP-even scalars has exactly the same interactions as the SM Higgs and corresponds to $\cos(\beta-\alpha) =0$.

\begin{figure}[h]
\centering
\includegraphics[scale = 0.175]{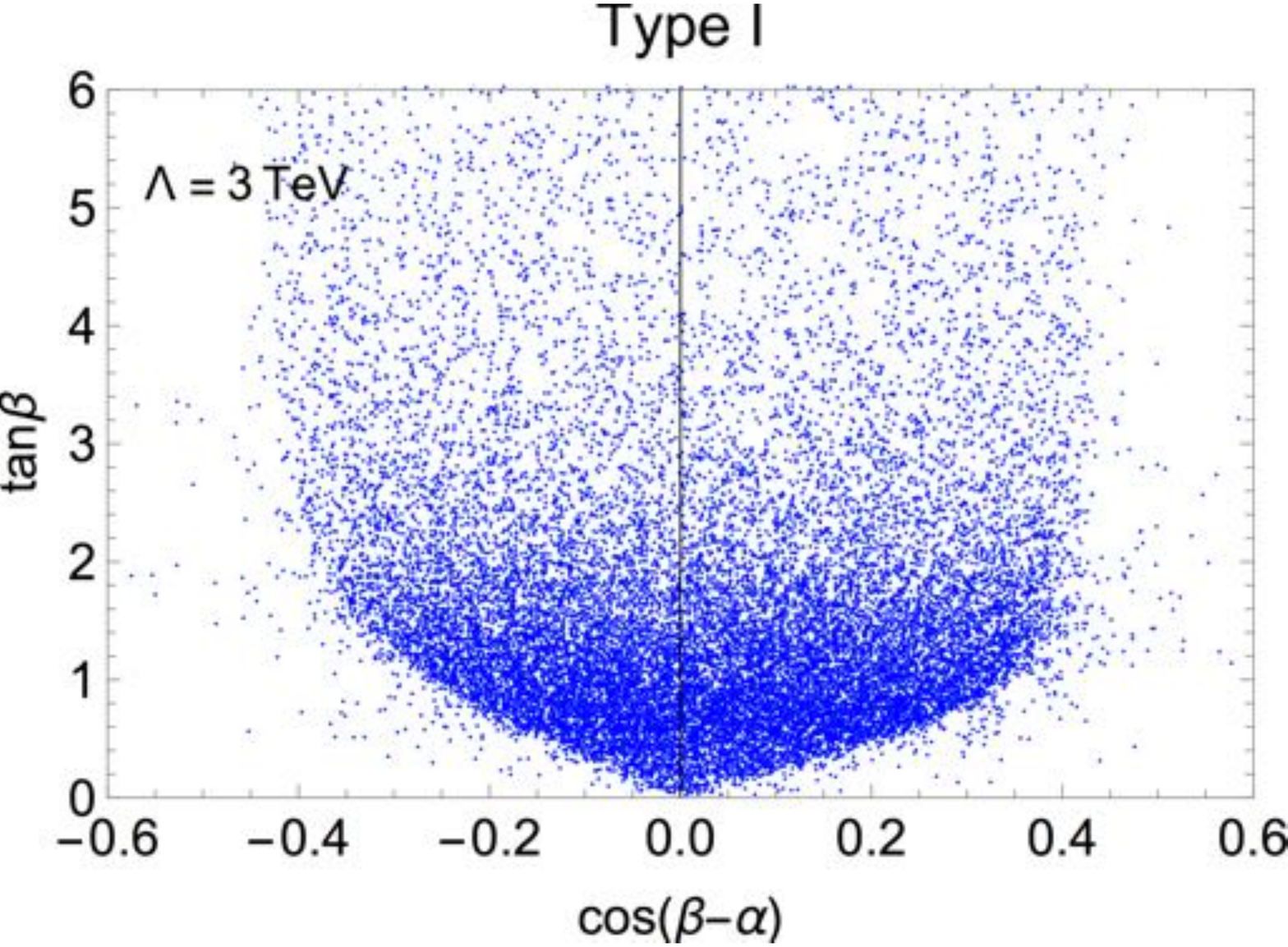} 
\includegraphics[scale = 0.175]{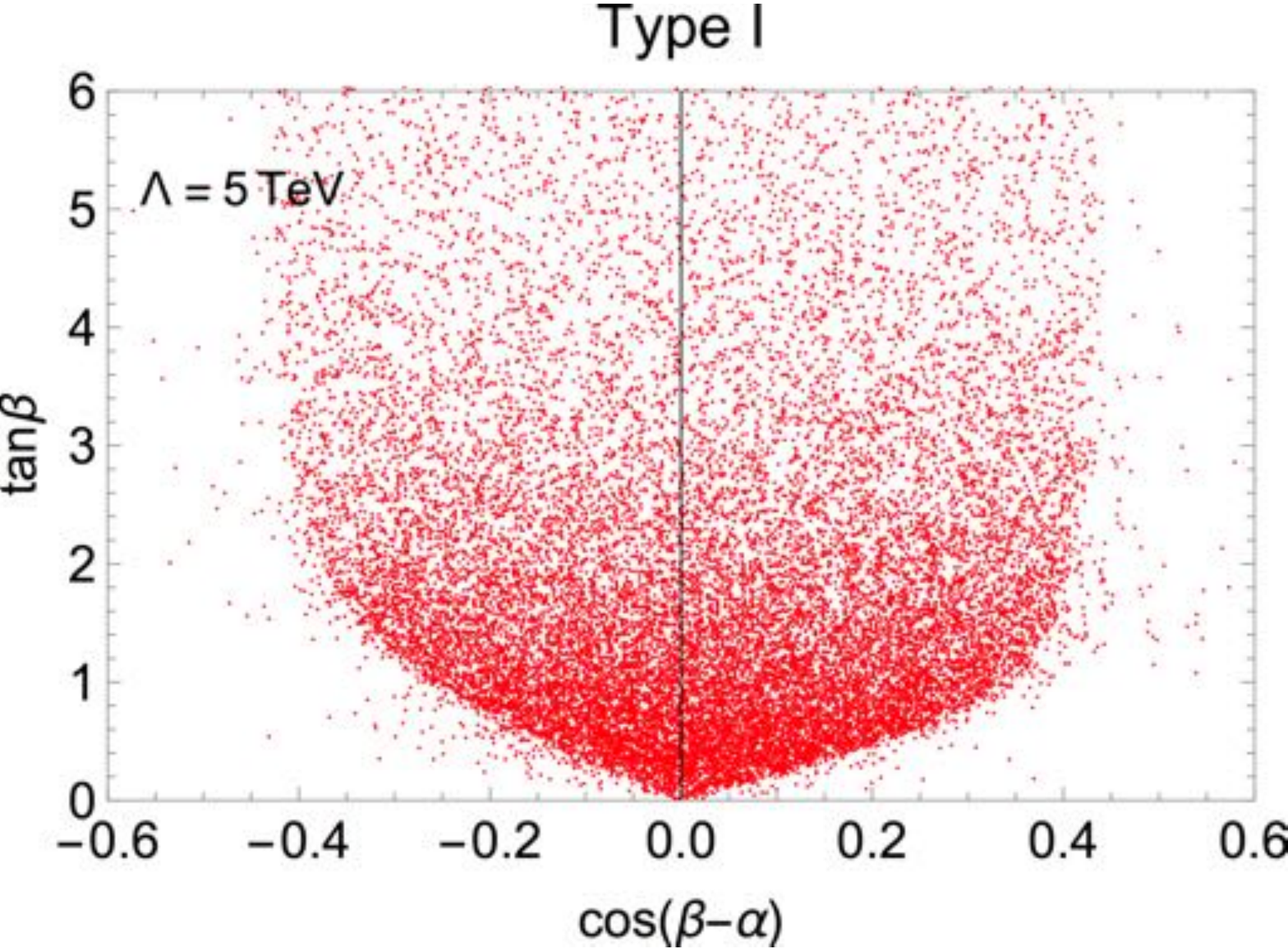} 
\includegraphics[scale = 0.175]{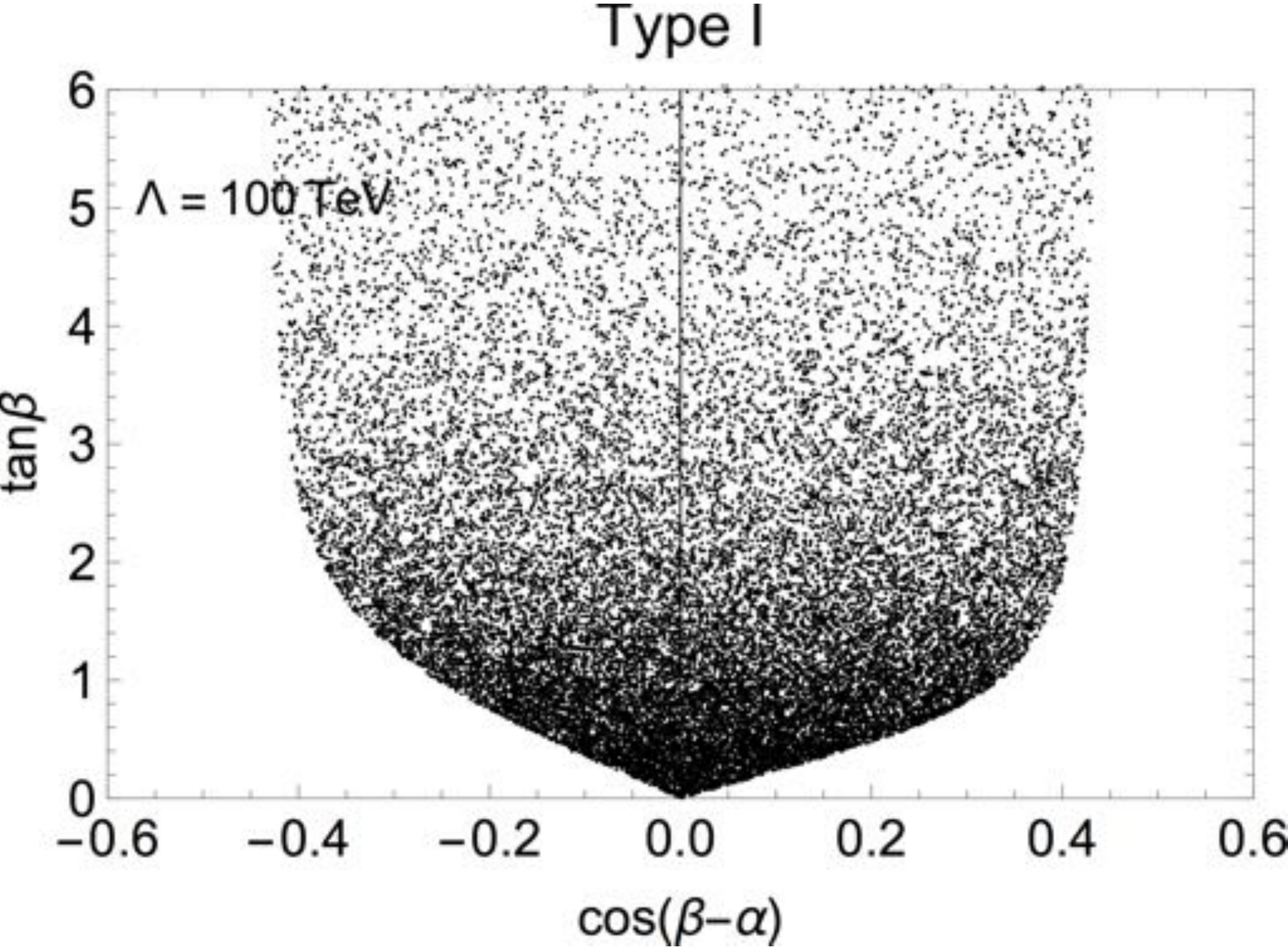} 
\hspace{5 mm}
\includegraphics[scale = 0.175]{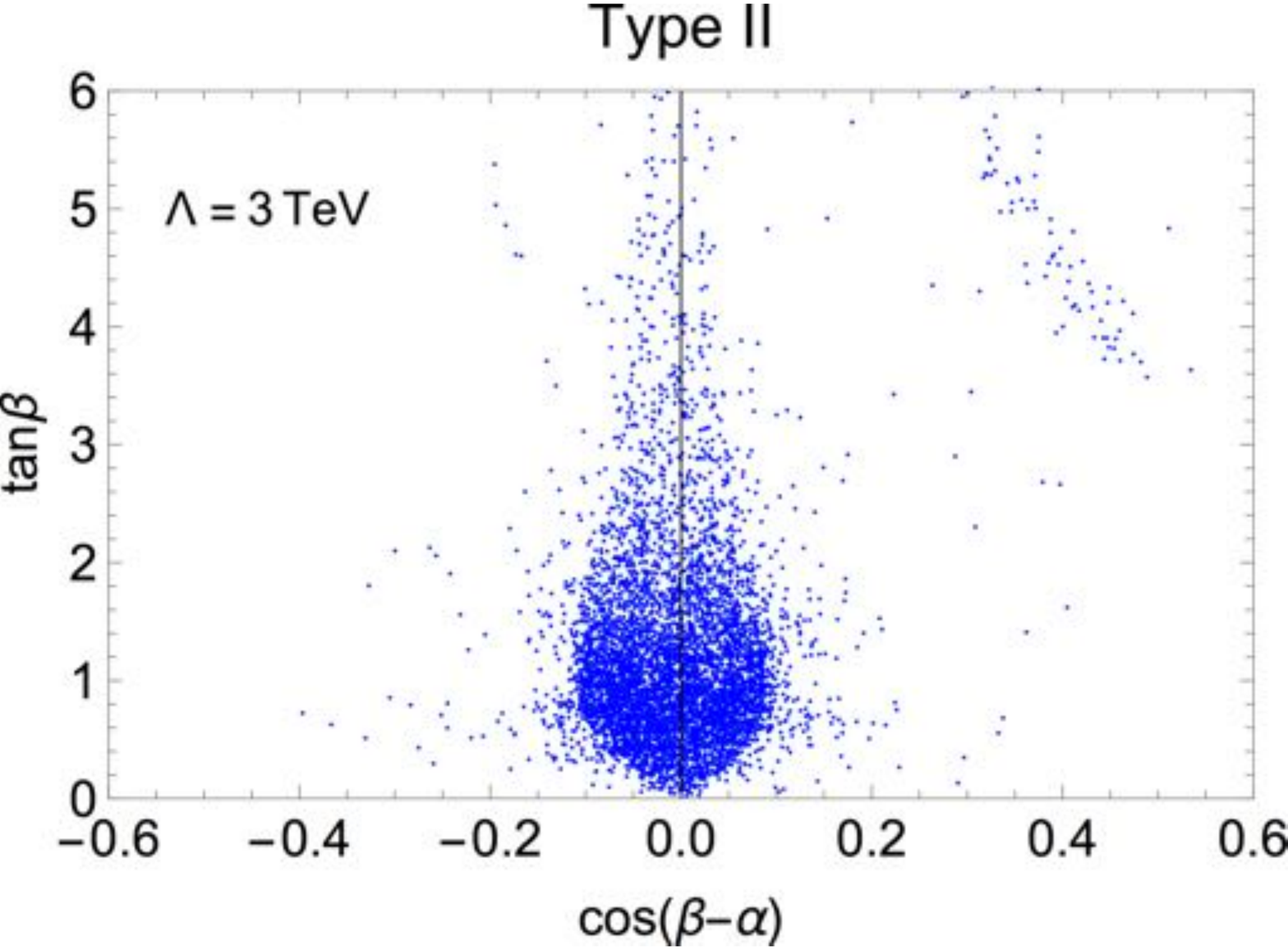}
\includegraphics[scale = 0.175]{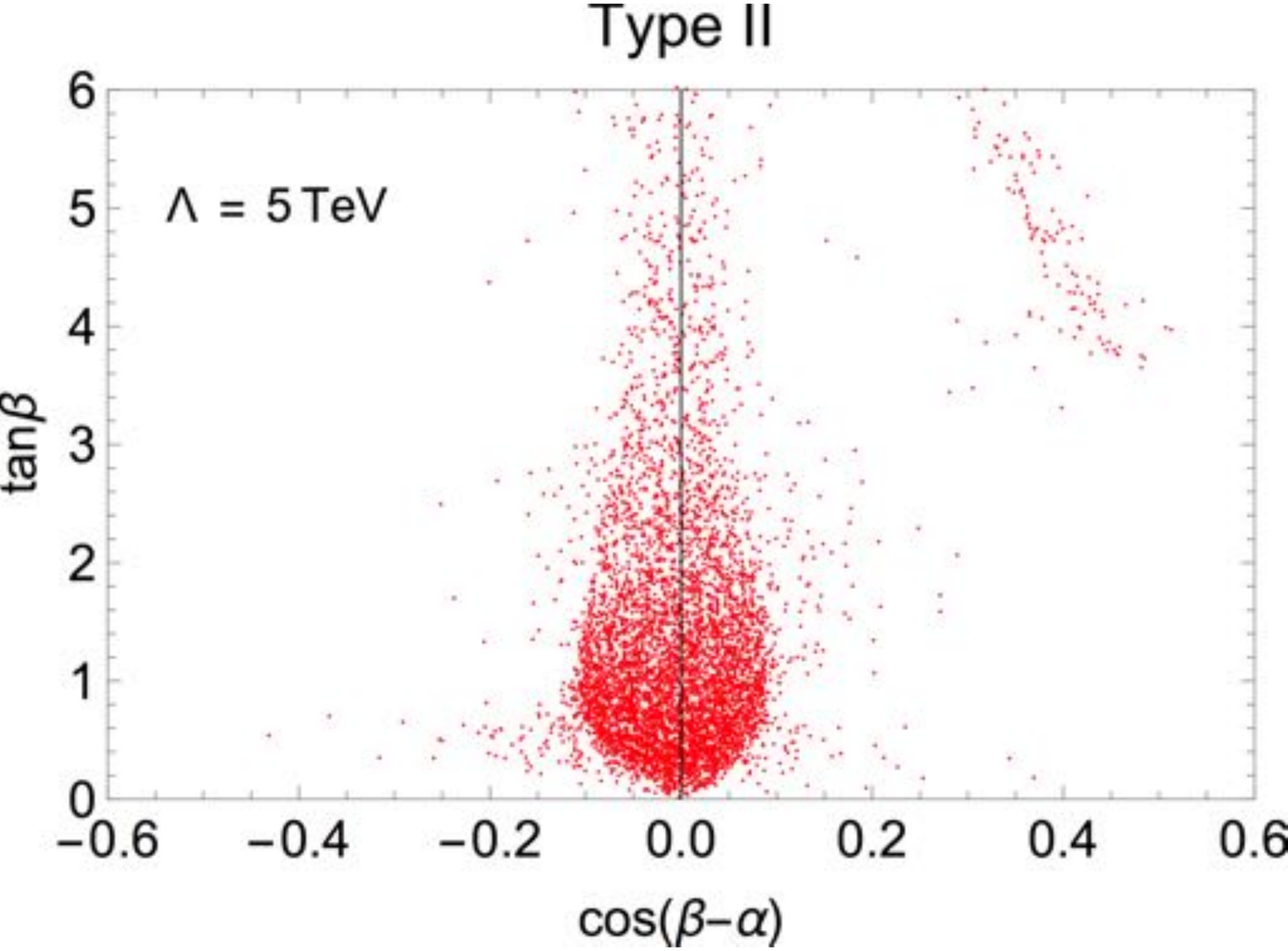}
\includegraphics[scale = 0.175]{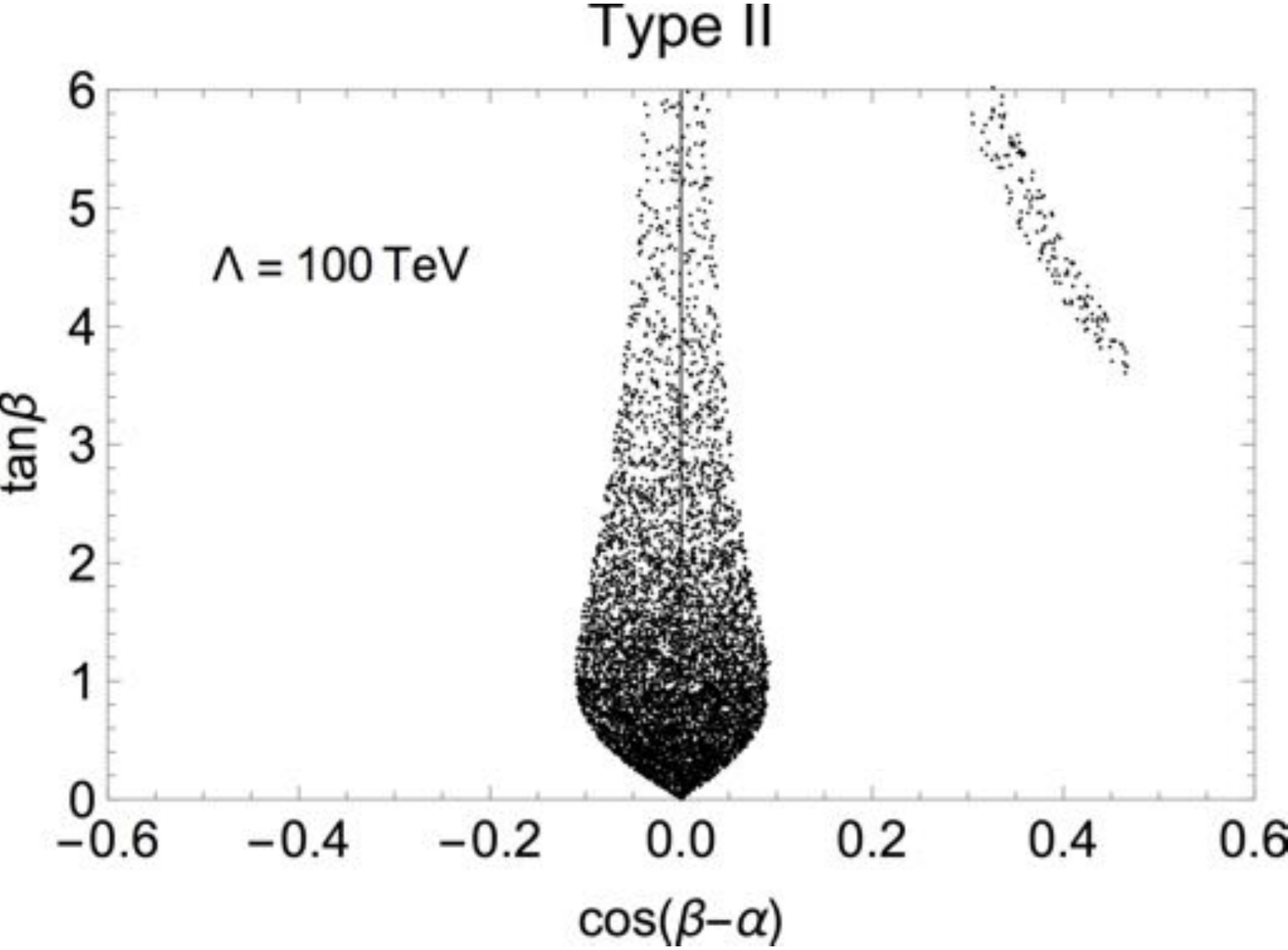} 
\caption{The top plots show the allowed regions for the type-I model and the bottom plots show the allowed regions in the type-II model. The blue (red, black) points shown are used for the $\Lambda = 3 (5, 100)$ $\TeV$ cases.  Values of the curvature scalar couplings, $\xi_1,\xi_2$ were allowed to range between $[-4,4]$.  We have varied the radion and heavy Higgs masses over the range $200$ to $1000$ GeV.}
\label{fig::2hdm_pspace}
\end{figure}

In this section we perform an analysis on the effects Higgs-radion mixing has on the 2HDM parameter space, $\cos(\beta - \alpha)$ and $\tan \beta$. 
We use a chi-square test to fit the model to the data presented in  Appendix \ref{appendixB}  and find the region in the 2HDM parameter space allowed by current LHC data on the SM-like Higgs boson, $h$.  By definition the chi-square function to be minimized is written as 
\begin{equation}
\chi^2 = \sum_{i} \frac{(R^p_i - R_i^{m})^2}{(\sigma_i)^2},
\end{equation}
where $R^P_i$ is the signal strength predicted by the model, $R_i^{m}$ is the measured signal strength and $\sigma_i$ is the corresponding standard deviation of the measured signal strength. Asymmetric uncertainties are averaged in quadrature $\sigma =\sqrt{\frac{\sigma_+^2+\sigma_-^2}{2}}$. The expected signal strengths are defined as the production cross section times branching ratio of a particular decay channel $ff$ normalized to the standard model prediction, i.e.,
\begin{equation}
R^p_f \equiv \frac{\sigma(pp \rightarrow h) BR(h \rightarrow ff)}{\sigma(pp \rightarrow h_{SM}) BR(h_{SM} \rightarrow ff)}.
\end{equation}
Directly obtaining analytical expressions for the mass eigenstates   is challenging therefore we resort to numerical techniques.
The analysis was carried out using two benchmarks for the radion vev, $\Lambda = 3,5\mbox{ TeV}$. We generated random values for 2HDM mixing angles, $(\alpha, \beta)$, the curvature scalar couplings $(\xi_1, \xi_2)$ and the scalar mass parameters before radion mixing $(m_h, m_H, m_r)$ amounting to seven degrees of freedom. By imposing the field $h$ has a mass of $125.09 \pm 0.5$ GeV one degree of freedom is removed leaving us with six degrees of freedom in our chi-square analysis. 
We also constrained the radion and heavy Higgs physical masses to lie in the range $[ 200, 1000 ] \ \GeV$. We plot the points allowed by the LHC data in Fig. \ref{fig::2hdm_pspace} at a 95\% confidence level for the type-I and type-II models.

No signficant difference can be observed between the $\Lambda = 3 $ $ \TeV$ and $\Lambda = 5 $ $\TeV$ plots for each type of model. Therefore it seems that a curvature-scalar mixing has no significant effect on the 2HDM parameter space.  One can understand this by looking at the off-diagonal elements of the mass matrix, equation \eqref{massmatrix}, which are $3\gamma K_{\phi}/Z \sim 1/1000 $ times the diagonal elements. This is a reasonable approximation since we assume natural values for the curvature-scalar mixing parameters, $\xi \sim \mathcal{O}(1)$ and therefore the unitary matrix that diagonalizes \eqref{massmatrix} is nearly diagonal which implies that the couplings of the SM-like Higgs to a pair of gauge bosons and fermions receive very small corrections and are nearly given by the corresponding couplings in the 2HDM, i.e.,
\begin{equation}
g_{hVV} =U_{hh}\sin(\beta-\alpha) + U_{Hh}\cos(\beta-\alpha) + U_{rh}\gamma(1-3\frac{m_V^2 k y_c}{\Lambda^2}) \approx \sin(\beta-\alpha),
\end{equation}
\begin{equation}
g_{hff} = U_{hh} \xi_h^f  + U_{Hh} \xi_{h}^f +  U_{rh} \gamma(c_L - c_R) \approx \xi_h^f  ,
\end{equation}
where $U_{ij}$ are the elements of the non-unitary transformation.  
The general shape of the regions is understood by looking at the behavior of the couplings. In the type-I model  $\xi_h^t = \cos\alpha/\sin\beta$ and in the large $\tan\beta$ limit the production cross section is suppressed, allowing the parameter space to grow. For type-II model the coupling to a pair of $b$ quarks is $\xi_h^b =- \sin\alpha/\cos\beta$ and therefore the production cross section is enhanced by the $b$ quark loop squeezing the parameter space. 
 
 \begin{figure}[htb]
\centering
\hspace*{0.4cm} 
\minipage{0.31\textwidth}
  \includegraphics[width=4.1 cm]{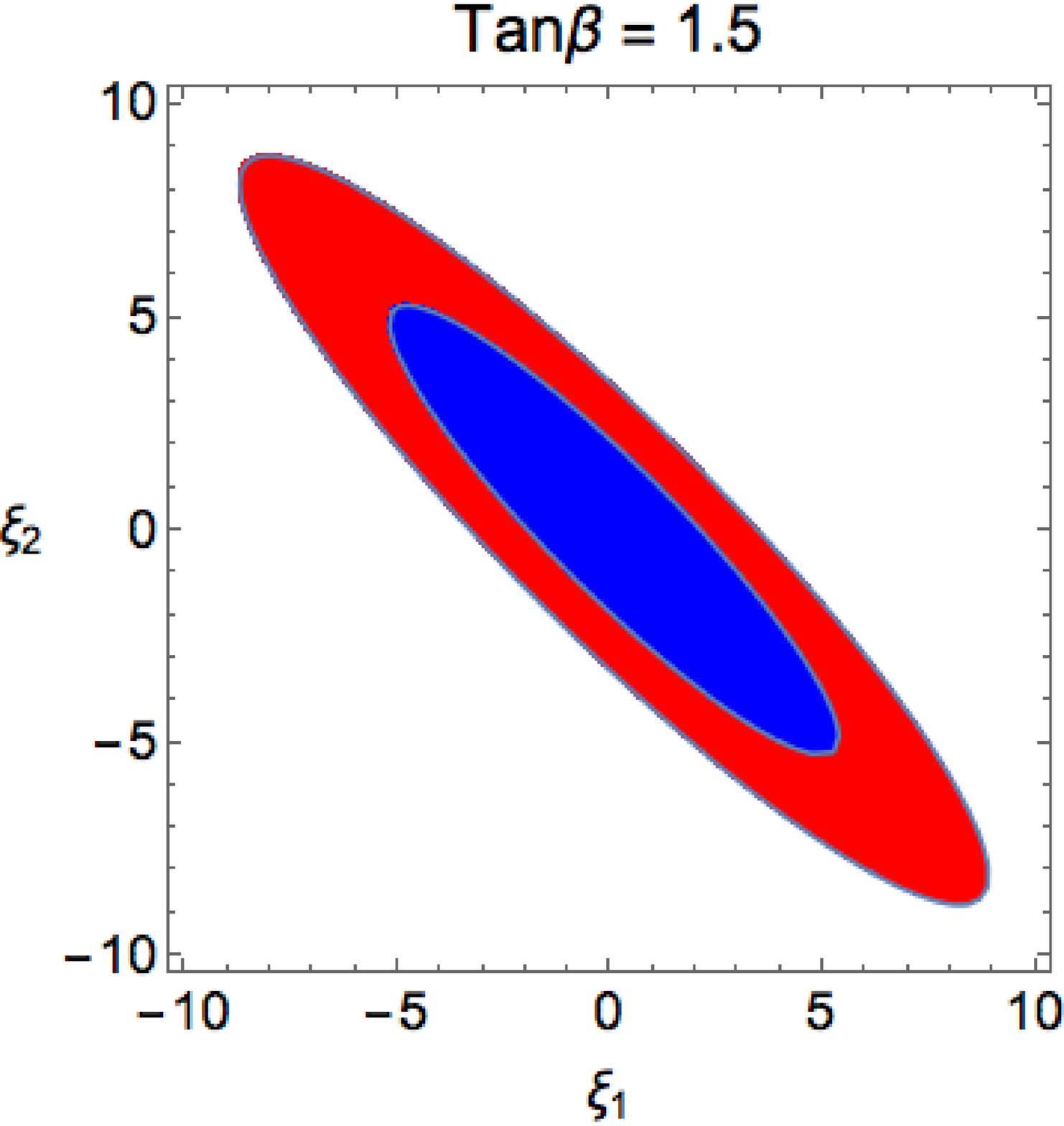}
 \endminipage\hfill
\minipage{0.31\textwidth}
  \hspace*{-0.23 cm} 
  \includegraphics[width=4.1 cm]{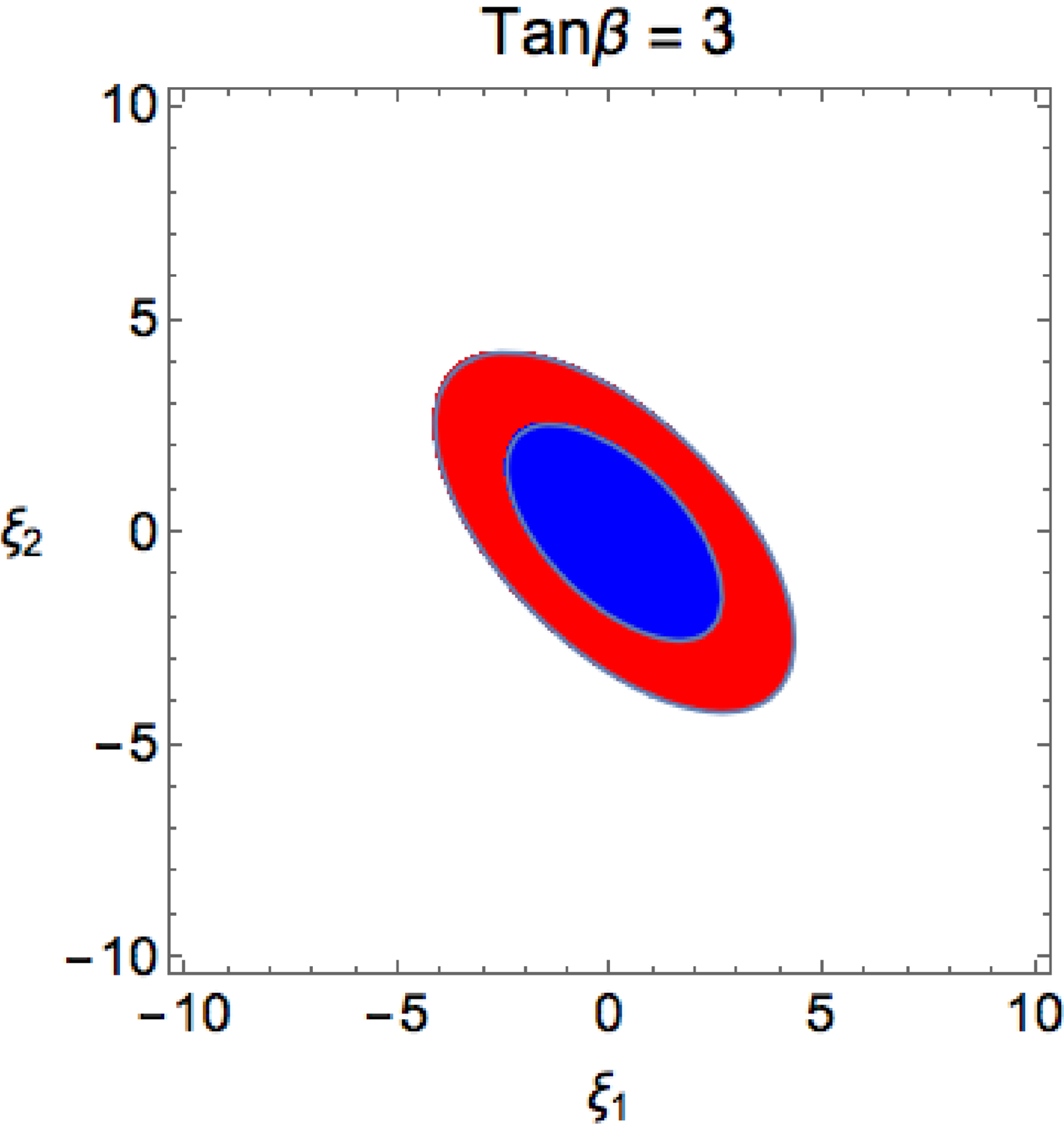}
  \endminipage\hfill
\minipage{0.31\textwidth}%
  \hspace*{-0.5 cm} 
  \includegraphics[width=4.1 cm]{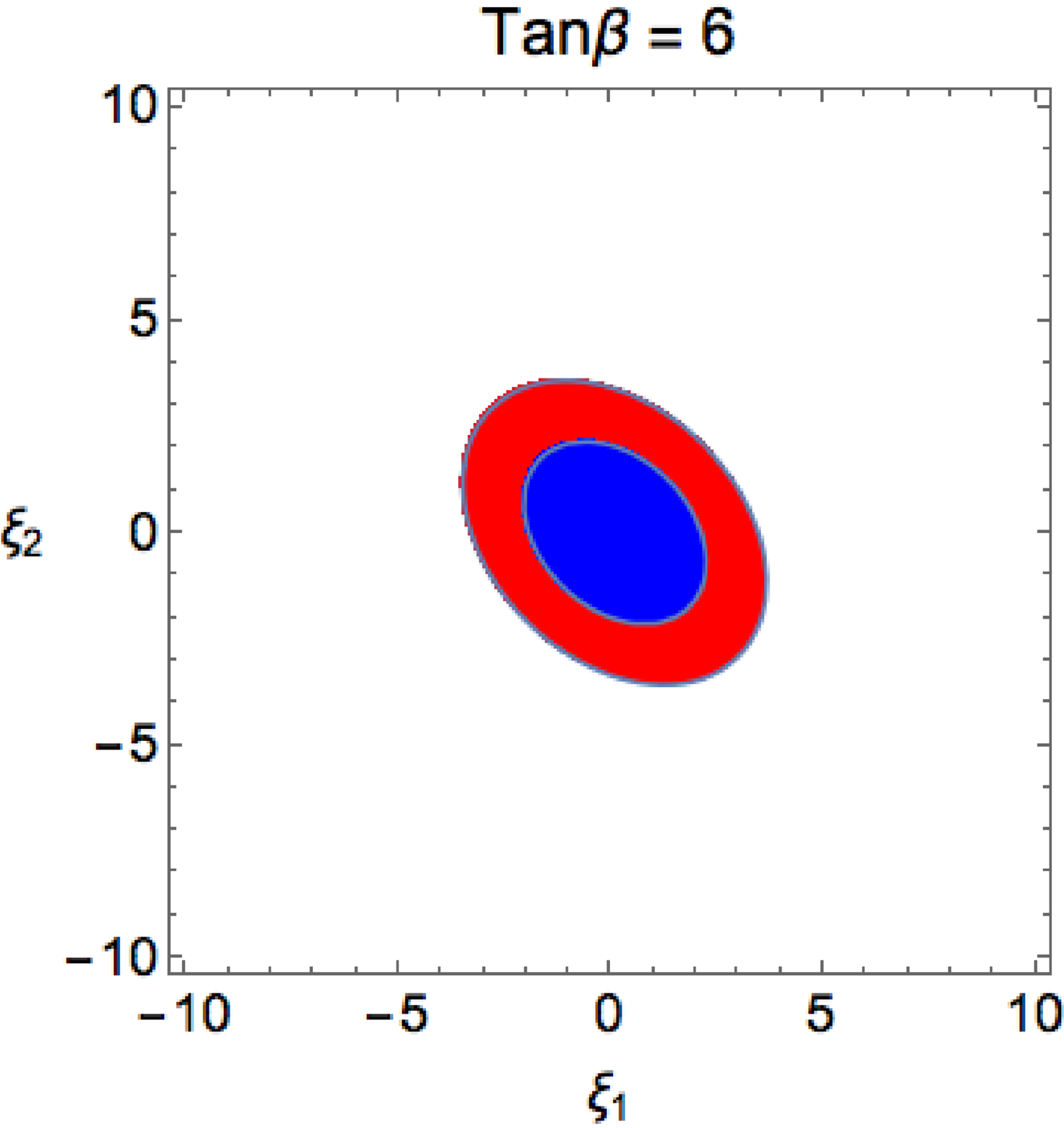}
\endminipage
\vspace{0.2 cm}
\caption{Theoretically allowed $\xi_1$-$\xi_2$ parameter space for different values of $\tan\beta$. The blue (red) region is for $\Lambda=3(5) \TeV$. } \label{fig2}
\end{figure}
 
\begin{figure}[h]
 \centering
\includegraphics[scale = 0.30]{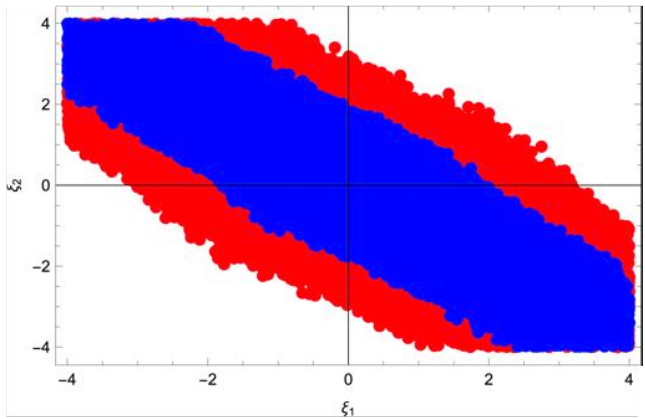} 
\caption{The parameter space of $\xi_1$ and $\xi_2$ allowed by the chi-square goodness of fit. The blue and red points correspond to $\Lambda=3  $ $\TeV$ and $\Lambda=5 $ $\TeV$ respectively.}
\label{fig::x1x2_pspace}
\end{figure}

The allowed region of the curvature-scalar parameter space is constrained by the requirement that the determinant of the kinetic mixing matrix, Eq. \eqref{Z}, be positive.    This condition was discussed in the last section.   We can examine the constraint in the $\xi_2-\xi_1$ plane.
  This depends only on $\tan\beta$ and $\gamma$ and is given, for $\Lambda = 3,5 $ TeV, in Figure \ref{fig2}.   However, large values of the $\xi_i$ can require some fine-tuning, and we have found that the density of points in a scatterplot drops substantially once $\xi_i$ is greater than 4 and less than -4.   As a result, restricting the mixing parameters to the range between $-4 \leq \xi_i \leq 4$ will not substantially affect our scatterplots below.    In that range, the region of the curvature-scalar parameter space allowed by the chi-square test is shown in Fig. \ref{fig::x1x2_pspace}.  The region shrinks by reducing the value of $\Lambda$.    Since the relationship between the $K_h,K_H$ parameters and the $\xi_1,\xi_2$ parameters depends on $\alpha$ and $\beta$, the $S,T$ constraints in the last subsection will not substantially reduce the allowed region (especially in view of the cautionary remarks at the end of the last subsection).

\subsection{Collider Signals}

Let us now consider some predictions of this model accessible to the LHC and how one may distinguish this model from some other multi-Higgs model. One feature of a multi-Higgs model is that the sum of the CP-even scalar couplings to Z bosons in quadrature should total to the square of the SM Higgs coupling to the Z bosons, namely
\begin{equation}
g^{-2}_{h_{SM}ZZ} \sum\limits_i^n g^2_{\phi_i ZZ} = 1.
\label{eq::sum-weak}
\end{equation}
Due to the bulk couplings of the radion to the bulk gauge bosons we find that the sum of the neutral scalar couplings in quadrature normalized to the $h_{SM}ZZ$ coupling gives $1+ \gamma^2(1-3 m_Z^2 k y_c/ \Lambda^2)^2$ being bounded from below by $1$ and setting it apart from other multi-Higgs models. However, this deviation from unity may be quite small. For $\Lambda_\phi = 3$ TeV one finds Eq. \ref{eq::sum-weak} gives 1.0054 and the deviation from unity vanishes in the limit $\Lambda_\phi \to \infty$. It is unlikely that the LHC will be able to measure such a small deviation, but such a measurement may be possible at the future ILC.

Another strategy to distinguish the heavy scalar state H from a radion is to measure the ratio of the widths of the heavy scalars to $b \bar{b}$ and $ZZ$ pairs,
\begin{equation}
R^\Phi_{bb/ZZ} \equiv 
\frac{\Gamma(\Phi \to \bar{b}b)}{\Gamma(\Phi \to ZZ)},
\mbox{   for  } \Phi = {r,\mbox{ }H}.
\end{equation}
The mass eigenstates, $H$ and $r$ are primarily aligned with the unmixed states. This means that couplings of $H$ to the Z boson and $b$ quark should be dominated by the corresponding expressions in a 2HDM . Then for $H$, $R^H_{bb/ZZ}$ should mostly scale like 
$\left( \frac{\sin\alpha}{\sin\beta} \frac{1}{\cos(\beta-\alpha)}\right)^2$ for the type-I model and 
$\left(\frac{\cos\alpha}{\cos\beta} \frac{1}{\cos(\beta-\alpha)}\right)^2$ for the type-II model. In either case this ratio becomes quite large in the neighborhood of $\cos(\beta-\alpha) = 0$. For the radion, in the limit that its fully aligned with the unmixed radion, $R^r_{bb/ZZ} \propto \frac{(c_L-c_R)^2}{\left(1-3\frac{m_Z^2 k y_c}{\Lambda^2}\right)^2} \approx (c_L-c_R)^2$. This is typically less than one and thus measurement of this ratio might distinguish $r$ from $H$.

 As an example, consider the benchmark point with $\tan\beta = 1$, $\cos(\beta - \alpha) = 0.01$,  $\Lambda = 5 \ \TeV $ and moderate mixing $\xi_1 = 2 $ and $\xi_2 =-3$. The values of the masses before mixing are fixed to $m_r = 540 \ \GeV$, $m_h = 125 \ \GeV$ and $m_H = 600 \  \GeV$ which yield the mass eigenvalues $m_r \approx m_H  \approx 600 \ \GeV$, $m_h = 125 \ \GeV$ and $R^r_{bb/ZZ} \approx 0.4$ and $R^H_{bb/ZZ} \approx 5540$.   This is a huge, five order of magnitude difference and would be easily detectable.


\subsection{Constraints From Heavy Higgs searches}

The radion interactions with the scalar sector come from the following sources:
\begin{enumerate}
\item[1]The quartic interactions in the 2HDM potential 
\begin{equation}
V(\Phi_1, \Phi_2 ) \supseteq \frac{\lambda_1}{2}(\Phi_1^\dagger \Phi_1)^2 + \frac{\lambda_2}{2}(\Phi_2^\dagger \Phi_2)^2 + \lambda_3 \Phi_1^\dagger \Phi_1 \Phi_2^\dagger \Phi_2 + \frac{\lambda_4}{2}(\Phi_1^\dagger \Phi_2 +\Phi_2^\dagger \Phi_1)^2.
\end{equation}
\item[2] The coupling of the radion with the trace of the energy momentum tensor
\begin{equation}
\mathcal{L} \supseteq  -\frac{r}{\Lambda}((\partial_\mu h)^2 -2m_h^2 h^2 + ...).
\end{equation}
\item[3] The curvature-scalar mixing term $\mathcal{L} = -\xi_{ab}\mathcal{R} \Phi_a^\dagger \Phi_b$, where we expand the Ricci scalar up to second order in $\gamma$:
\begin{equation}
\mathcal{R} \supseteq -\frac{\gamma}{v} \Box  r +  2 \frac{\gamma^2}{v^2} r \Box r + \frac{\gamma^2}{v^2}(\partial_\mu r)^2 + \mathcal{O}(\gamma^3).
\end{equation}
\item[4] There is a model dependent contribution coming from the potential of the GW scalar field that one can consider however we will assume this interaction to be small as it is proven in \cite{Dominici:2002jv} that addition of this extra term doesn't affect the phenomenology.
\item[5] Non-zero mixing will also induce tree-level interactions of the radion with a gauge field and a scalar, namely $rW^{\pm}H^{\mp}$ and $rZA$ coming from a direct expansion of the kinetic term in equation \eqref{kinetic term}.
\end{enumerate} 

In this model the amount of kinetic mixing between the Higgs field and the radion is parametrized by the parameter $K_h$ of equation \eqref{kh}. Similarly the amount of kinetic mixing between the heavy Higgs state and the radion is encoded in the parameter $K_H$ given in equation \eqref{kH}. We use the most recent LHC direct searches for a heavy scalar decaying into a pair of SM Higgs bosons \cite{Aaboud:2018knk,Sirunyan:2017guj}, into  $WW $ bosons \cite{Aaboud:2017gsl} and into a pair of $ZZ$ bosons  \cite{Sirunyan:2018qlb}  to find bounds on the amount of mixing. The most relevant decay channels, when kinematically accesible, are $\phi_i \rightarrow hh, \phi_j \phi_j, h\phi_j, bb, tt, WW, ZZ, gg, AA, H^+ H^-, ZA, W^{\pm}H^{\mp}$ with $\phi_i=r, H$.   The trilinear interactions coming from the 2HDM potential have a dependence on the pseudoscalar mass $m_A$ and on the quartic coupling of the potential $\lambda_4$.  
\begin{figure}[H]
\centering
\includegraphics[scale = 0.250]{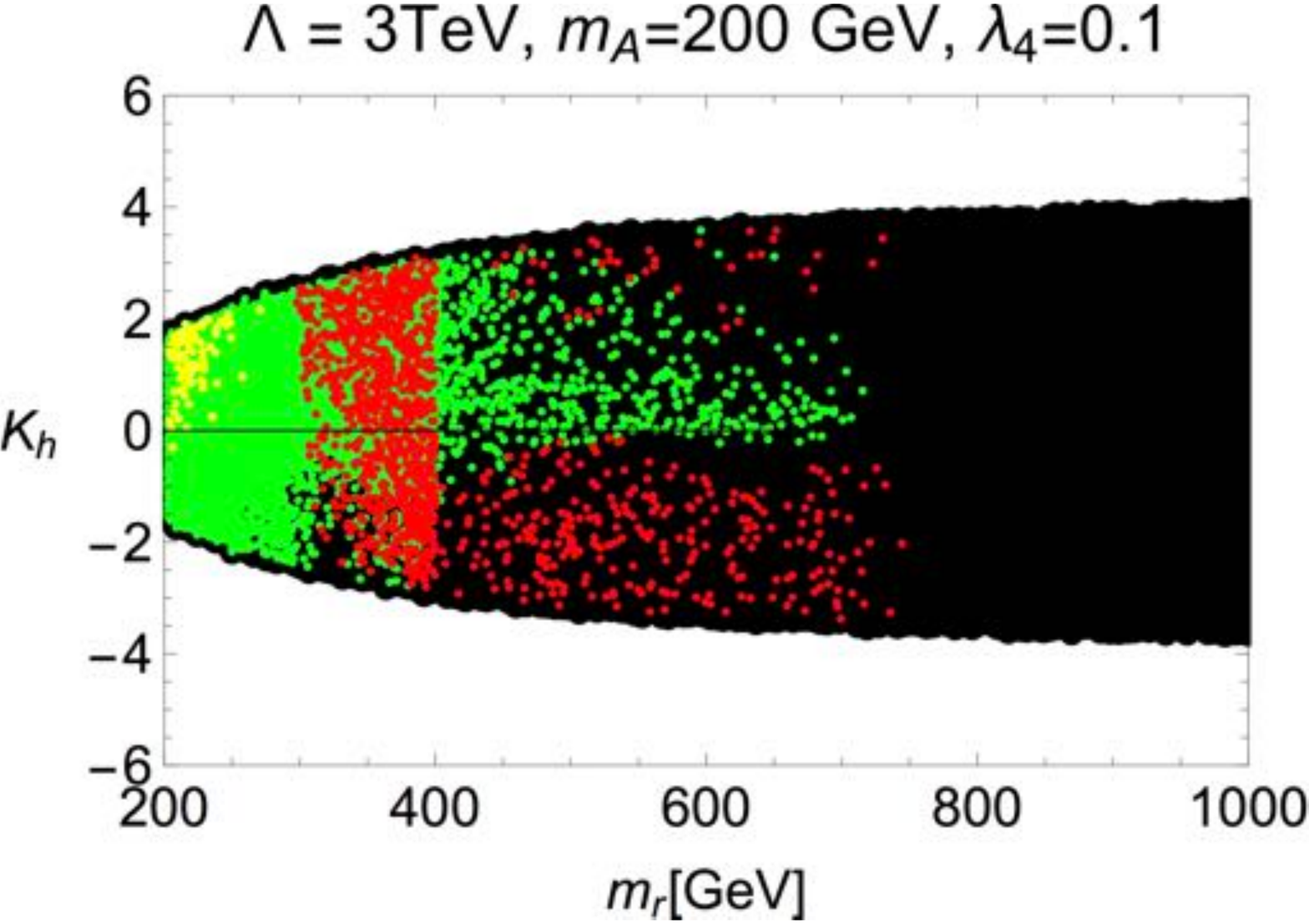} 
\hspace{5mm}
\includegraphics[scale = 0.25]{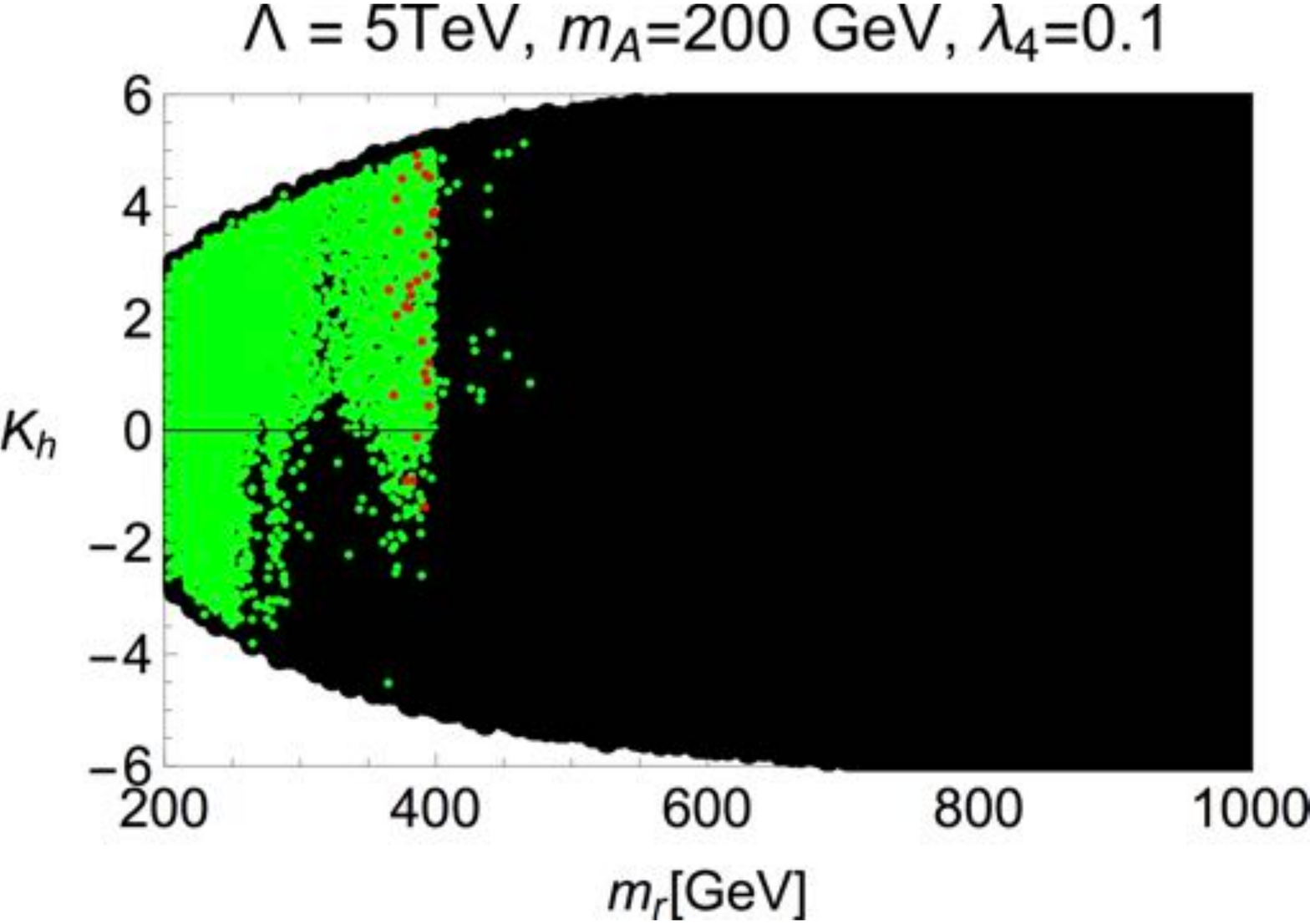} 
\hspace{30mm}
\includegraphics[scale = 0.25]{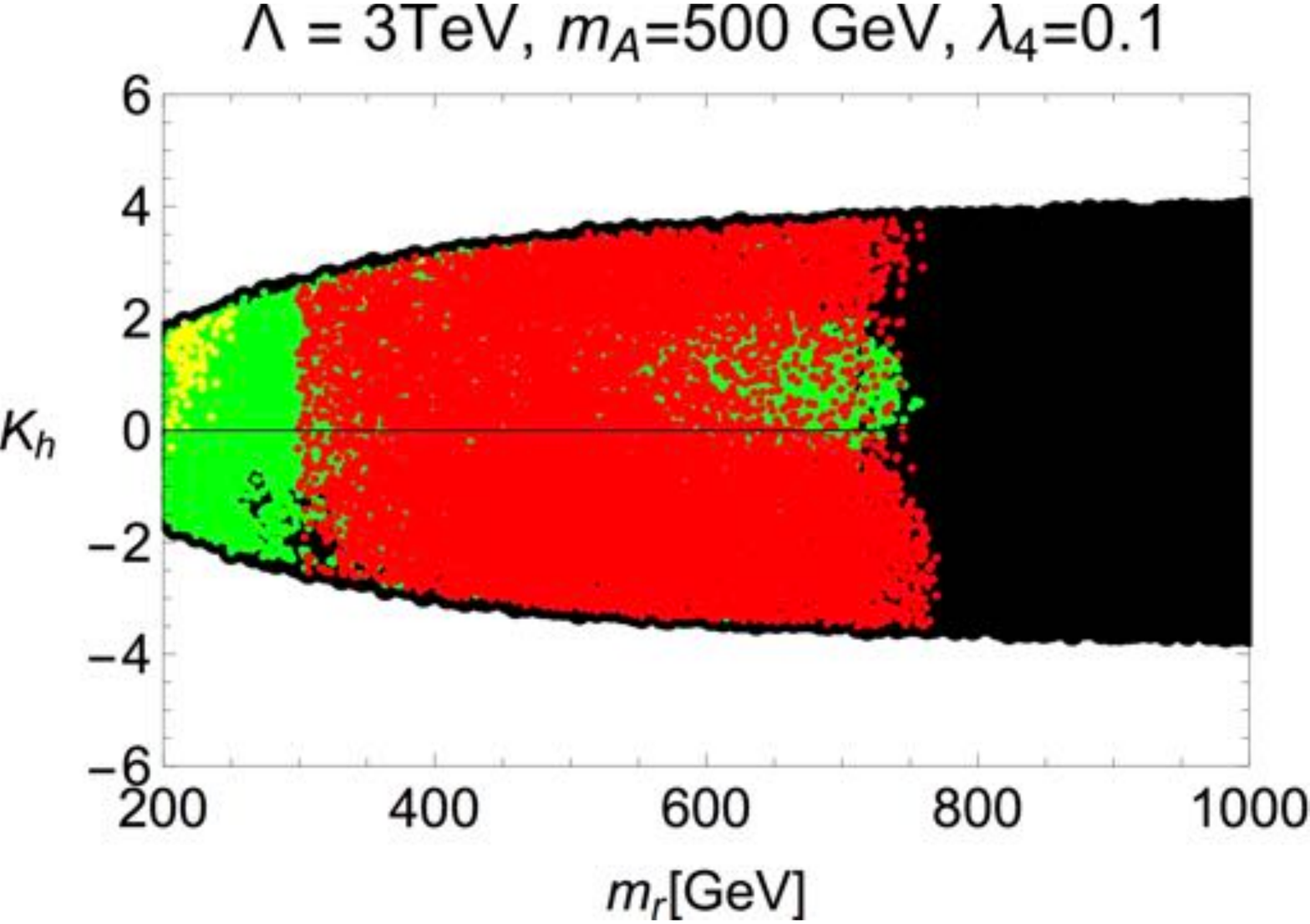} 
\hspace{5mm}
\includegraphics[scale = 0.25]{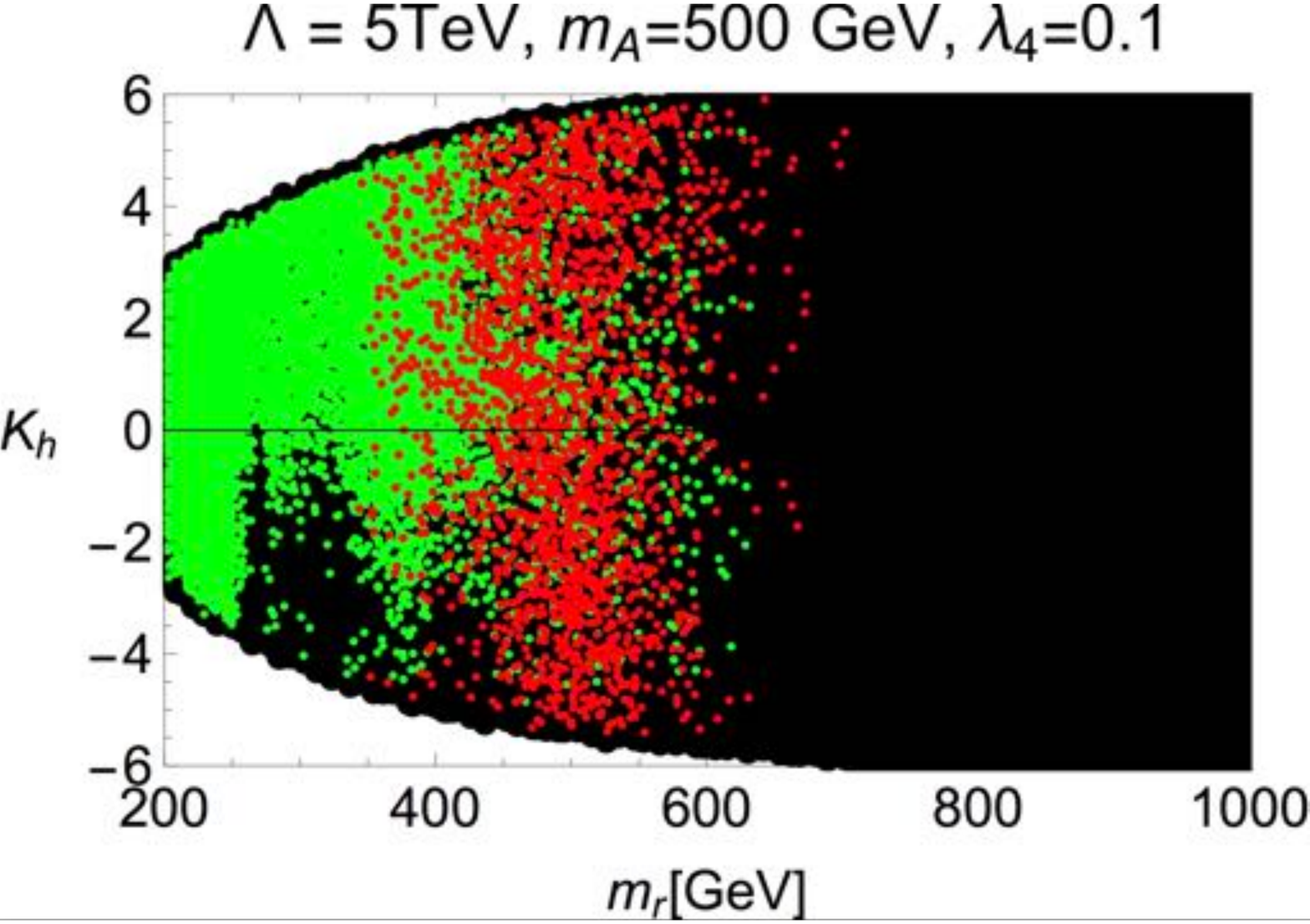}
\hspace{30mm}
\includegraphics[scale = 0.25]{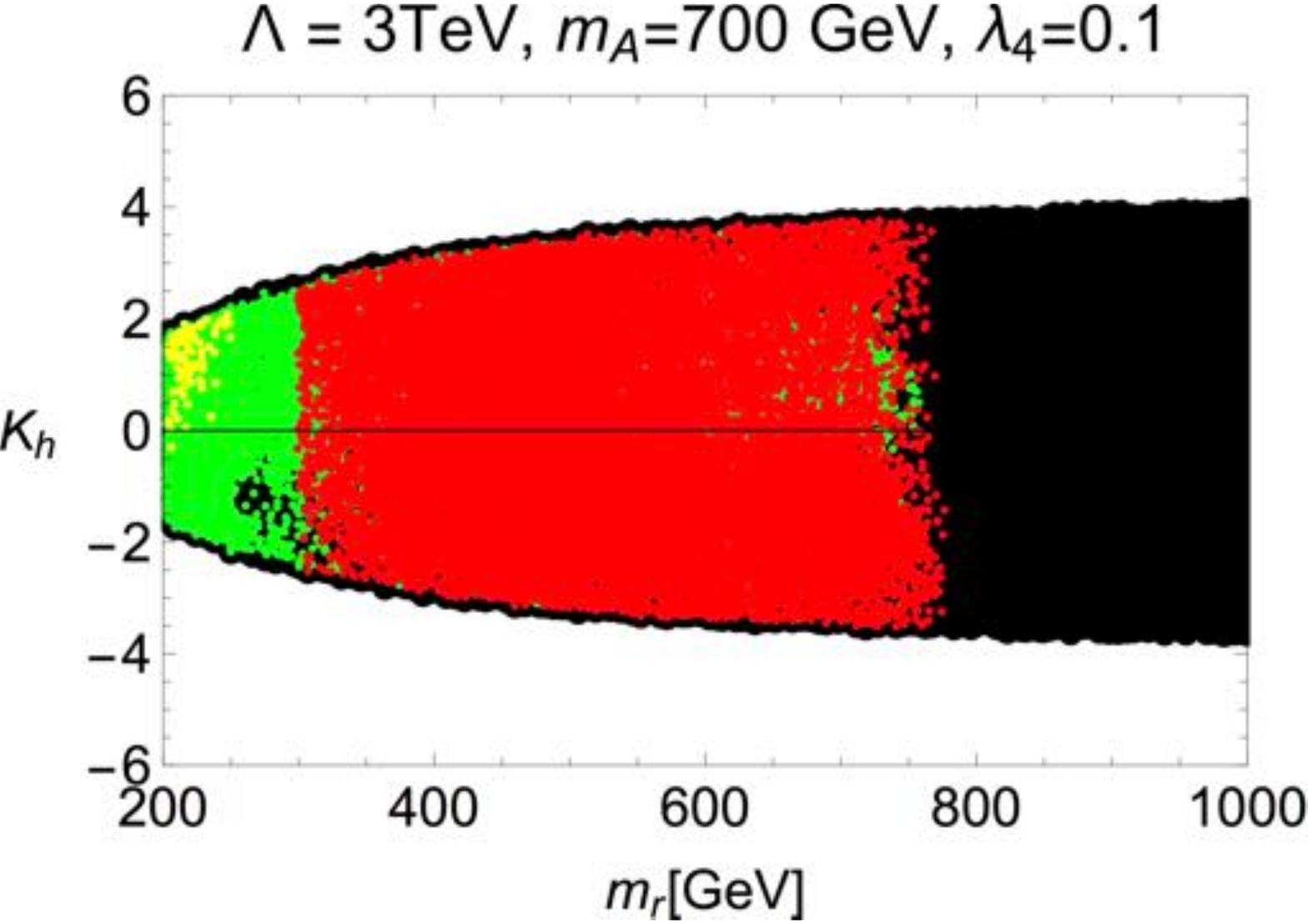} 
\hspace{5mm}
\includegraphics[scale = 0.25]{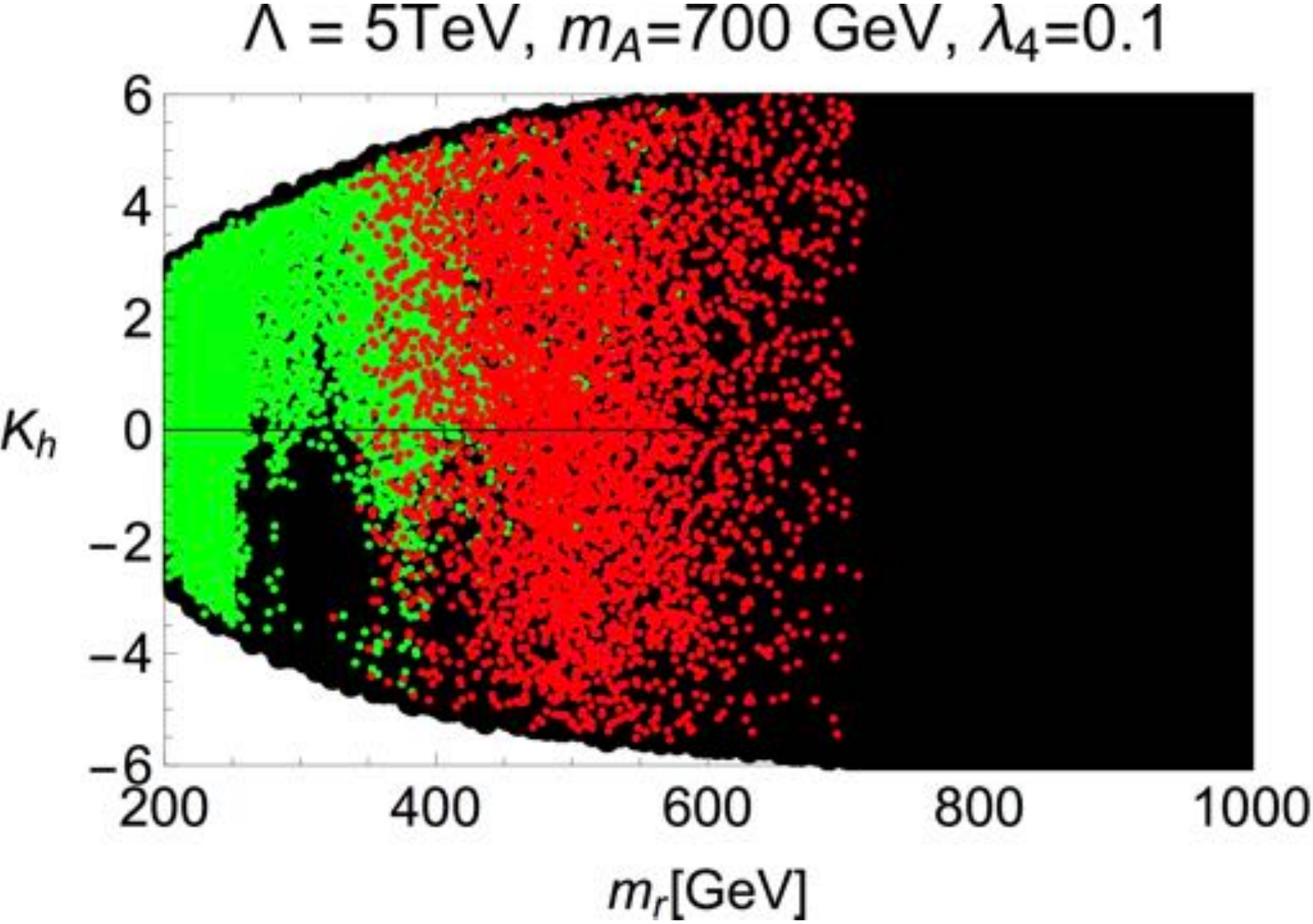} 
\caption{ Scatter plots of the amount of mixing between the Higgs and the radion, $K_h$ defined in equation \eqref{kh}, as function of the radion mass for the type-I 2HDM. The black region is theoretically allowed and the points colored yellow, green and red are forbidden by heavy scalar searches in the $WW$, $ZZ$ and $hh$ channels respectively. The benchmark point $\Lambda = 3  (5)  \TeV$ was used on the left (right).  Due to the custodial symmetry, the charged scalar mass is identical to the pseudoscalar mass, whose value is given above each figure.   The heavy neutral Higgs mass, $m_H$, is varied from $200$ to $1000$ GeV.} \label{mixingparameter}
\end{figure}
We scanned over all the parameters and chose as benchmark values $\Lambda=3, 5 \ \TeV$, $m_A=200, 500, 700 \ \GeV$  and fixed $\lambda_4=0.1$. Changing the value of the quartic coupling does not affect significantly the results. The results are presented as scattered plots in figures \ref{mixingparameter} and  \ref{mixingparameter1} where we show the allowed region in $m_r$-$K_h$ and $m_H$-$K_H$ parameter space for the type-I 2HDM (for the type-II the results are not dramatically different and therefore we do not show them here). In those figures the background black points correspond to the points that are both theoretically allowed and that survived the chi-square analysis of the previous subsection while the points colored yellow, green and red correspond to regions that are forbidden by LHC searches of a heavy scalar decaying in the $WW$, $ZZ$ and $HH$ channels respectively. No bounds were found from Higgs resonant production searches in \cite{Sirunyan:2017guj}.  
One can immediately notice that direct searches in the $WW$ and $ZZ$ channel forbid mainly the low mass region $m_{r}=200-400 \ \GeV$ with the bounds from thee $WW$ being weaker than those from the $ZZ$ channel and no bounds at all from the $WW$ channel were found for the heavy Higgs. The di-Higgs search channels put constraints mostly in the intermediate mass region $m_{r/H} = 300-800 \ \GeV$.

 From the figure  we can notice that as the pseudoscalar mass increases the bounds coming from the di-Higgs boson and $ZZ$ channels become more stringent. This is reasonable since an increase in the pseudoscalar mass corresponds, via the 2HDM potential, to an increase in the trilinear coupling of the radion to a pair of SM Higgs fields and the branching fraction becomes bigger.

The LHC has also searched for a CP-odd Higgs scalar in the processes $pp \rightarrow H/A \rightarrow Z A/H $  \cite{ Khachatryan:2016are, CMS:2016qxc,Aaboud:2018eoy} where the final state $Z$ boson decays into two oppositely charged electrons or muons and the scalar, either $H$ or $A$, is assumed to decay into a pair of $b$ quarks. These final states were motivated by the large branching fractions predicted in a 2HDM with type-II Yukawa structure and the benchmark values $\tan\beta=0.5 $-$1.5$ and $\cos(\beta- \alpha) =0.01$ are used in those references. In those papers, the charged Higgs boson masses were kept equal to the highest mass involved in the benchmark signal, namely $m^2_{H^\pm} \approx m_H^2$ for $H \rightarrow ZA$ or $m^2_{H^\pm} \approx m_A^2$ for $A \rightarrow ZH$.

  Due to the custodial symmetry imposed in the 2HDM potential we can only account for the latter triplet mass degeneracy but we can consider both decay topologies. To the best of our knowledge there has been no search for the signal $H \rightarrow ZA$ with $m_{H^\pm}\approx m_A$. If such a search appears in the literature we would expect more stringent bounds since the branching fraction $BR(H \rightarrow ZA)$ would be reduced by the opening of the channels $H^+ H^-$ and $W^\pm H^\mp$.

\begin{figure}[H]
\centering
\includegraphics[scale = 0.25]{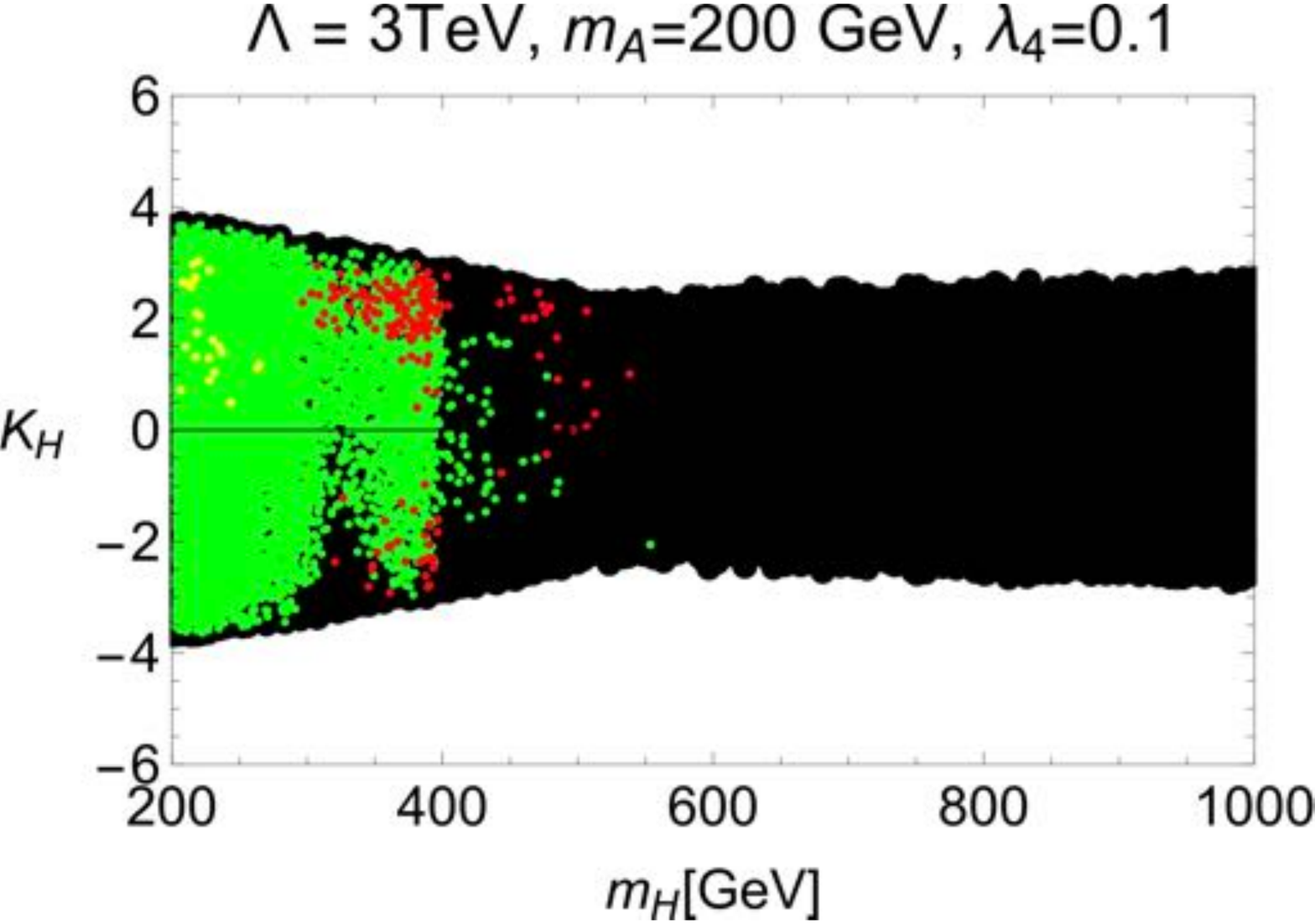} 
\hspace{5mm}
\includegraphics[scale = 0.25]{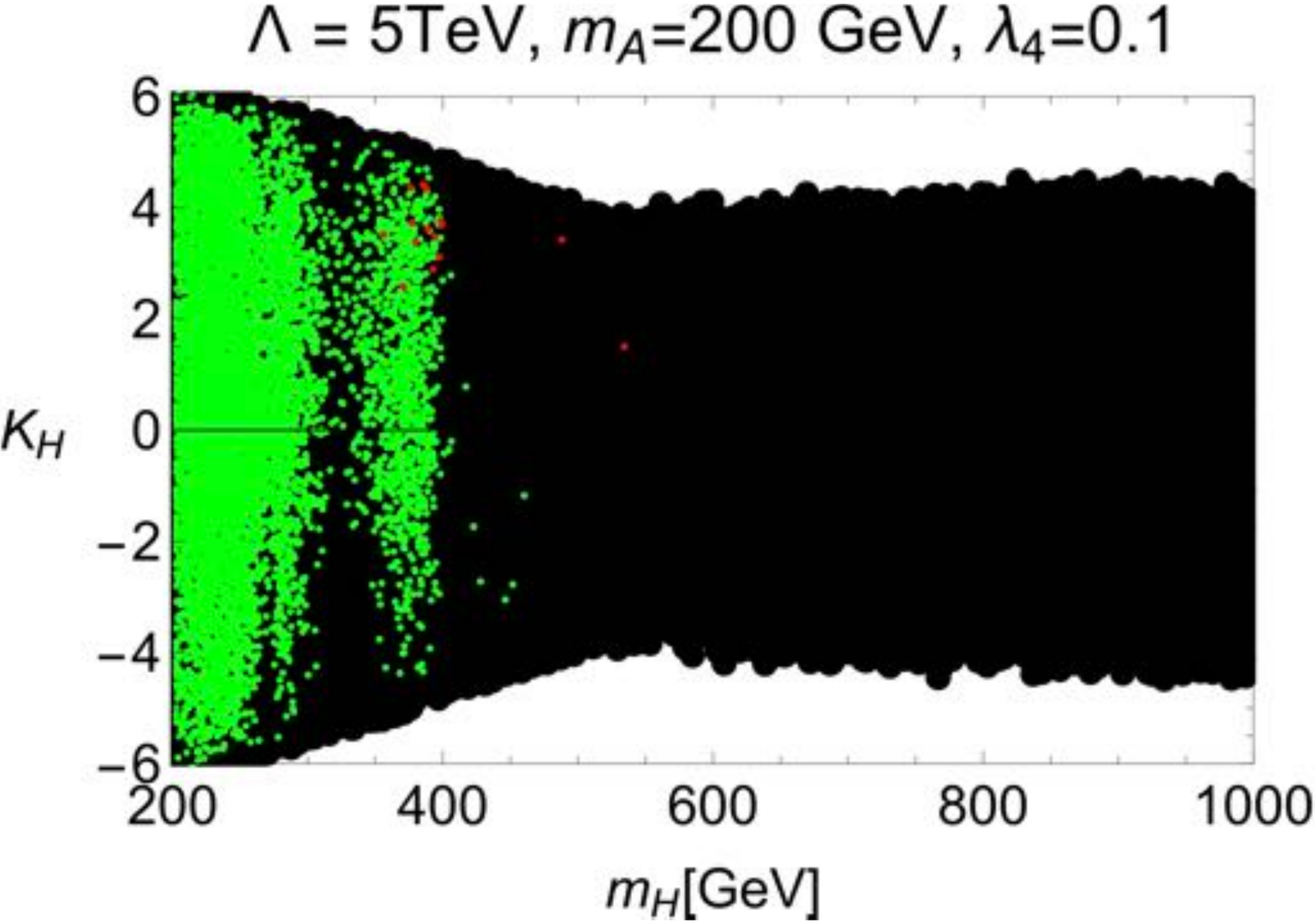} 
\hspace{5mm}
\includegraphics[scale = 0.25]{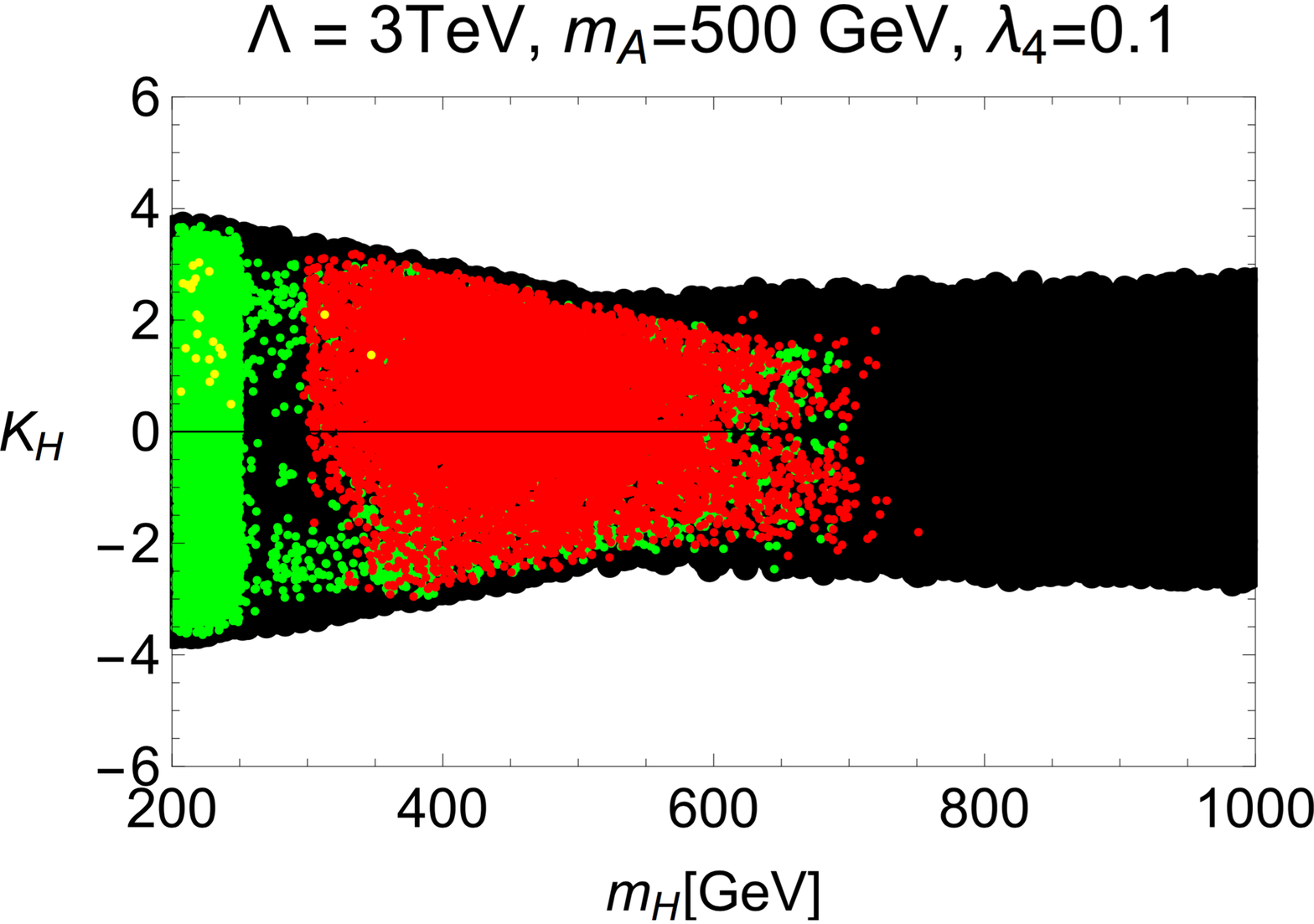} 
\hspace{5mm}
\includegraphics[scale = 0.25]{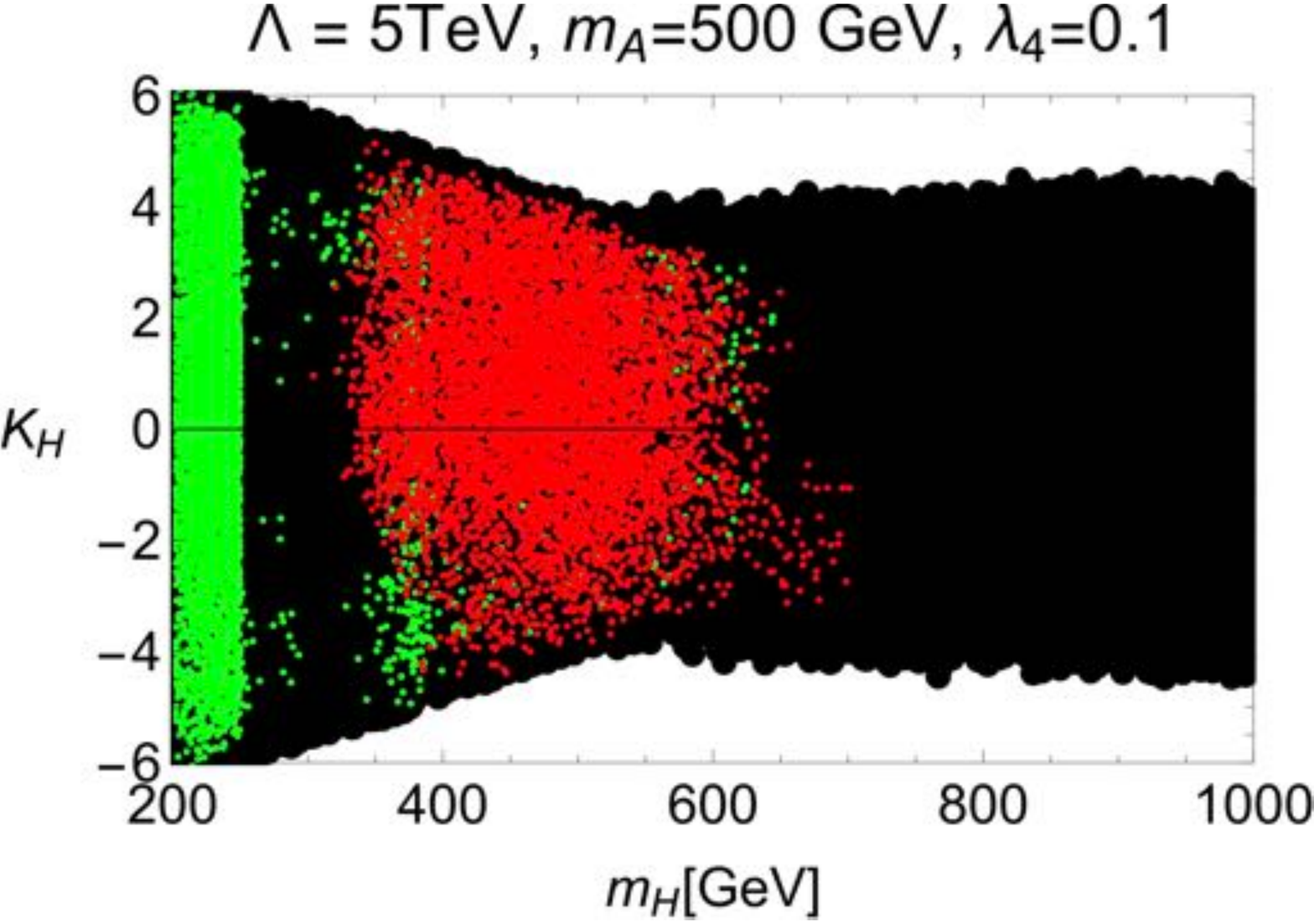} 
\hspace{5mm}
\includegraphics[scale = 0.25]{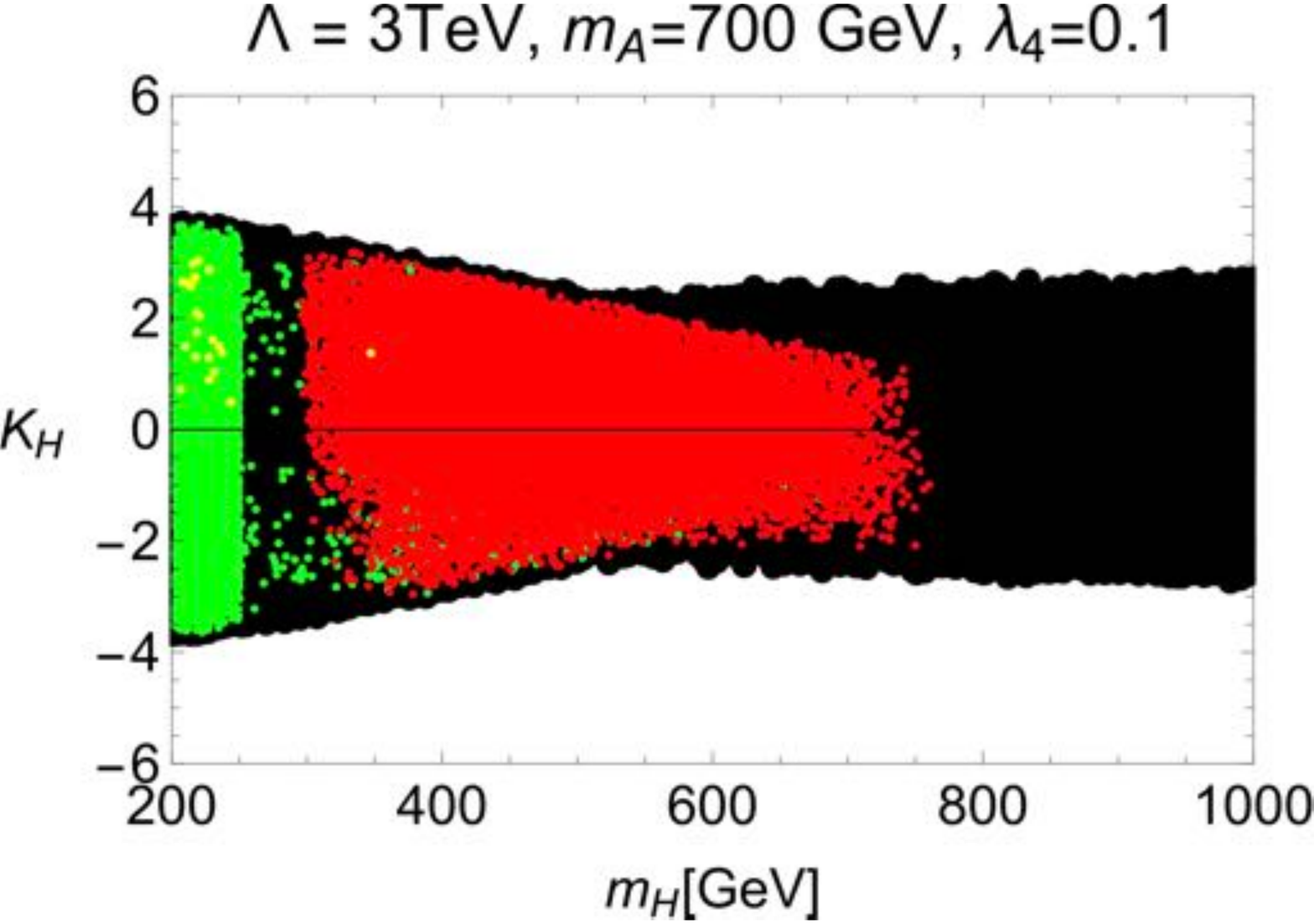} 
\hspace{5mm}
\includegraphics[scale = 0.25]{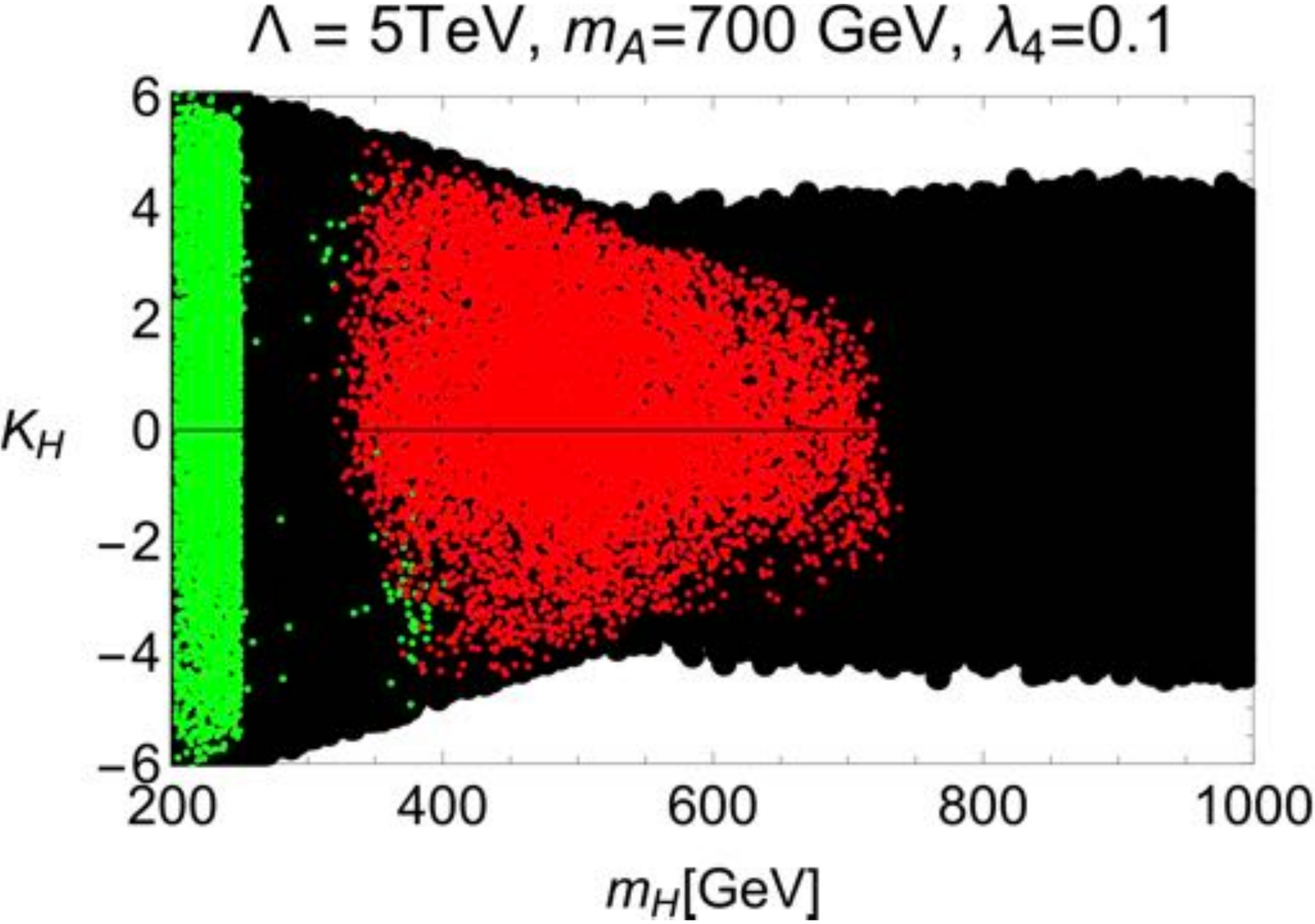} 
\hspace{28mm}  
\caption{ Scatter plots of the amount of mixing between the heavy Higgs and the radion, $K_H$ defined in equation \eqref{kH}, as function of the heavy Higgs mass for the type-I 2HDM. The black region is theoretically allowed and the points colored yellow, green and red are forbidden by heavy scalar searches in the $WW$, $ZZ$ and $hh$ channels respectively. The benchmark point $\Lambda = 3  (5)  \TeV$ was used on the left (right). Due to the custodial symmetry, the charged scalar mass is identical to the pseudoscalar mass, whose value is given above each figure.   The radion mass, $m_r$, is varied from $200$ to $1000$ GeV.} \label{mixingparameter1}
\end{figure}

\begin{figure}[h]
  \centering
\includegraphics[scale=0.25]{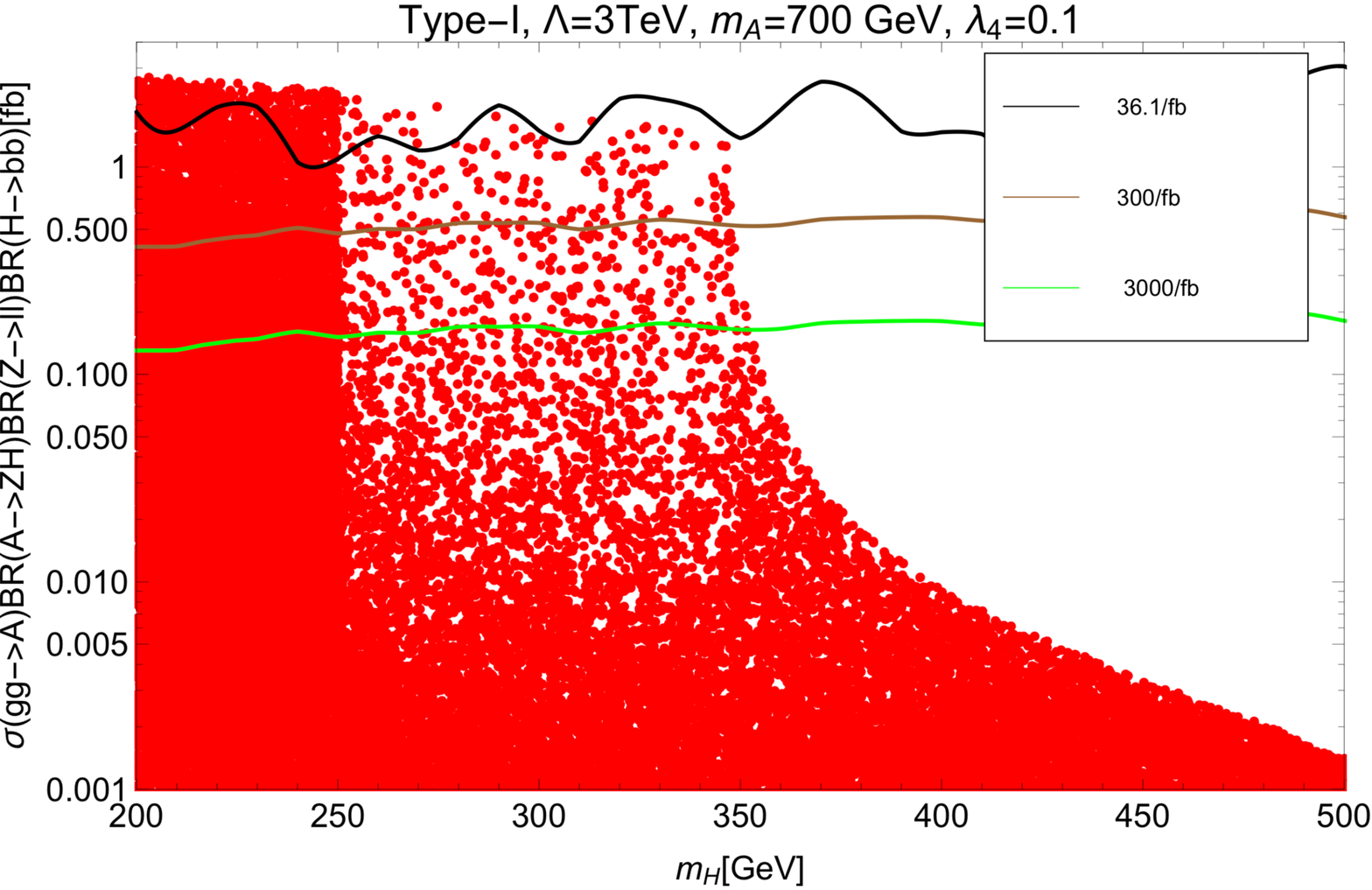}
\hspace{5mm}
\includegraphics[scale=0.25]{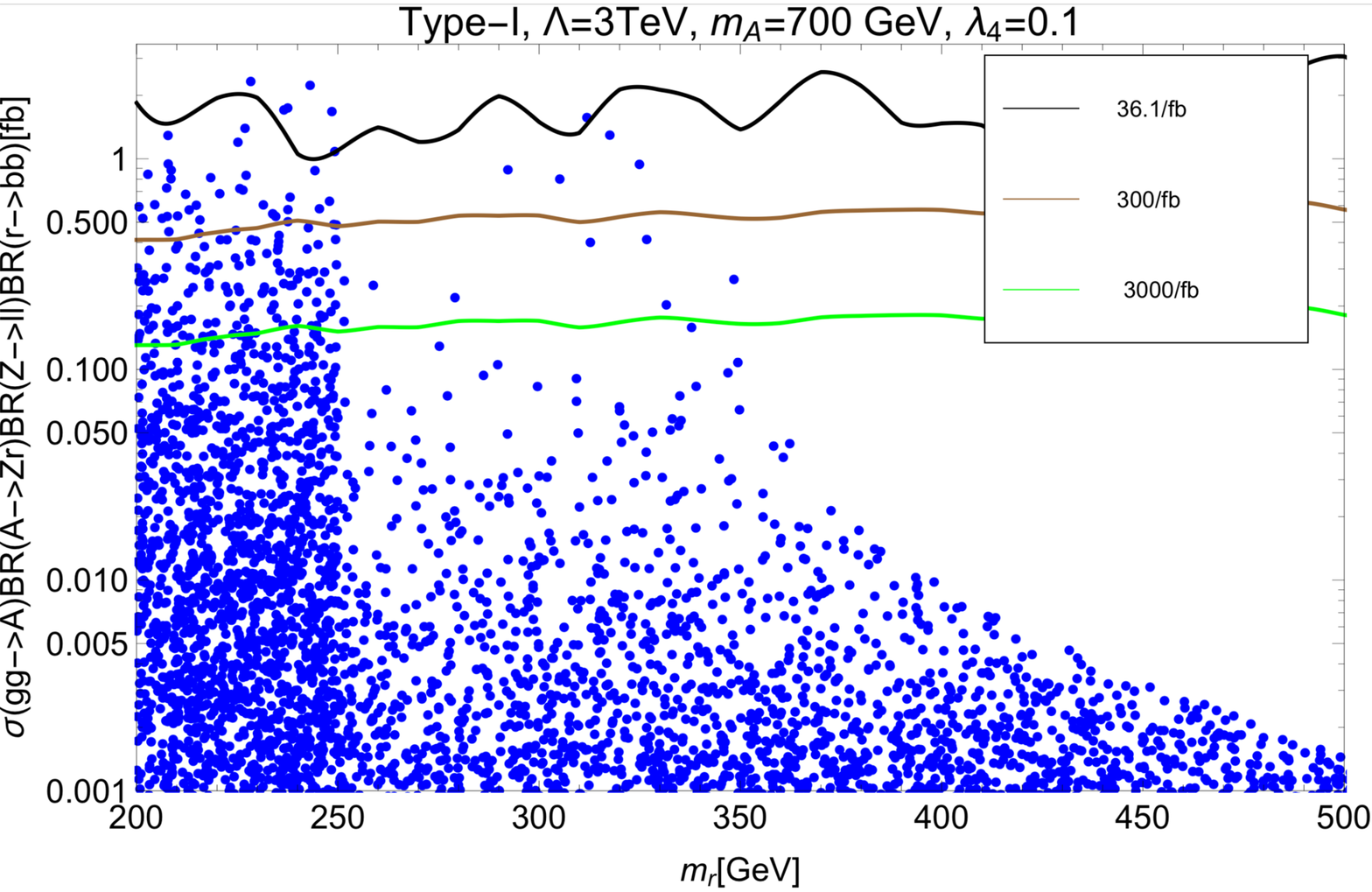}
\hspace{5mm}
\includegraphics[scale=0.25]{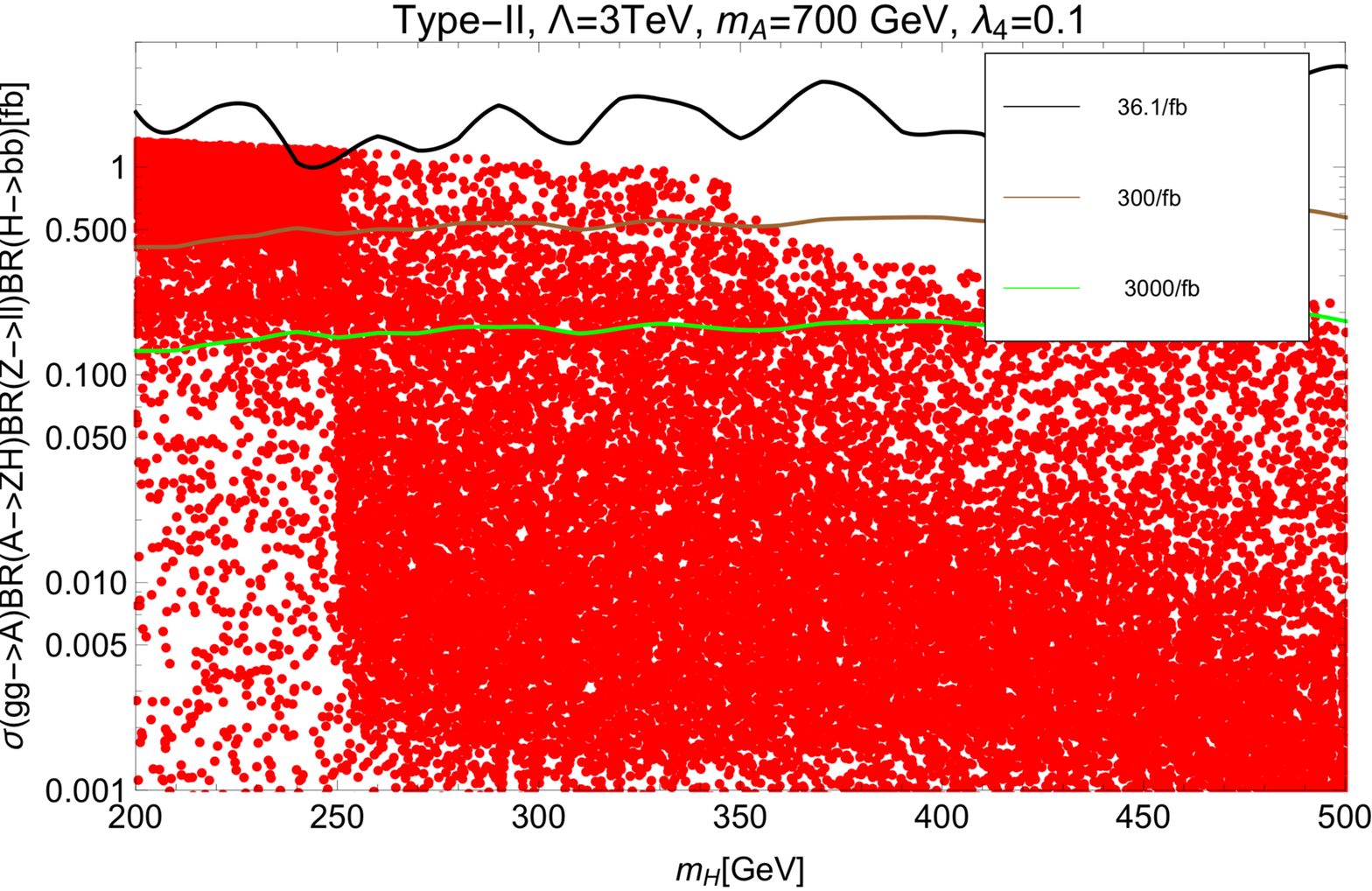}
\hspace{5mm}
\includegraphics[scale=0.25]{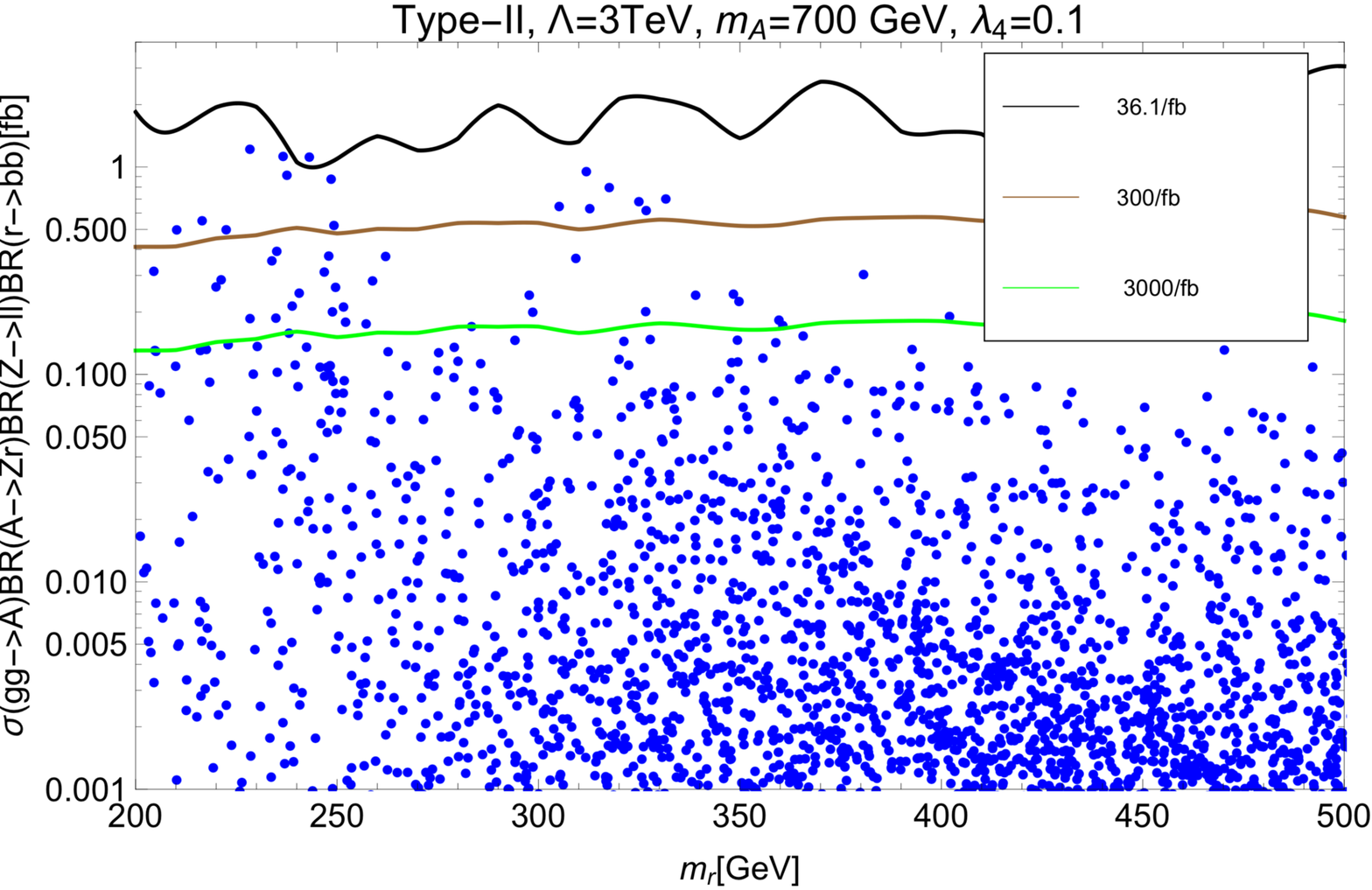}
\caption{The observable $\sigma(gg \rightarrow A \rightarrow ZX) BR(Z \rightarrow l^+ l^-) BR(X \rightarrow b \bar{b})$ as a function of the resonance mass with $X= H$(red), $r$(blue) for type-I (top) and type-II (bottom) models. We fixed $\Lambda = 3 \ \TeV$,  $m_A=700\  \GeV$ and $\lambda_4 = 0.1$. Due to the custodial symmetry, the charged scalar mass is identical to the pseudoscalar mass, whose value is given above each figure.   The heavy neutral Higgs (radion) mass is varied from $200$ to $1000$ GeV in the right (left) figures and the values of $\alpha$ and $\beta$ are chosen to be consistent with the constraints of Figure \ref{fig::2hdm_pspace}. The solid lines represent current and future upper bounds at the LHC.}  \label{ZAR}
\end{figure}

In figure \ref{ZAR} we show the production cross section, via gluon fusion, for $A$ times the branching fractions  $BR(A \rightarrow ZX)BR(Z \rightarrow l^+ l^-)BR(X \rightarrow b \bar{b})$ in the type-I (top) and type-II model (bottom) as a function of the mass $m_X$ where $X=H$(red), $r$(blue). The values $m_A = 700 \ \GeV$ and $\lambda_4=0.1$ were fixed. 

\begin{figure}[h]
 \centering
\includegraphics[scale=0.25]{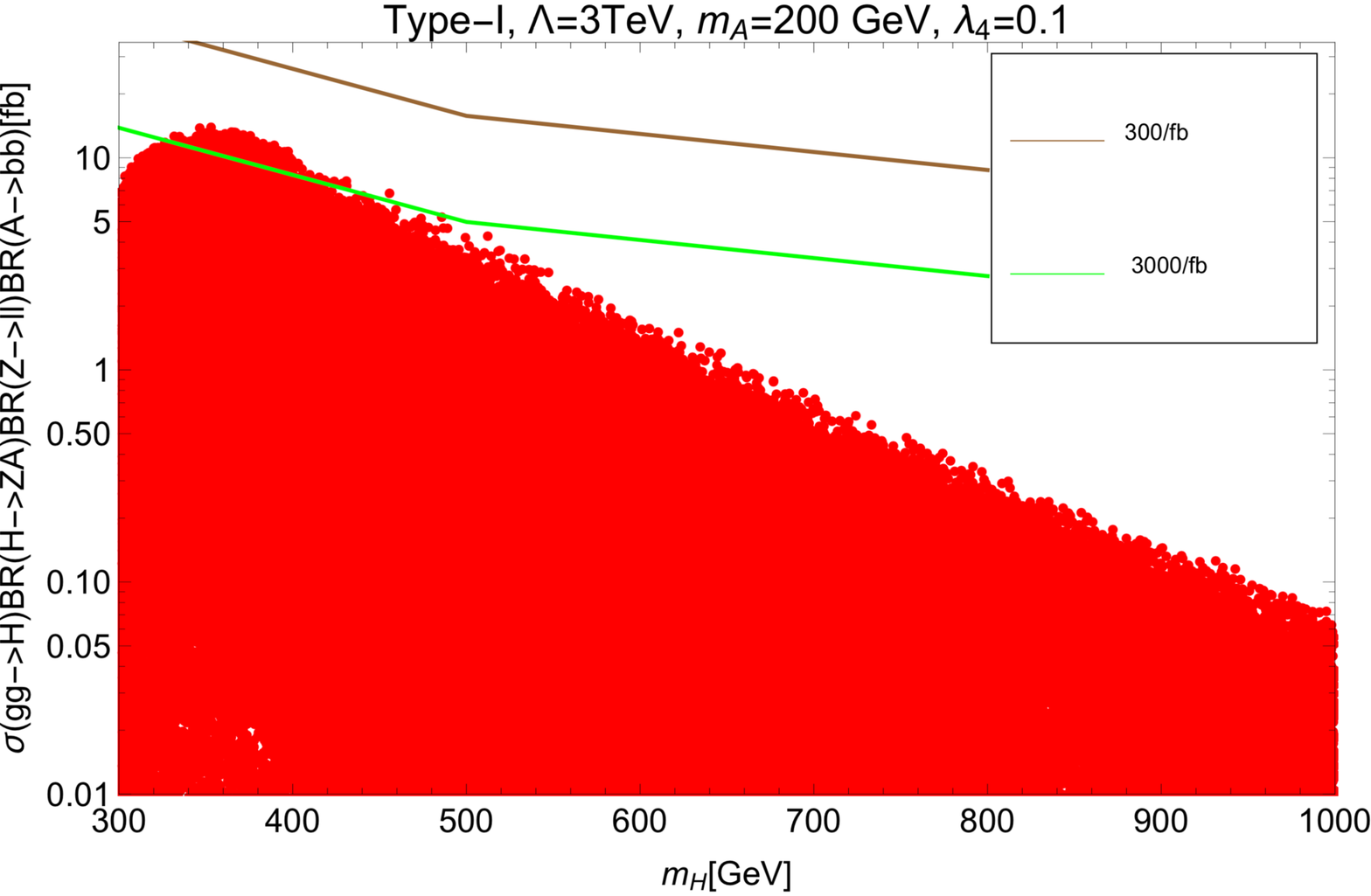}
\hspace{5mm}
\includegraphics[scale=0.25]{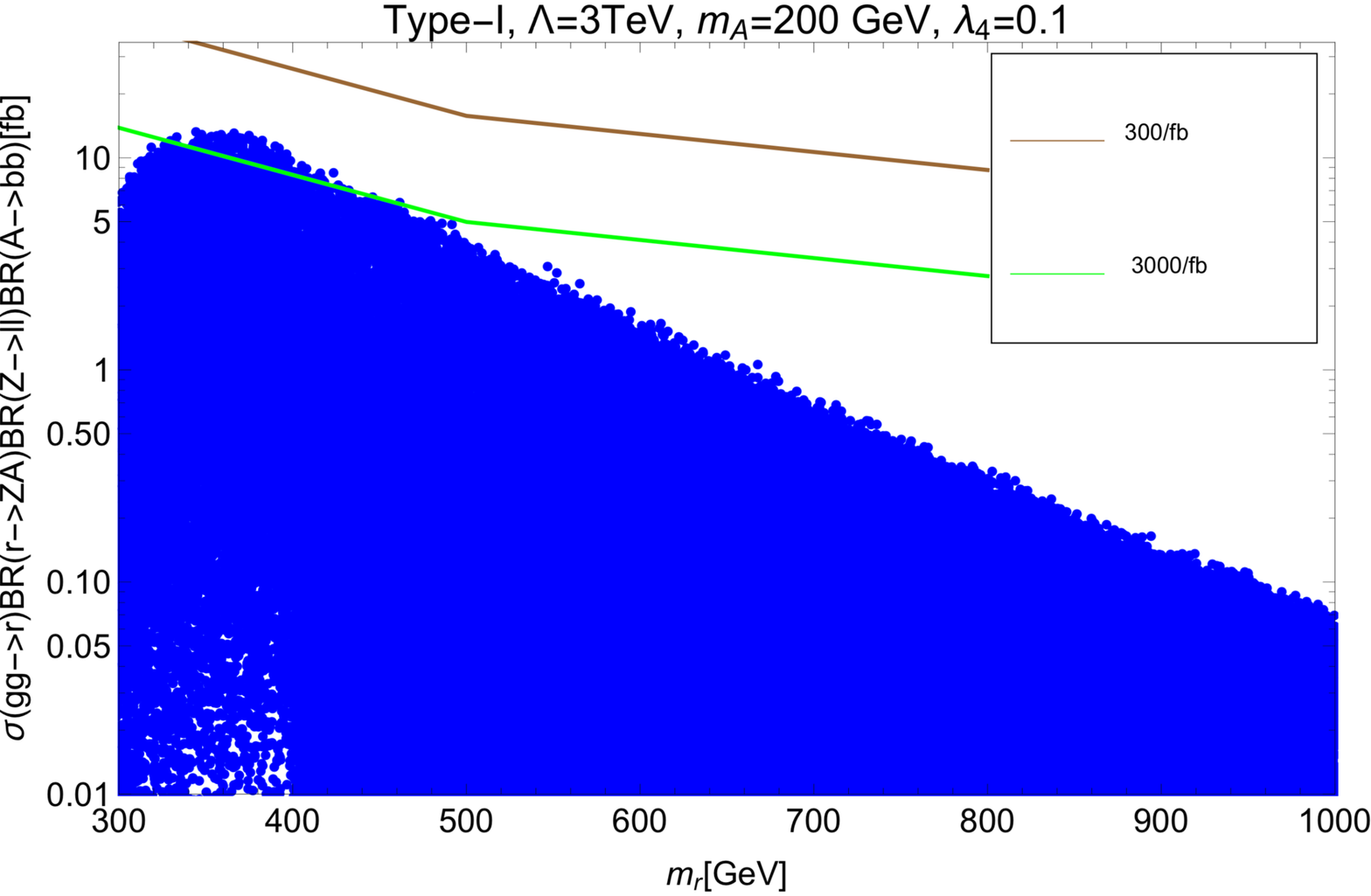}
\hspace{5mm}
\includegraphics[scale=0.25]{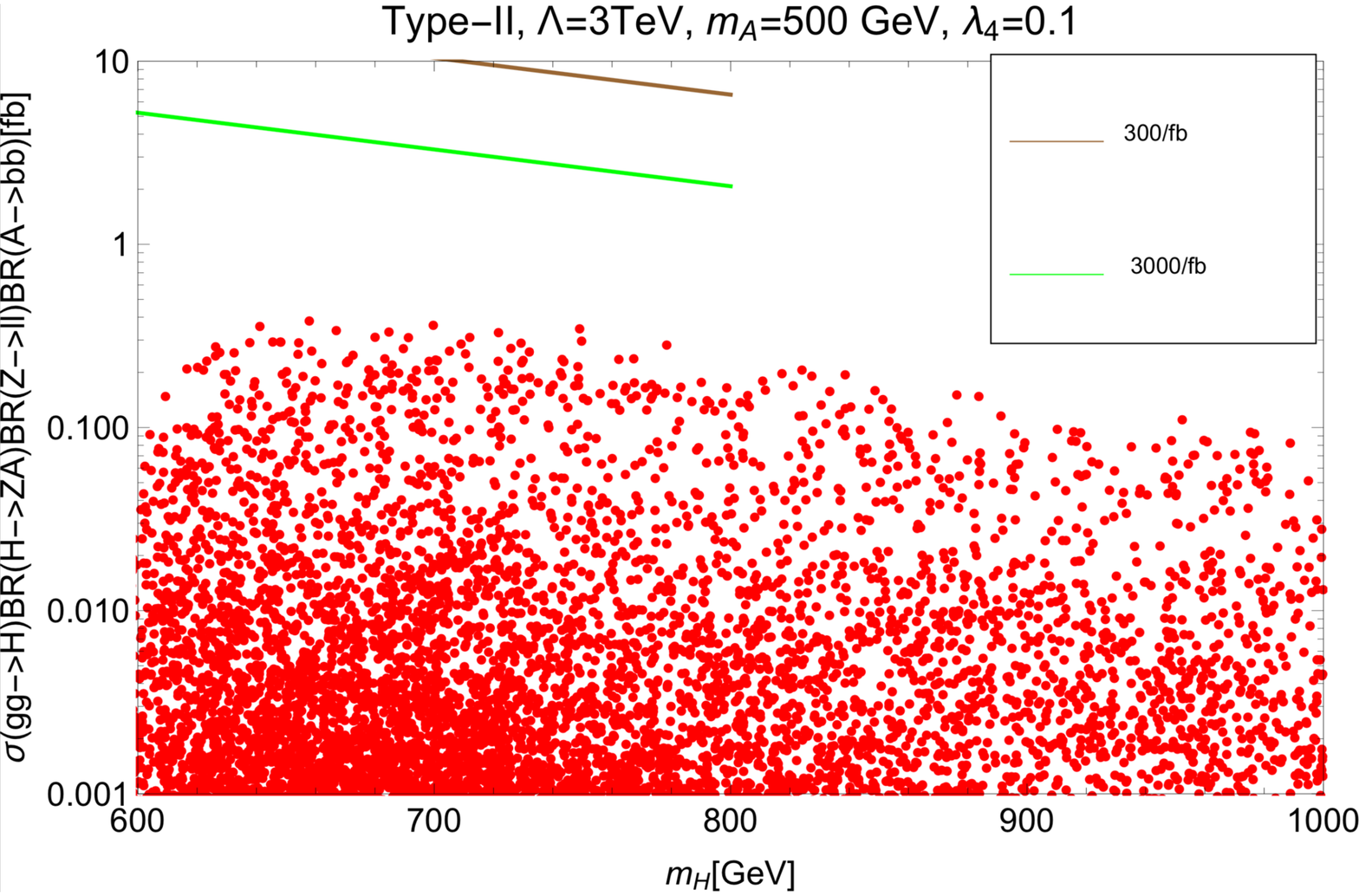}
\hspace{5mm}
\includegraphics[scale=0.25]{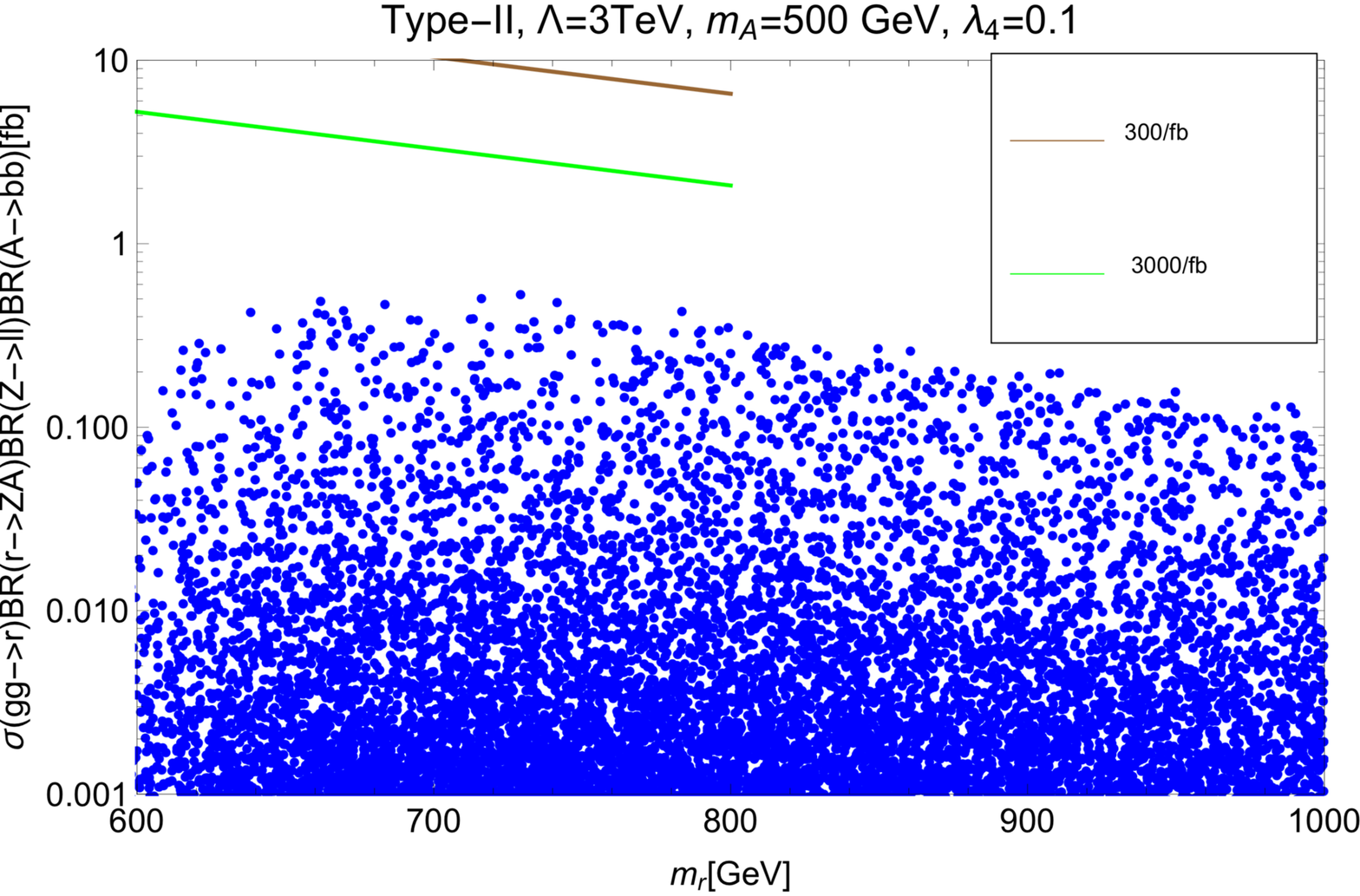}
\caption{The observable $\sigma(gg \rightarrow X \rightarrow ZA) BR(Z \rightarrow l^+ l^-) BR(A \rightarrow b \bar{b})$ as a function of the resonance mass with $X= H$(red), $r$(blue) in the type-I (top) and type-II (bottom) models. We fixed $\Lambda = 3 \TeV$, $m_A= 200 \GeV$($m_A=500 \GeV$) on top (bottom) and $\lambda_4 = 0.1$. Due to the custodial symmetry, the charged scalar mass is identical to the pseudoscalar mass, whose value is given above each figure.   The heavy neutral Higgs (radion) mass is varied from $200$ to $1000$ GeV in the right (left) figures and the values of $\alpha$ and $\beta$ are chosen to be consistent with the constraints of Figure \ref{fig::2hdm_pspace}. The solid lines represent future upper bounds at the LHC} \label{XZA}
\end{figure}

 The $95 \%$ CL upper limits from ATLAS \cite{Aaboud:2018eoy}, after multiplying by $BR(Z\rightarrow l^+ l^-) \approx 0.0336$ \cite{Patrignani:2016xqp}, for $m_A=700\  \GeV$ are shown in Fig. \ref{ZAR}. We have also shown the expected limits for $300 \ {\rm fb}^{-1}$ and $3000\ {\rm fb}^{-1}$  \footnote{Since the limits are background limited, we are assuming in Figs. \ref{ZAR} and \ref{XZA} that the bounds will scale as $1/ \sqrt{N}$.}. It is clear that the LHC will only be able to cover a small range of parameter space, however discovery of the process for $m_H> 400 \ \GeV$ in the near future would rule out the model. In any event the hadronic decay mode ($b \bar{b}$ or $t \bar{t}$) will dominate the pseudoscalar decays.

In figure \ref{XZA} we show the production cross section via gluon fusion of a heavy Higgs boson (red) and a radion (blue)  times the branching fractions  $BR(X \rightarrow ZA)BR(Z \rightarrow l^+ l^-)BR(A \rightarrow b \bar{b})$ as a function of the mass $m_X$ and with $X= H$, $r$ for the type-I (top) and type-II (bottom) models. For type-I model we fixed $m_A = 200 \ \GeV$ and in the type-II, due to lower bounds on the charged Higgs \cite{Misiak:2017bgg}, we fixed $m_A=500 \ \GeV$.

Current upper limits from CMS \cite{Khachatryan:2016are, CMS:2016qxc} are out of the range of the figures. Extrapolations of the expected reach for $300 \ {\rm fb}^{-1}$ and $3000 \ {\rm fb}^{-1}$ are given by the brown and green lines, respectively, in figure \ref{XZA}.

 We can see from figure \ref{XZA} that for this decay our predictions are not in reach for the LHC except at the very edge of the parameter space in the type-I 2HDM.  Note that discovery of this decay mode in the near future would rule out these models. The primary decays of the radion would be into pairs of Higgs bosons or Z's depending on its mass and scalar trilinear coupling. The decays of $H$ might also be into these final states as well as $b \bar{b}$ and  $t \bar{t}$ depending on its mass and scalar trilinear coupling.


\section{Conclusions}

In this work we considered two Higgs doublets coupling to the Ricci scalar in the $\TeV$-brane of an RS model.  Assuming CP-conservation, the inclusion of this term causes kinetic mixing between the CP-even scalars of the 2HDM and the radion field of the RS model.

The most up to date LHC measurements of the signal strengths of the  SM Higgs boson were used to fit the model and the allowed $\cos(\beta - \alpha)$-$\tan\beta$ parameter space for type-I and type-II 2HDM were presented.  

We have discussed two possible ways to differentiate this model from other scenarios with similar scalar states. One possibility is to look at the sum of squared couplings of the scalars to gauge bosons. This model predicts a small deviation of about $0.5\%$ from the SM value which could be measured at a future ILC. The other possibility is to look at the ratio of decay widths to a pair of $b$ quarks and $Z$ bosons for both scalars. Future experiments might distinguish the scalars by determining the value of the mixing angles $\alpha$ and $\beta$.

Throughout this work we have taken the mass of the extra scalars to be in the range of $200 $-$1000 \ \GeV$ and we study the constraints that LHC searches of heavy resonances impose on the amount of mixing. The most stringent bounds arise if we take $\Lambda = 3\  \TeV$ and $m_A= 700\  \GeV$ where a radion is disfavored in the mass range $m_r <780 \  \GeV$ while a heavy Higgs is disfavored in the mass range $300\ \GeV<m_H< 750 \ \GeV$ and $m_H<  250 \ \GeV$ and kinetic mixing for both, radion and Higgs, is constrained to  $-4<K_h,\ K_H<4$. These constraints relax significantly by reducing $m_A$ and increasing the value $\Lambda$.

Finally we showed how improvements of the experimental analysis for the decay topologies $X \rightarrow ZA$ and $A \rightarrow ZX$ where $X= r$ or $H$ could further constrain the parameter space of, or possibly eliminate, the model

\begin{acknowledgments}
This work was supported by the National Science Foundation under Grant PHY-1519644. MM also acknowledges support from CONACYT.  
\end{acknowledgments}

\begin{appendices}

\section{Scalar Couplings After Mixing}  \label{interactions}

The interactions of the physical scalars to SM fields can be obtained by substituting the transformation of equation \eqref{diagonalization} into the unmixed couplings. A summary is given by 
 \begin{equation}
 g_{\phi VV}  = U_{2 \phi }\sin(\beta-\alpha) + U_{3 \phi}\cos(\beta - \alpha) + U_{1 \phi}\gamma \left( 1-3\frac{m_v^2 ky_c}{\Lambda^2}\right) \quad  \phi=r, h, H, 
 \end{equation}

 \begin{equation}
  g_{\phi ff}  = U_{2\phi} \xi_h^f + U_{3\phi} \xi_{H}^f +  U_{1\phi } \gamma (c^f_L - c^f_R), \quad  \phi=r, h, H,
 \end{equation}
 
 \begin{equation}
 g_{\phi gg} = \left(\frac{2\pi}{\alpha_s k y_c} + 7 \right) U_{1\phi }\gamma + \sum_{q} F_q (\xi_h^q U_{2\phi} + \xi_{H}^q U_{3 \phi} + \gamma U_{1\phi})   \quad  \phi=r, h, H.
 \end{equation}
  
The trilinear interactions between scalar eigenstates $r$, $h$, and $H$ are given by 
\begin{equation}
\mathcal{L} \supseteq y_1 r \partial^\mu h \partial_\mu H +y_2 r \Box h H + y_3 r h \Box H  + g_{rhH}rhH,
\end{equation}
where 
\begin{align}  
y_1 = &\frac{2}{v} \gamma   \left\lbrace -6 \gamma  \left[ \text{$\xi_1 $} \sin ( \beta-\alpha )  + \text{$\xi_2 $} \cos (\alpha +\beta ) \right] (U_{11} U_{12} U_{23}+U_{11} U_{13} U_{22}+U_{12} U_{13} U_{21})   \right.\nonumber \\
& -6 \gamma \left[ \text{$\xi_1 $} \cos (\beta -\alpha) + \text{$\xi_2 $} \sin (\alpha +\beta ) \right](U_{11} U_{12} U_{33}+U_{11} U_{13} U_{32}+U_{12} U_{13} U_{31}) \nonumber \\
& +6 \text{$\xi_2 $} U_{11} \left[ \sin (2 \alpha ) (U_{32} U_{33}-U_{22} U_{23})  +\cos (2 \alpha ) (U_{22} U_{33}+U_{23} U_{32}) \right]  +6 \text{$\xi_1 $} U_{11} (U_{22} U_{23} \nonumber \\
& +U_{32} U_{33}) \left.-U_{11} U_{22} U_{23}-U_{11} U_{32} U_{33}+U_{12} U_{21} U_{23}+U_{12} U_{31} U_{33}+U_{13} U_{21} U_{22} \right. \nonumber \\
&+\left.U_{13} U_{31} U_{32}  \right\rbrace ,
\end{align}
\begin{align}
y_2 = & \frac{2}{v} \gamma  \left\lbrace 3 U_{11} (U_{22} U_{23} + U_{32} U_{33}) \xi_1 + U_{13}(U_{21} U_{22} + U_{31} U_{32}) (1 + 3 \xi_1)  \right. \nonumber  \\
        &+  3 (U_{13} U_{22} U_{31} + U_{13} U_{21} U_{32} + U_{11} U_{23} U_{32} +   U_{11} U_{22} U_{33}) \xi_2 \cos(2 \alpha)  \nonumber \\
        &- 6 U_{13} (U_{12} U_{31} +  2 U_{11} U_{32}) \gamma \xi_1 \cos(\alpha - \beta) - 6 U_{13} (U_{12}  U_{21}       + 2 U_{11} U_{22}) \gamma \xi_2 \cos(\alpha+ \beta) \nonumber \\      
        & +  3 (-U_{13} U_{21} U_{22} - U_{11} U_{22} U_{23} + U_{13} U_{31} U_{32} +  U_{11} U_{32} U_{33}) \xi_2 \sin(2 \alpha) \nonumber  \\
      & +  6 U_{13} (U_{12} U_{21} + 2 U_{11} U_{22}) \gamma \xi_1 \sin(\alpha - \beta) \nonumber \\ 
      &\left.-  6 U_{13} (U_{12} U_{31} +  2 U_{11} U_{32}) \gamma \xi_2 \sin(\alpha + \beta) \right\rbrace,
\end{align}

\begin{align}
y_3 = &\frac{2 \gamma}{v} (U_{12} (U_{21} U_{23} + U_{31} U_{33}) + 3 (U_{13} U_{21} U_{22} + U_{11} U_{22} U_{23} + U_{13} U_{31} U_{32} \nonumber \\
          & + U_{11} U_{32} U_{33}) \xi_1 +    3 (U_{13} U_{22} U_{31} + U_{13} U_{21} U_{32} + U_{11} U_{23} U_{32}  \nonumber \\
          & + U_{11} U_{22} U_{33}) \xi_2 \cos(2 \alpha) -  6 (U_{12} U_{13} U_{31} + U_{11} U_{13} U_{32} \nonumber \\
          &+   U_{11} U_{12} U_{33}) \gamma \xi_1 \cos(\beta- \alpha ) -  6 (U_{12} U_{13} U_{21} + U_{11} U_{13} U_{22}  \nonumber \\
          &  +  U_{11} U_{12} U_{23}) \gamma \xi_2 \cos(\alpha + \beta)+ 3 (-U_{13} U_{21} U_{22} - U_{11} U_{22} U_{23} + U_{13} U_{31} U_{32}   \nonumber \\
          & +  U_{11} U_{32} U_{33}) \xi_2 \sin(2 \alpha) +  6 (U_{12} U_{13} U_{21} + U_{11} U_{13} U_{22} \nonumber \\
           &+   U_{11} U_{12} U_{23}) \gamma \xi_1 \sin(\alpha - \beta) - 6 (U_{12} U_{13} U_{31} + U_{11} U_{13} U_{32}\nonumber \\
           & + U_{11} U_{12} U_{33}) \gamma \xi_2 \sin(\alpha + \beta)).
\end{align}

The tree-level coupling has two contributions, one from the trace of the energy-momentum tensor and another one from the 2HDM potential, i.e. $g_{rhH} = g_{rhH}^{trace} + g_{rhH}^{2HDM}$ where 
\begin{align}
g_{rhH}^{2HDM} =  &\frac{1}{2v}(\cos\beta (U_{33} \cos\alpha - U_{23} \sin\alpha) (U_{21} \cos\alpha +  U_{31} \sin\alpha) (U_{22} \cos\alpha   \nonumber \\
                                &+  U_{32} \sin\alpha) (m_A^2- v^2\lambda_4- (m_h^2 - m_H^2) \cos\alpha \csc\beta \sec\beta \sin\alpha)   \nonumber \\
                                &+ \cos\beta (U_{32} \cos\alpha - U_{22} \sin\alpha) (U_{21} \cos\alpha+ U_{31} \sin\alpha) (U_{23} \cos\alpha  \nonumber \\
                                &+  U_{33} \sin\alpha) (m_A^2 - v^2\lambda_4 - (m_h^2 -   m_H^2) \cos\alpha \csc\beta \sec\beta \sin\alpha)  \nonumber \\
                                & +  \cos\beta (U_{31} \cos\alpha - U_{21} \sin\alpha) (U_{22} \cos\alpha +  U_{32} \sin\alpha) (U_{23} \cos\alpha   \nonumber \\
                                  &+  U_{33} \sin\alpha) (m_A^2- v^2\lambda_4- (m_h^2 - m_H^2) \cos\alpha \csc\beta \sec\beta \sin\alpha)     \nonumber\\
                                  &+ (U_{32} \cos\alpha - U_{22} \sin\alpha) (U_{33} \cos\alpha -  U_{23} \sin\alpha) (U_{21} \cos\alpha   \nonumber \\
                                   & +   U_{31} \sin\alpha)(m_A^2- v^2\lambda_4- (m_h^2 - m_H^2) \cos\alpha \csc\beta \sec\beta \sin\alpha) \sin\beta    \nonumber \\
                                    &+ (U_{31} \cos\alpha - U_{21} \sin\alpha)(U_{33} \cos\alpha - U_{23} \sin\alpha) (U_{22} \cos\alpha +U_{32} \sin\alpha) (m_A^2     \nonumber \\
                                    &- v^2\lambda_4 - (m_h^2 - m_H^2) \cos\alpha \csc\beta  \sec\beta \sin\alpha) \sin\beta + (U_{31} \cos\alpha  \nonumber \\
                                    &  -  U_{21} \sin\alpha) (U_{32} \cos\alpha- U_{22} \sin\alpha)(U_{23} \cos\alpha + U_{33} \sin\alpha) (m_A^2 - v^2\lambda_4 \nonumber \\
                                &- (m_h^2 - m_H^2) \cos\alpha \csc\beta \sec\beta \sin\alpha) \sin\beta -  6 (U_{21} \cos\alpha   \nonumber \\
                                &+ U_{31} \sin\alpha) (U_{22} \cos\alpha+ U_{32}  \sin\alpha) (U_{23} \cos\alpha +  U_{33} \sin\alpha) ((m_A^2         \nonumber \\
                                &+ v^2\lambda_4) \cot^2 \beta -\csc^2 \beta  (m_h^2 \cos^2 \alpha + m_H^2 \sin^2\alpha))  \sin{\beta} +  6 \sec\beta (U_{31} \cos\alpha     \nonumber \\
                                &- U_{21} \sin\alpha) (U_{32} \cos\alpha -  U_{22} \sin\alpha) (U_{33} \cos\alpha - U_{23} \sin\alpha) (m_H^2 \cos\alpha^2 +  m_h^2 \sin\alpha^2   \nonumber \\
                                &- (m_A^2 + v^2\lambda_4) \sin^2\beta)+2 v^2\lambda_4 ((U_{23} U_{32} + U_{22} U_{33}) \cos(2\alpha) + (-U_{22} U_{23}  \nonumber \\
                                &+ U_{32} U_{33}) \sin(2\alpha)) (U_{21} \cos(\alpha + \beta)+U_{31} \sin(\alpha + \beta))+ 2 v^2\lambda_4 ((U_{23} U_{31} \nonumber \\
                                & + U_{21} U_{33}) \cos(2\alpha)+ (-U_{21} U_{23}  + U_{31} U_{33}) \sin(2\alpha)) (U_{22} \cos(\alpha + \beta)   \nonumber \\
                                & + U_{32}\sin(\alpha + \beta)) + 2 v^2\lambda_4 ((U_{22} U_{31} + U_{21} U_{32}) \cos(2\alpha) (-U_{21} U_{22}  \nonumber \\ 
                                &+ U_{31} U_{32}) \sin(2\alpha ) (U_{23} \cos(\alpha + \beta) + U_{33} \sin(\alpha + \beta))), 
 \end{align}
\begin{align}
g_{rhH}^{trace} = & 4 \frac{\gamma}{v} (m_h^2 (U_{13} U_{21} U_{22} + U_{12} U_{21} U_{23} + U_{11} U_{22} U_{23}) \nonumber \\
&+  m_H^2 (U_{13} U_{31} U_{32} + U_{12} U_{31} U_{33} + U_{11} U_{32} U_{33})).
\end{align}
The other interactions like $rhh$, $rHH$, etc. can be similarly obtained and are not illustrated here.

\newpage

\section{LHC Data}  \label{appendixB}

\begin{center}
\begin{table}[h]
\centering
\renewcommand*{\arraystretch}{1.2}
\scalebox{1}{
\begin{tabular}{  | c   | c  | c  |  }   
\hline  
Decay & Production & Measured Signal Strength $R_m$\\ 
\hline 
$\gamma \gamma$    &
\begin{tabular}{c}
 ggF+tth\\[0.15cm]
VBF +Vh\\[0.15cm]
ggF \\[0.15cm]
VBF \\[0.15cm]
Vh \\[0.15cm]
\end{tabular}
   & \begin{tabular}{c}
  $1.19^{+0.20}_{-0.18}$ [CMS]  \cite{CMS:2017rli}\\[0.15cm]
$1.01^{+0.57}_{-0.51}$ [CMS]  \cite{CMS:2017rli} \\[0.15cm]
$0.8^{+0.19}_{-0.18}$ [ATLAS] \cite{ATLAS:2017myr} \\[0.15cm]
$2.1^{+0.6}_{-0.6}$  [ATLAS] \cite{ATLAS:2017myr}\\[0.15cm]
$0.7^{+0.9}_{-0.8}$ [ATLAS]  \cite{ATLAS:2017myr} \\[0.15cm]
\end{tabular}   \\  
\hline
WW* & \begin{tabular}{c}
ggF  \\[0.15cm]
VBF \\[0.15cm]
ggF \\[0.15cm]
VBF\\[0.15cm]
Wh\\[0.15cm]
\end{tabular} & \begin{tabular}{c}
$1.02^{+0.29}_{-0.26}$ [ATLAS] \cite{ATLAS:2014aga} \\[0.15cm]
$1.27^{+0.53}_{-0.45}$  [ATLAS] \cite{ATLAS:2014aga}\\[0.15cm]
$0.76 \pm 0.21$  [CMS] \cite{Chatrchyan:2013iaa}\\[0.15cm]
$1.7^{+1.1}_{-0.9}$ [ATLAS] \cite{ATLAS:2016gld}\\[0.15cm]
$3.2^{+4.4}_{-4.2}$ [ATLAS] \cite{ATLAS:2016gld}\\[0.15cm]
\end{tabular}\\ 
\hline
ZZ* & \begin{tabular}{c}
ggF \\[0.1cm]
VBF + Vh\\[0.1cm]
ggF\\[0.1cm]
VBF\\[0.1cm]
\end{tabular} & \begin{tabular}{c}
$1.7^{+0.5}_{-0.4}$ [ATLAS] \cite{Aad:2014eva} \\[0.15cm]
$0.3^{+1.6}_{-0.9}$ [ATLAS] \cite{Aad:2014eva}\\[0.15cm]
$1.20^{+0.35}_{-0.31} $ [CMS] \cite{CMS:2017jkd}\\[0.15cm]
$0.00^{+1.37}_{-0.00}$ [CMS] \cite{CMS:2017jkd}\\[0.15cm]
\end{tabular} \\ [0.15cm]
\hline
bb  & \begin{tabular}{c}
VBF \\[0.15cm]
Vh \\[0.15cm]
Vh 
\end{tabular} &\begin{tabular}{c}
$-3.7^{+2.4}_{-2.5}$ [CMS] \cite{CMS:2016mmc} \\[0.15cm]
$1.20^{+0.42}_{-0.36}$ [ATLAS] \cite{Aaboud:2017xsd} \\[0.15cm]
$1.2 \pm 0.4 $ [CMS] \cite{Sirunyan:2017elk}
\end{tabular}\\ [0.15cm]
\hline
$\tau \tau$  & \begin{tabular}{c}
VBF \\[0.15cm]
ggF\\[0.15 cm]
VBF + Vh \\[0.15 cm]
WH \\[0.15 cm]
tth
\end{tabular} &  \begin{tabular}{c}
$1.2 \pm 0.4$ [ATLAS] \cite{Aad:2015vsa} \\[0.15cm]
$2.0^{+1.5}_{-1.2}$ [ATLAS] \cite{Aad:2015gba}\\[0.15 cm]
$1.24^{+0.59}_{-0.54}$ [ATLAS] \cite{Aad:2015gba}\\[0.15 cm]
$2.3 \pm 1.6$ [ATLAS] \cite{Aad:2015zrx} \\[0.15 cm]
$1.5^{+1.2}_{-1.0}$ [ATLAS] \cite{Aaboud:2017jvq}
\end{tabular}\\
\hline  
\end{tabular}}\caption{Measured Higgs Signal Strengths}\label{dedq}
\end{table}
\end{center}

\section{Contributions to S and T from the scalar sector.} \label{STapp}

The relevant contributions are below.  (b) and (c) refer to the diagrams of Fig. \ref{diagrams}

  In any new physics model (NP), any field that couples to the SM gauge bosons $\gamma$, $W^{\pm}$ and $Z$ will contribute to their vacuum polarization diagrams and will generate the tensor structure
\begin{equation}
\Pi_{VV}^{\mu \nu} = \Pi_{VV}(p^2)\eta^{\mu \nu} + \tilde{\Pi}_{VV}(p^2)p^\mu p^\nu
\end{equation}
where $p^\mu$ is the 4-momentum of the gauge boson.

These corrections can be parametrized by the oblique parameters S and T \cite{Peskin:1991sw}
\begin{equation}
S \equiv \frac{4 c_W^2 s_W^2}{\alpha} \left( \frac{\Pi_{ZZ}(m_Z^2)}{m_Z^2} - \frac{\Pi_{ZZ}(0)}{m_Z^2}    - \frac{c_W^2 - s_W^2}{c_W s_W } \frac{\Pi_{Z \gamma}(m_Z^2)}{m_Z^2} - \frac{\Pi_{\gamma \gamma}(m_Z^2)}{m_Z^2}    \right)
\end{equation}
\begin{equation}
\alpha T \equiv \frac{\Pi_{WW}(0)}{m_W^2} - \frac{\Pi_{ZZ}(0)}{m_Z^2}  
\end{equation}
which are defined relative to the SM contributions so that $S=T=0$ in the SM for some reference value of the Higgs mass.

\begin{align}
\Pi_{ZZ}^{new(b)}(m_Z^2)  = \frac{g^2}{16 \pi^2 c_W^2} \sum_{\phi= r,h,H} & \left\lbrace \left[ \gamma U_{r\phi} + s_{\beta-\alpha} U_{h\phi} +c_{\beta-\alpha} U_{H\phi} \right]^2 B_{22}(m_Z^2; m_Z^2, m_\phi^2)  \right.   \nonumber \\
                                       &  \left. + \left[  s_{\beta-\alpha} U_{H\phi} -c_{\beta-\alpha} U_{h\phi} \right]^2 B_{22}(m_Z^2; m_A^2, m_\phi^2)  \right\rbrace 
\end{align}

\begin{align}
\Pi_{ZZ}^{new(c)}(m_Z^2)  =- \frac{g^2 m_Z^2}{16 \pi^2 c_W^2} \sum_{\phi= r,h,H} & \left[ c_{\beta-\alpha} U_{H\phi} + s_{\beta - \alpha}U_{h\phi} - \gamma \left( 1 - 3 \frac{m_Z^2 k y_c}{\Lambda^2} \right)U_{r\phi}  \right] ^2  \nonumber \\
& \times B_0(m_Z^2; m_Z^2, m_\phi^2)
\end{align}

\begin{align}
\Pi_{WW}^{new(b)}(0)  = \frac{g^2 }{16 \pi^2 } \sum_{\phi= r,h,H} &    \left\lbrace \left[ c_{\beta- \alpha}U_{H \phi} + s_{\beta- \alpha} U_{h \phi} - \gamma U_{r \phi} \right]^2 B_{22}(0;m_W^2,m_\phi^2) \right. \nonumber \\
& \left.  + \left[ c_{\beta-\alpha} U_{h \phi} + s_{\beta-\alpha} U_{H \phi} \right]^2 B_{22}( 0 ; m_A^2,m_\phi^2)  \right\rbrace
\end{align}

\begin{align}
\Pi_{WW}^{new(c)}(0)  =- \frac{g^2 m_Z^2}{16 \pi^2 } \sum_{\phi= r,h,H} & \left[ c_{\beta-\alpha} U_{H\phi} + s_{\beta - \alpha}U_{h\phi} - \gamma \left( 1 - 3 \frac{m_W^2 k y_c}{\Lambda^2} \right)U_{r\phi}  \right] ^2  \nonumber \\
& \times B_0(0; m_W^2, m_\phi^2)
\end{align}

\end{appendices}

\providecommand{\href}[2]{#2}\begingroup\raggedright 

\endgroup

\end{document}